\documentclass[11pt]{article}
\usepackage{amssymb}
\usepackage{mathtools}
\usepackage{amsmath}
\usepackage{amstext}
\usepackage{graphicx,epsfig}
\usepackage{epsfig}
\usepackage{verbatim} 
\usepackage{caption}
\usepackage{fancybox}
\usepackage{slashed}
\usepackage{color}
\usepackage{ulem}
\usepackage{enumitem}
\usepackage{subfigure}
\usepackage{bbm}
\usepackage{parskip}
\usepackage{dsfont}
\usepackage{tabu}
\usepackage[numbers,sort&compress]{natbib}

\newcommand{\Comment}[1]{{}}
\definecolor{darkblue}{rgb}{0.15,0.35,0.55}
\definecolor{reddish}{rgb}{0.65, 0.2, 0.2}
\usepackage[linktocpage=true]{hyperref}
\hypersetup{
colorlinks=true,
citecolor=darkblue,
linkcolor=reddish,
urlcolor=darkblue,
pdfauthor={},
pdftitle={},
pdfsubject={}
}

\newcommand{\be}{\begin{equation}}
\newcommand{\ee}{\end{equation}}
\newcommand{\bea}{\begin{align}}
\newcommand{\eea}{\end{align}}

\newcommand{\rd}{{\rm d}}

\newcommand{\Tr}{\, {\rm Tr} \, }

\newcommand*\classicalEHvertex{\includegraphics[scale=.025]{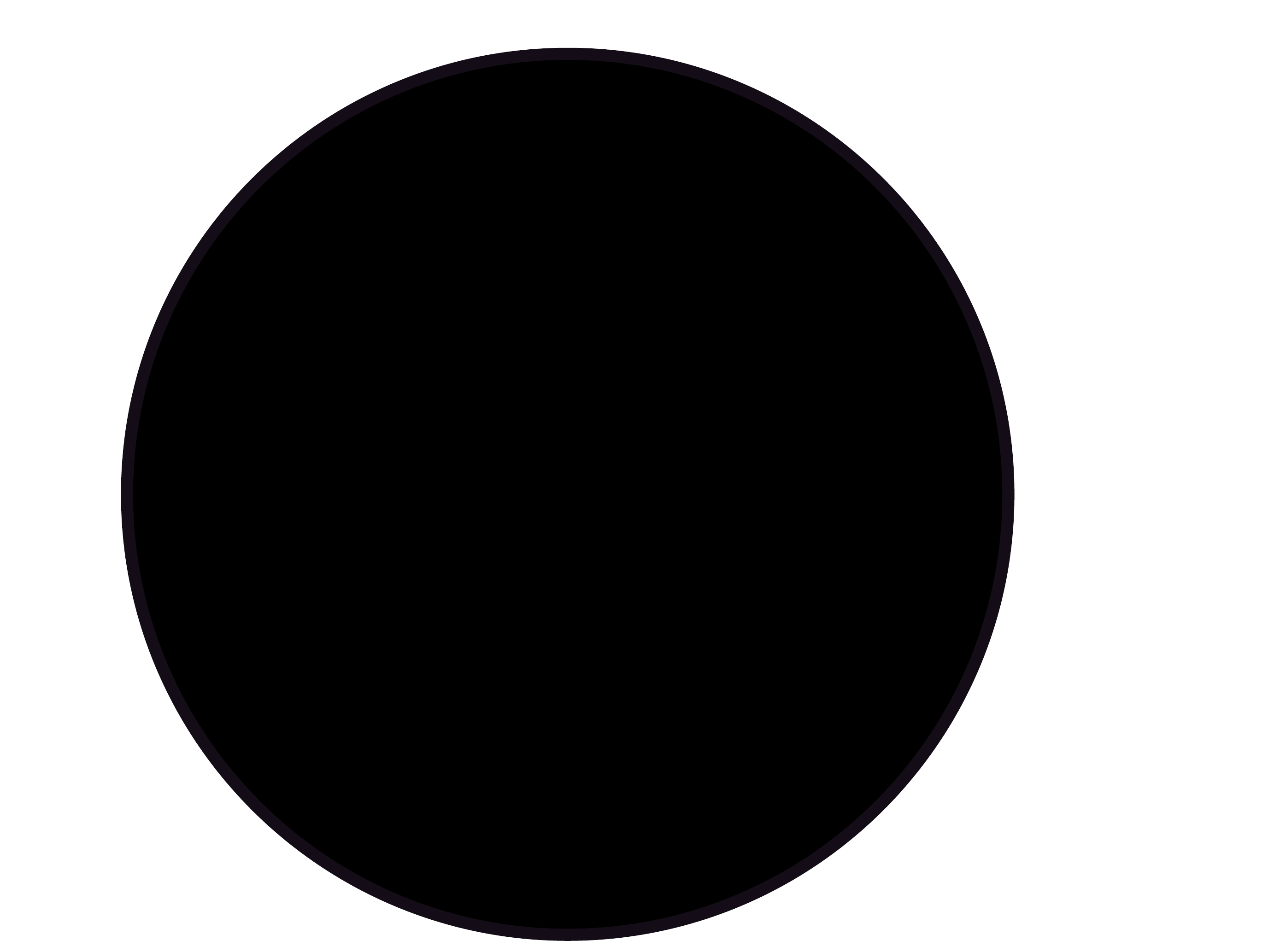}}
\newcommand*\classicalFFvertex{\includegraphics[scale=.025]{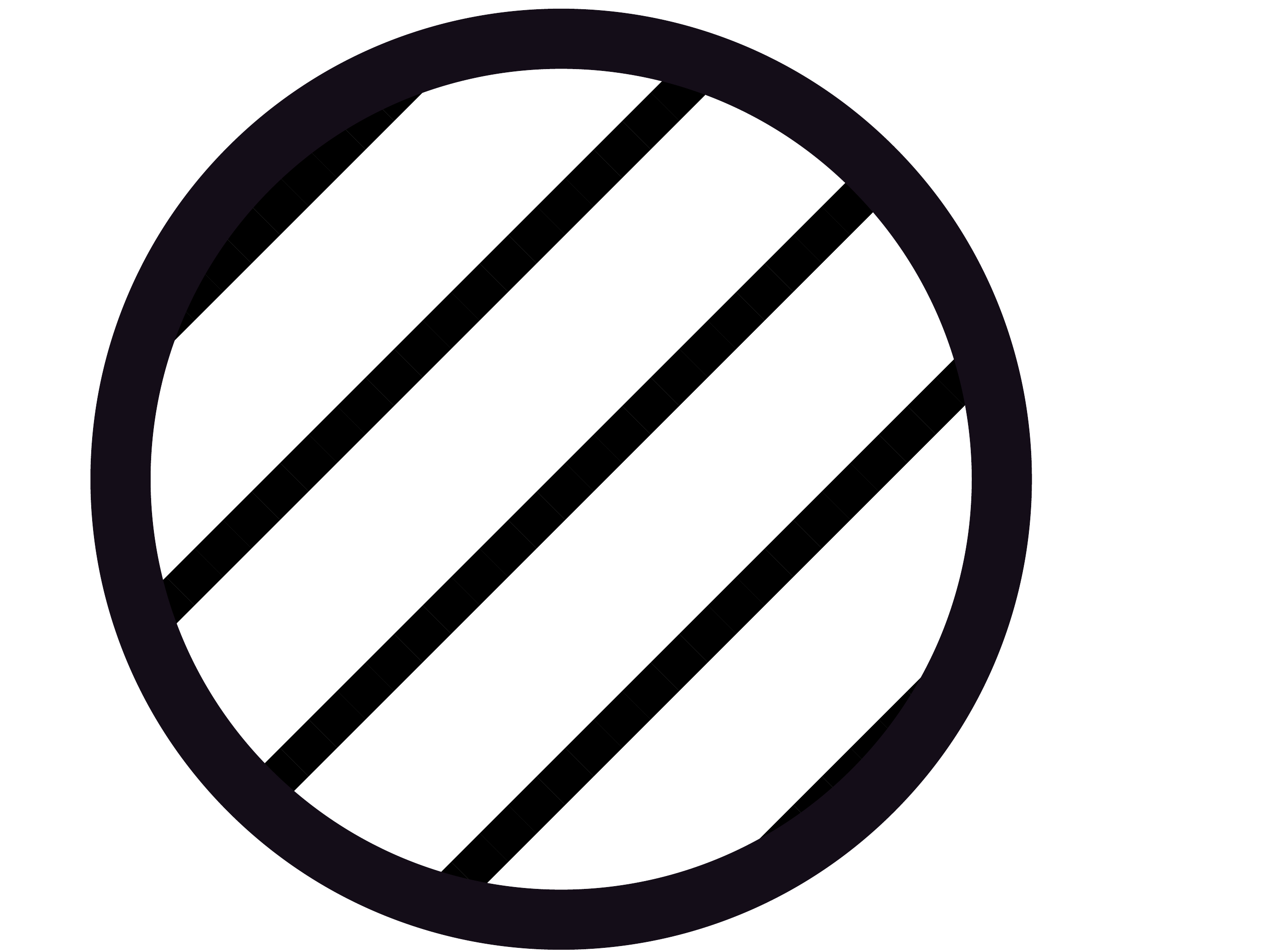}}
\newcommand*\photonline{\includegraphics[scale=.025]{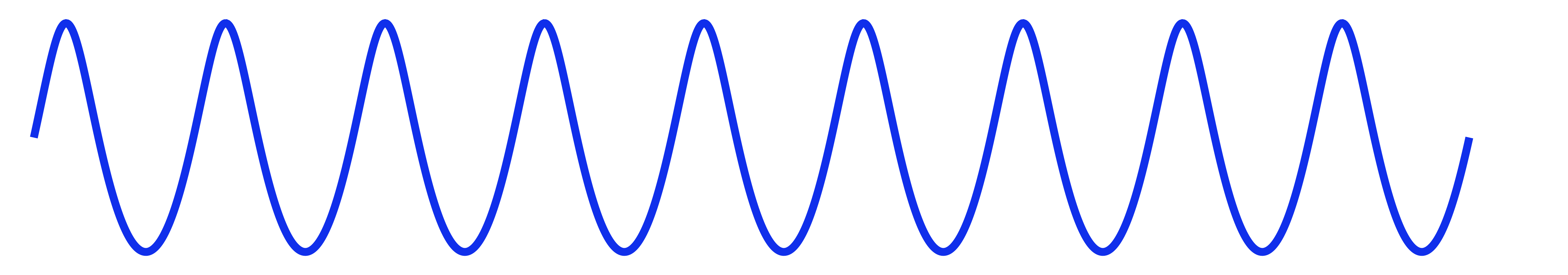}}
\newcommand*\gravitonline{\includegraphics[scale=.025]{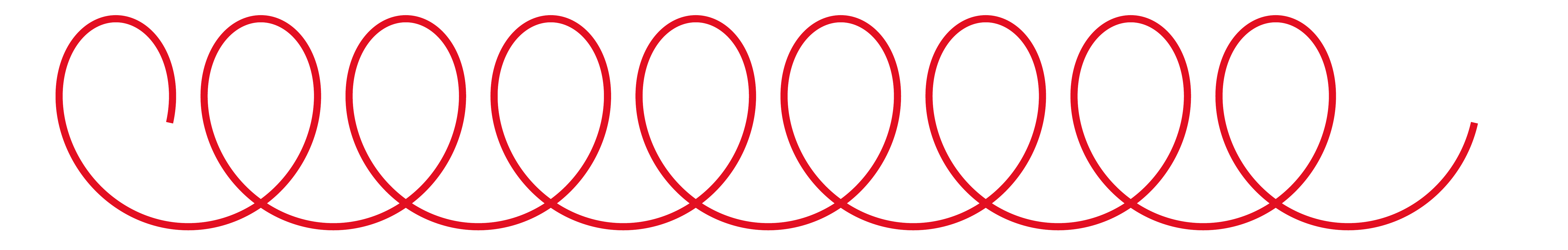}}
\newcommand*\quantumgravityvertex{\includegraphics[scale=.025]{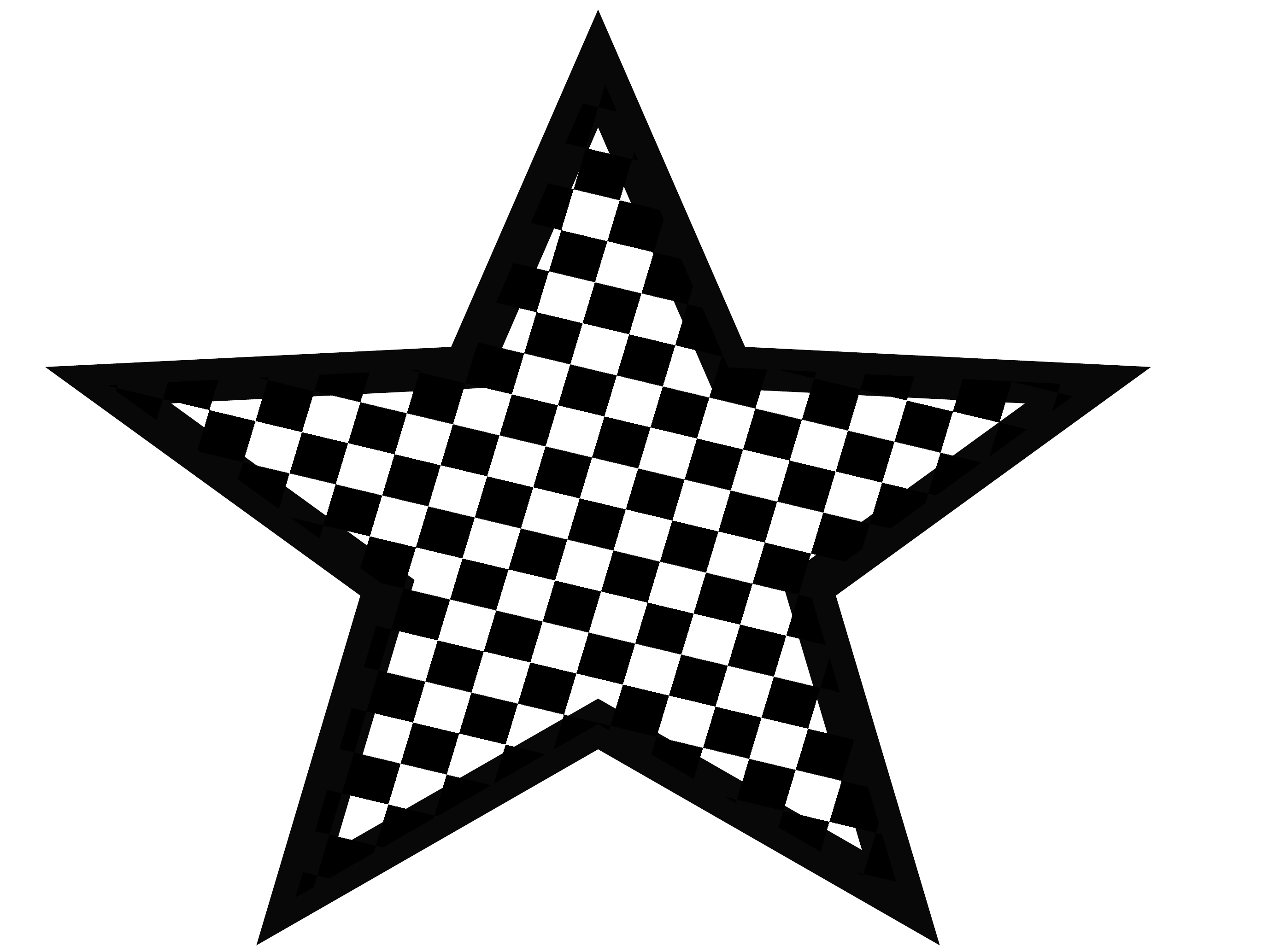}}

\newcommand{\nn}{ \nonumber\\}

\title{}
\author{}

\numberwithin{equation}{section}

\begin{document}

~
\vspace{1truecm}
\renewcommand{\thefootnote}{\fnsymbol{footnote}}
\begin{center}
{\huge \bf{Superluminality, Black Holes and EFT}}
\end{center} 

\vspace{1truecm}
\thispagestyle{empty}
\centerline{\Large Garrett Goon,${}^{\rm a,}$\footnote{\href{mailto:gg399@cam.ac.uk}{\tt gg399@cam.ac.uk}} Kurt Hinterbichler,${}^{\rm b,}$\footnote{\href{mailto:kurt.hinterbichler@case.edu}{\tt kurt.hinterbichler@case.edu}}}
\vspace{.7cm}

\centerline{\it ${}^{\rm a}$Department of Applied Mathematics and Theoretical Physics}
\centerline{\it Cambridge University, Cambridge, CB3 0WA, UK}

\vspace{.3cm}

\centerline{\it ${}^{\rm b}$CERCA, Department of Physics, Case Western Reserve University, }
\centerline{\it 10900 Euclid Ave, Cleveland, OH 44106, USA}

\vspace{.5cm}
\begin{abstract}
\vspace{.03cm}
\noindent
Under the assumption that a UV theory does not display superluminal behavior, we ask what constraints on superluminality are satisfied in the effective field theory (EFT).  We study two examples of effective theories: quantum electrodynamics (QED) coupled to gravity after the electron is integrated out, and the flat-space galileon.  The first is realized in nature, the second is more speculative, but they both exhibit apparent superluminality around non-trivial backgrounds.  In the QED case, we attempt, and fail, to find backgrounds for which the superluminal signal advance can be made larger than the putative resolving power of the EFT.  In contrast, in the galileon case it is easy to find such backgrounds, indicating that if the UV completion of the galileon is (sub)luminal, quantum corrections must become important at distance scales of order the Vainshtein radius of the background configuration, much larger than the naive EFT strong coupling distance scale.  Such corrections would be reminiscent of the non-perturbative Schwarzschild scale quantum effects that are expected to resolve the black hole information problem.  Finally, a byproduct of our analysis is a calculation of how perturbative quantum effects alter charged Reissner-Nordstrom black holes. 

\end{abstract}

\newpage

\tableofcontents

\newpage
\renewcommand*{\thefootnote}{\arabic{footnote}}
\setcounter{footnote}{0}

\section{Introduction and Summary}

It is expected that a reasonable physical theory should pass a few baseline ``consistency" tests.  One often invoked criteria is freedom from superluminalities; no signal should travel with a velocity exceeding the speed of light.  Nevertheless, there exist interesting effective field theories (EFTs), including some which we know to be realized in nature, which display apparent superluminal behavior.  No true consistency condition should rule out a theory realized in nature, so if freedom from superluminalities is indeed such a condition, the effect must be spurious, i.e.$\!\,$ outside the regime of validity of the theory.
The goal of this paper to gain a better understanding of when, or whether, superluminality can be acceptable in the context of an EFT.

Though often touted as a failure of ``consistency," or as ``acausality," one should keep in mind that superluminality does not necessarily imply closed time-like curves (time machines) \cite{Babichev:2007dw,Geroch:2010da,Burrage:2011cr,Papallo:2015rna}, and even closed time-like curves do not necessarily imply inconsistency \cite{PhysRevD.42.1915}.  Nevertheless, we may still proceed under the conservative assumption that a fundamental UV theory should not allow superluminal signaling, an assumption nature has not yet shown us a violation of, and ask what this implies for the effective theory.  
In this paper, this topic is studied in the context of two specific EFTs, one realized in nature and the other speculative.  

Our example realized in nature will be quantum electrodynamics (QED) coupled to gravity.  
The UV\footnote{UV here means valid up to the Planck scale, not truly UV, but we know this must be UV completed in some way, since it's realized in nature.} action is a minimally coupled Dirac fermion\footnote{Of course, precision tests confirm to high accuracy the $\{A_{\mu},\psi\}$ sector of the theory and the classic tests of GR confirm the Einstein-Hilbert term, but little is known about possible non-minimal couplings between $\{A_{\mu},\psi\}$ and $g_{\mu\nu}$ and other higher order interactions. We assume these are negligible in the UV action.}
\begin{align}
S_{\rm QED}&=\int\rd^{4}x\sqrt{-g}\, \left [\frac{M_{p}^{2}}{2}R-\frac{1}{4e^{2}}F_{\mu\nu}^{2}+\bar{\psi}\left (i\slashed D-m_{e}\right )\psi\right ]\ , \label{QEDfullintroa}
\end{align}
where $m_{e}$ is the mass of the fermion.
Integrating out the fermion generates an EFT for a photon which is non-minimally coupled to gravity:
\begin{align}
S&=\int\rd^{4}x\sqrt{-g}\, \left [\frac{M_{p}^{2}}{2}R-\frac{1}{4e^{2}}F_{\mu\nu}^{2}+\frac{1}{4\pi^{2}}\frac{1}{360 m_{e}^{2}}R_{\mu\nu\rho\sigma}F^{\mu\nu}F^{\rho\sigma}+\ldots\right ]\ . \label{QEDEFTIntro}
\end{align}
The higher derivative operators are suppressed by the electron mass $m_e$, corresponding to the strong coupling distance scale $\sim m_{e}^{-1}$.

\begin{figure}[h!] 
  \captionsetup{width=0.9\textwidth}
   \centering
     \includegraphics[width=4in]{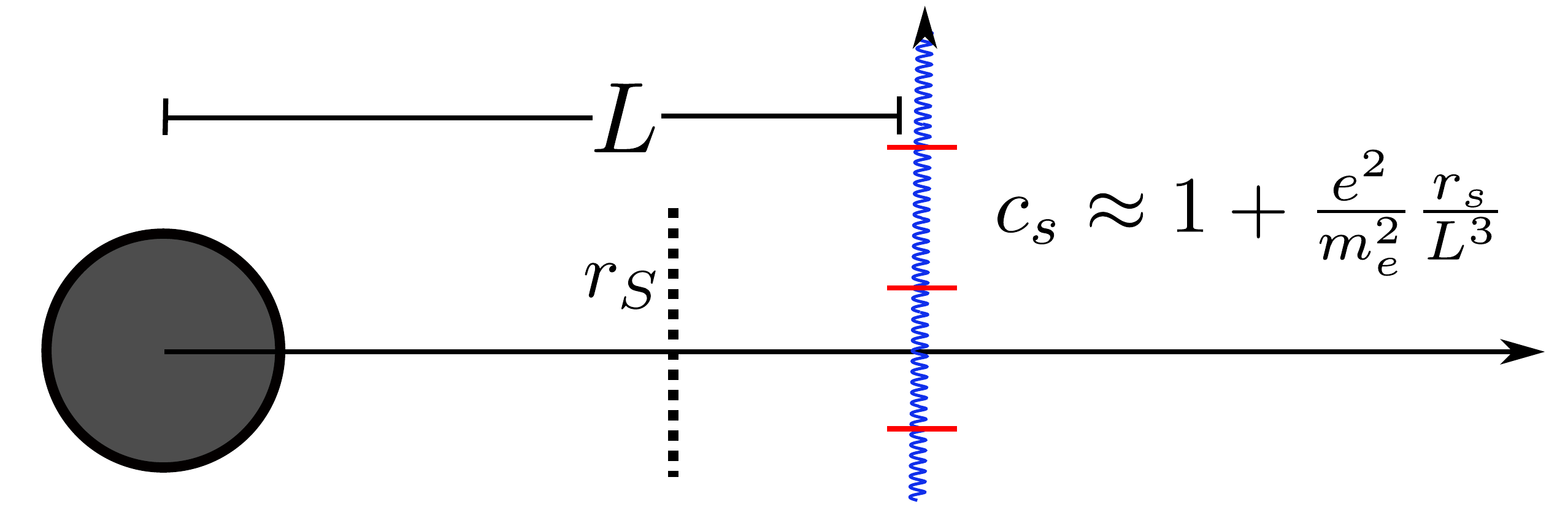}
   \caption{Sketch of the Drummond-Hathrell problem \cite{Drummond:1979pp}.  A photon passes a distance $L$ from a Schwarzschild BH of radius $r_{s}$.  If the polarization is pointing radially outwards, as indicated by the red lines, the EFT \eqref{QEDEFTIntro} gives a superluminal speed.}
   \label{fig:DrummondHathrell}
\end{figure}

These derivative couplings can alter photon (and graviton) propagation on non-trivial backgrounds.
In a seminal paper, Drummond and Hathrell \cite{Drummond:1979pp} demonstrated that in the EFT \eqref{QEDEFTIntro} {photons} can propagate on black hole (BH) backgrounds with a speed\footnote{Used in the context of photon propagation, ``superluminal" is perhaps not the best word.  ``Superluminal" here means the photon travels faster than some hypothetical massless test particle which is coupled minimally to the theory, or equivalently, that the photon travels outside the light-cone of the background metric.} $c_{s}>1$.   
The setup is shown in Fig.$\!$ \ref{fig:DrummondHathrell}.  Consider a photon traveling in the angular direction at an impact parameter $L$ from a Schwarzschild BH with Schwarzschild radius $r_s$.  The photon's polarization is pointing radially.  From the EFT \eqref{QEDEFTIntro}, the photon's speed $c_{s}=1+\delta c_{s}$ can be estimated to be of order
\begin{align}
\delta c_{s}\sim \frac{\bar{R}_{\mu\nu\rho\sigma}}{m_{e}^{2}}\approx \frac{e^{2}}{m_{e}^{2}}\frac{r_{s}}{L^{3}} \ .
\end{align}
Only for these kinematics do we get $c_{s}>1$.  When the polarization vector points azimuthally the speed is subluminal $c_{s}<1$ and radially propagating photons have $c_{s}=1$ regardless of polarization.

This effect is a bit of a longstanding oddity.  The expectation is that full QED \eqref{QEDfullintroa} should not allow for superluminal propagation \cite{Dubovsky:2007ac}, so why is it displayed in the effective theory?  There have been many studies of the problem from an array of angles, coming to a variety of conclusions (see e.g. \cite{Shore:1995fz,Shore:2003zc,Shore:2007um,Khriplovich:1994qj,Mohanty:1998qq,Preti:2010zz,Akhoury:2010hi}\footnote{Prominent in the literature is the work of Shore, a former student of Drummond, who, with Daniels, extended the calculation to Reissner-Nordstrom \cite{Daniels:1993yi} and Kerr \cite{Daniels:1995yw} and, with Hollowood and collaborators, studied the nature of QED photon trajectories with an emphasis on carefully studying the fate of the effect in the full UV theory \cite{Hollowood:2007kt,Hollowood:2007ku,Hollowood:2008kq, Hollowood:2009qz,Hollowood:2010bd,Hollowood:2010xh,Hollowood:2011yh,Hollowood:2012as,Hollowood:2015elj,Hollowood:2016ryc}. }).

One possible resolution was already pointed out in the original paper \cite{Drummond:1979pp}: this unexpected effect is \textit{tiny}.  Specifically, as the photon traverses its entire path across the black hole, the effect generates a cumulative distance advance of order\footnote{For example, taking a Standard Model electron and solar mass black hole, the distance advance is at least as small as $\Delta d\lesssim 10^{-31}{\rm m}$, much smaller than the cutoff $m_{e}^{-1}\sim 10^{-13}{\rm m}$ and not so far from the Planck length $l_{p}\sim 10^{-34}{\rm m}$.}
\begin{align}
\Delta d\approx \delta c_{s}\times L\approx m_{e}^{-1}\left (\frac{e^{2}}{m_{e}}\frac{r_{s}}{L^{2}}\right ) \ll m_{e}^{-1}\ ,\label{QEDDistanceAdvanceApproximation}
\end{align}
i.e.$\!\,$ a length parametrically \textit{smaller} than the cutoff distance of the theory (the inverse electron mass $m_{e}^{-1}$) for any valid choices of $e$, $r_{s}$ and $L$.    Because this small distance advance is below the resolving power of the EFT, the superluminality cannot be said to be a ``real" effect, at least in this particular setup.   This is an indication that the apparent superluminality is simply an artifact of the approximations made when using the effective theory.

If the superluminality is an artifact of the EFT expansion, it must be cured in the full theory\footnote{EFTs also protect themselves against other apparent pathologies, such as ghosts arising from higher derivative operators in the EFT \cite{Simon:1990ic,Jaen:1986iz,Burgess:2014lwa}.}.  One way this can happen is as follows: the velocities we are implicitly talking about in the effective field theory are group velocities, $v_g=\frac{\rd\omega}{\rd k}$ (which happen to be same as the phase velocities $v_p=\frac{\omega}{k}$ for the massless theories we are talking about since the dispersion relations are, to lowest order, linear  $\omega\propto k$).  However, the speed at which actual information carrying signals can be sent is instead given by the front velocity which tracks the movement of the sharp boundary between regions of zero and non-zero signal \cite{BrillouinBook,MilonniBook,deRham:2014zqa}.  A perfect description of this discontinuous surface is inaccessible to the perturbative EFT\footnote{Dispersion relations relate the front velocity to IR quantities, but their use in curved space is subtle \cite{Hollowood:2007ku}.}, since it cannot resolve spatial distances smaller than its strong coupling distance scale.
If the accumulated distance advance along any particle path on any background calculated using any EFT notion of velocity is smaller than the resolution of the EFT, we may attribute any discrepancy between the EFT velocity and some expected front velocity in the full theory to the inherent fuzziness of the EFT, and there is no cause for concern.

Thus the question is the following:  does $\Delta d \ll m_{e}^{-1}$ persist for \textit{all} possible backgrounds and setups?  Clearly, there are two possibilities: 
\begin{enumerate}
\item There exists \textit{no} setup describable within the QED EFT which generates a macroscopic distance advance, $\Delta d\gg m_{e}^{-1}$. 
\item If we work hard enough, we can construct a setup in QED which generates a macroscopic distance advance, $\Delta d\gg  m_{e}^{-1}$.
\end{enumerate}
If the first scenario were true, then our naive expectations about the EFT would be met: the full UV theory can be free of superluminalities and the effective description can be used and trusted all the way to distances $\sim m_e^{-1}$ without worrying about the spurious superluminality.
If the second scenario were true, then we would have a background with some scale $\Lambda^{-1}\gg m_{e}^{-1}$ over which we would have superluminality.
  In this case, under the assumption that the full UV theory is (sub)luminal, strong quantum effects or extra degrees of freedom must come in at the background-dependent scale $\Lambda$, sooner than the naive cutoff $m_{e}$, in order to cure the superluminality.

In either case, when studying an EFT with an unknown UV completion, the low-energy superluminality never acts as a ``consistency test" to rule out the effective theory.  Instead, it simply tells us when strong coupling or UV degrees of freedom must enter if the full theory is to be (sub)luminal.

We expect that the first scenario must be true for the QED effective theory.  Since the UV theory is known, we know that quantum effects and extra degrees of freedom should not become important until the distance $\sim m_e^{-1}$.  Thus it should be impossible to find a background or setup with $\Delta d\gg m_{e}^{-1}$.

In what follows, we find strong evidence for the first scenario: it is extremely difficult to generate $\Delta d>m_{e}^{-1}$ in QED.   Though we will not be able to analyze every possible scenario, and are therefore unable to elevate our results to the level of a theorem, we will build setups which go to great lengths to try to magnify the superluminal effect, yet still fall short of accomplishing $\Delta d>m_{e}^{-1}$.  Specifically, we attempt to build up the distance advance by passing the photon through an enormous number of black hole pairs.  The black holes are taken to be be nearly extremal Reissner-Nordstrom (RN), so that the only forces which destabilize the pairs of BHs are those generated from loops.

The construction provides a rich demonstration of the conspiracies which must occur in order to prevent the generation of macroscopic distance advances.  There are many competing scales to balance and effects to account for, and only when they are all included do we find that macroscopic superluminality in QED is avoided.  The thought experiments give an idea of how extreme and contrived any scenario generating $\Delta d>m_{e}^{-1}$ would likely need to be.

Our other example of a superluminal EFT, the more speculative one, is that of the galileon, a single scalar $\pi(x)$ whose defining property is a global shift symmetry $\pi(x)\to\pi(x)+b+c_{\mu}x^{\mu}$ with constant $b,c_\mu$ \cite{Nicolis:2008in}.  Galileons have been widely studied as a particularly interesting class of EFTs.  For example, they capture much of the interesting phenomenology of IR modified gravity theories including the Dvali-Gabadadze-Porrati (DGP) \cite{Dvali:2000hr,Luty:2003vm} braneworld model, the de Rham-Gabadadze-Tolley (dRGT) theory of massive General Relativity (GR) \cite{deRham:2010kj} (see \cite{Hinterbichler:2011tt,deRham:2012az,deRham:2014zqa} for reviews) and other brane-world setups \cite{deRham:2010eu,Hinterbichler:2010xn,Goon:2011qf,Goon:2011uw}.  They also possess many interesting properties in their own right, such as Vainshtein screening  \cite{Vainshtein:1972sx,Babichev:2013usa} and strong non-renormalization theorems \cite{Luty:2003vm,Goon:2016ihr}.

The galileons come in many different forms and generalizations (e.g. \cite{deRham:2010eu,Hinterbichler:2010xn,Goon:2011qf,Goon:2011uw,Goon:2011xf,Trodden:2011xh,Goon:2012dy}), but the simplest example is the cubic galileon
\begin{align}
 \mathcal{L}&=-\frac{1}{2}(\partial\pi)^{2}-\frac{1}{\Lambda^{3}}(\square\pi)(\partial\pi)^{2}+\frac{1}{M_{p}}\pi T^\mu_{\ \mu} \label{CubicGalileon}\ ,
 \end{align} 
 where $\Lambda$ is the strong coupling scale of the EFT.  We have coupled it with gravitational strength to a matter source\footnote{The matter coupling might not appear to be invariant under the galileon symmetry, but it is in the limit that the matter is non-dynamical.} which is the trace of the matter stress tensor.  This is the coupling that occurs in most IR modified gravity applications of the galileon.

In the presence of a static point mass $T^\mu_{\ \mu}(x)\sim M\delta^{3}(\vec{x})$, a non-trivial spherically symmetric field profile $\bar \pi(r)$ develops. This creates a potential, $V\sim \frac{1}{M_{p}}\bar{\pi}(r)$, felt by matter. Far from the source, the quadratic kinetic term of \eqref{CubicGalileon} dominates over the cubic term and we have $\bar{\pi}(r)\sim \frac{M}{M_{p}}\frac{1}{r}$, resulting in a gravitational strength fifth force $V\sim \frac{M}{M_{p}^{2}}\frac{1}{r}$.

  \begin{figure}[h!] 
   \captionsetup{width=0.9\textwidth}
   \centering
     \includegraphics[width=4in]{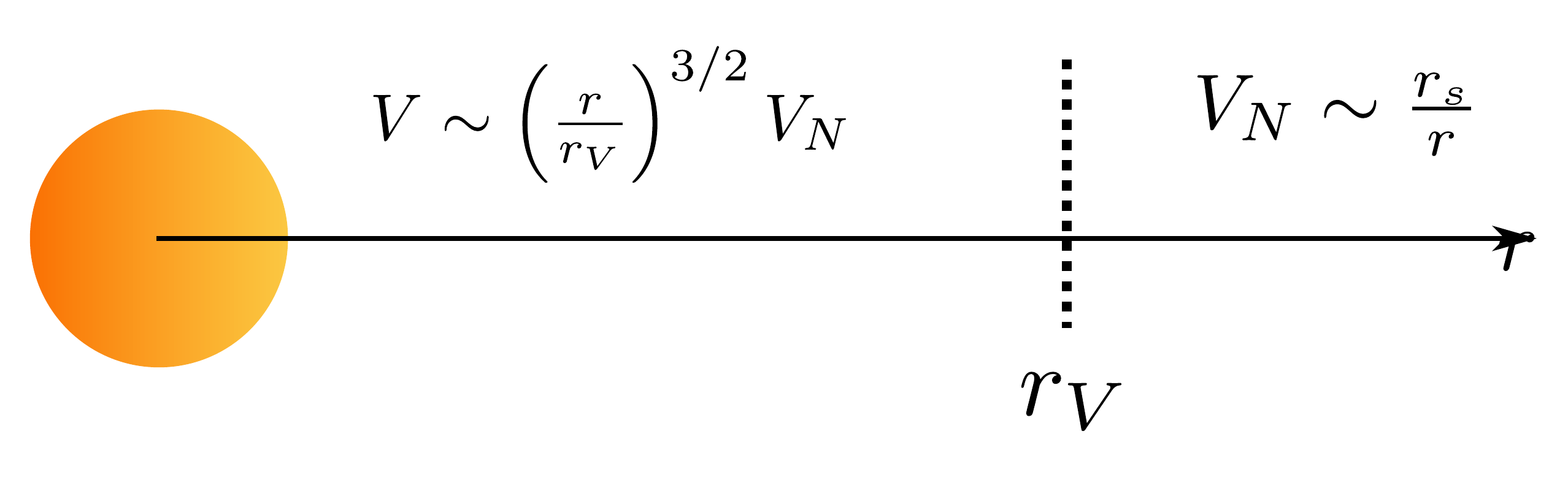}
   \caption{Sketch of the Vainshtein mechanism for the cubic galileon \eqref{CubicGalileon} around the Sun.  Far from a source, the cubic galileon generates a potential of Newtonian strength $V\sim V_{N}\sim r_{s}/r$.  Below the non-linear distance scale $r_{V}\sim \Lambda^{-1}(M/M_{p})^{1/3}$ screening becomes effective and the fifth force is suppressed by a factor of $(r/r_{V})^{3/2}$.}
   \label{fig:VainshteinMechanism}
\end{figure}
 If this force persisted at all distance scales, the model would be ruled out phenomenologically. However, the galileon has a highly efficient screening mechanism, known as the Vainshtein mechanism \cite{Vainshtein:1972sx} (see \cite{Babichev:2013usa} for a review), active in regions sufficiently close to the source.  There is a distance scale $r_{V}\equiv\Lambda^{-1}\left (M/M_{p}\right )^{1/3}$  
 the ``Vainshtein radius" of the source, where the cubic interaction in \eqref{CubicGalileon} becomes as important as the quadratic kinetic term and the field profile changes significantly.  At distances much smaller than the Vainshtein radius, the cubic term dominates and we have $ \bar{\pi}(r)\sim  \left (\frac{r}{r_{V}}\right )^{3/2}\frac{M}{M_{p}}\frac{1}{r} $,
greatly suppressing the potential 
$V\sim \frac{M}{M_{p}^{2}}\frac{1}{r}\left (\frac{r}{r_{V}}\right )^{3/2}$, see Fig.$\!$ \ref{fig:VainshteinMechanism}. This effect is crucial for the compatibility of galileon, DGP and dRGT theories with solar system test of gravity\footnote{Consider galileons in the Solar System.   In models where the size of the IR modification is chosen to account for the present accelerated expansion, one typically has $\Lambda^{-1}\sim\mathcal{O}(10^{3}{\rm km})\sim \mathcal{O}(10^{-11}{\rm pc})$, meaning that the Sun's Vainshtein radius is $r_{V}^{\odot}\sim \mathcal{O}(200 {\rm pc})$.  Since the Solar System's radius is $\sim \mathcal{O}(10^{-4}{\rm pc})$, any local galileon potential is suppressed by a factor of at least $\sim 10^{-9}$ relative to to the usual Newtonian result, making all effects minuscule, but still possibly detectible with precise enough measurements \cite{Dvali:2002vf}.}.

\begin{figure}[h!] 
  \captionsetup{width=0.9\textwidth}
   \centering
     \includegraphics[width=4in]{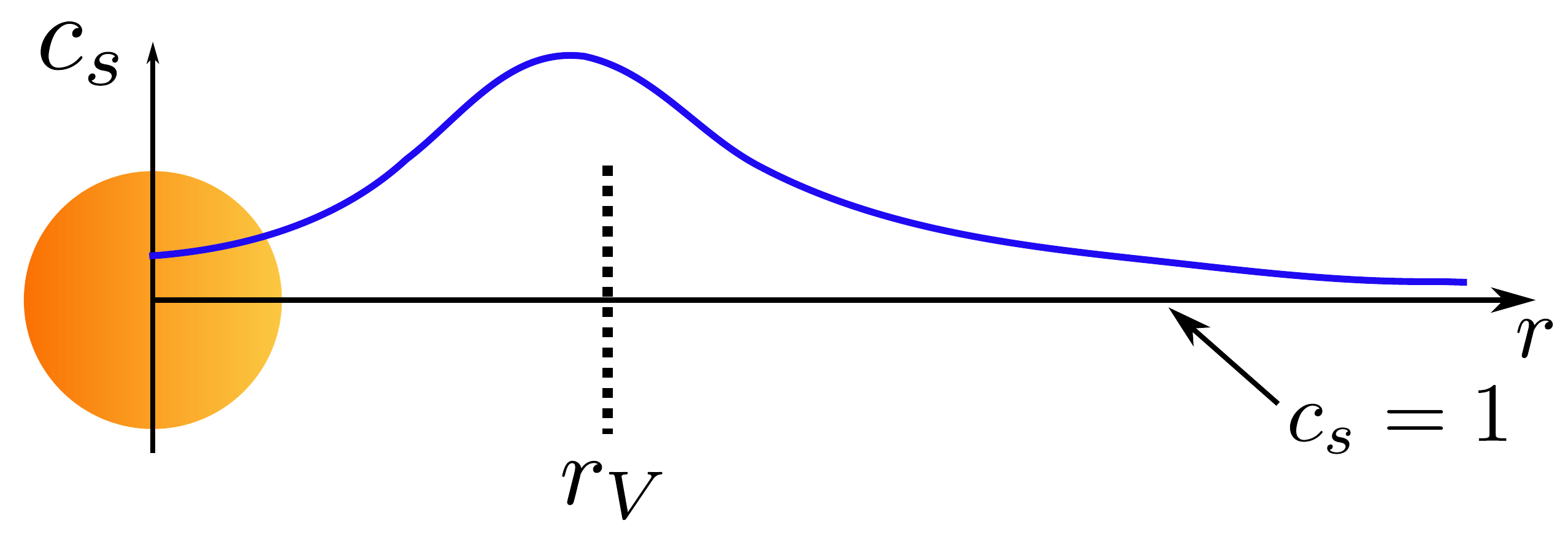}
   \caption{Sketch of superluminality induced by the Sun for the cubic galileon.  Radially moving perturbations travel with a position dependent speed of sound, indicated by the blue curve. The horizontal axis represents $c_{s}=1$.  For the purely cubic galileon, $c_{s}\ge 1$ when $r\lesssim r_{V}$, though $c_{s}\to 1$ at large $r$.  
  }
   \label{fig:VainshteinSuperluminality}
\end{figure}

The same non-linearities responsible for screening also generate superluminal sound speeds for perturbations about the $\bar{\pi}(r)$ background \cite{Adams:2006sv,Nicolis:2008in}.  This effect is quite generic to generalizations of the galileons \cite{Goon:2010xh,Andrews:2010km,Evslin:2011vh,Curtright:2012gx,deFromont:2013iwa,Garcia-Saenz:2013gya} and seems to be a generic feature of theories possessing Vainshtein screening (there are exceptions, however \cite{Berezhiani:2013dw,Gabadadze:2014gba}).
 Specifically, radially propagating perturbations around the background $\bar \pi(x)$ acquire a speed $c_{s}> 1$ at distances $r\gtrsim r_{V}$.   Expanding the cubic interaction \eqref{CubicGalileon} about the background allows us to read off the approximate expression for the sound speed $c_{s}=1+\delta c_{s}$,
 \begin{align}
 \delta c_{s}\sim \frac{\partial^{2}\bar{\pi}(r)}{\Lambda^{3}}\approx \left (\frac{r_{V}}{r}\right )^{3}\label{ApproximateCsGalileons}\ .
 \end{align}
 The sign of $\delta c_{s}$ turns out to be positive and \eqref{ApproximateCsGalileons} represents an $\mathcal{O}(1)$ effect near $r_{V}$, with $c_{s}$ settling back to unity as $r\to \infty$, see Fig.$\!$ \ref{fig:VainshteinSuperluminality}.  (By including higher order galileon operators it is possible to turn $c_{s}$ subluminal at distances close to the source so that significant superluminality only exists at $r\sim r_{V}$ \cite{Nicolis:2008in}.)  All other perturbations, i.e.$\!\,$ those in the angular directions, propagate subluminally.  
 
This superluminality has caused much worry, and is thought to imply similar superluminalities within the full DGP \cite{Hinterbichler:2009kq} and dRGT \cite{Deser:2015wta} theories\footnote{It has not been directly shown within the full massive theories, however, that there is superluminality within the naive regime of validity of the full massive theories.  In addition, there are indications that enforcing certain cosmological boundary conditions may eliminate superluminality outright \cite{Berezhiani:2013dw,Gabadadze:2014gba}.}.
In contrast to QED, the galileon superluminality generates \textit{macroscopic} distance advances, as compared to the galileon strong coupling distance scale $\Lambda^{-1}$.  Sending a galileon signal from near the Vainshtein radius to infinity, the distance advance is of order \cite{Creminelli:2014zxa}
\begin{align}
\Delta d\approx \int_{\sim r_{V}}^{\infty}\rd r \, \delta c_{s}\approx\int_{\sim r_{V}}^{\infty}\rd r \, \left (\frac{r_{V}}{r}\right )^{3}\approx r_{V}\approx \Lambda^{-1}\left (M/M_{p}\right )^{1/3} \ ,\label{GalileonDistanceAdvanceApproximation}
\end{align}
which is parametrically \textit{larger} than the scale $\Lambda^{-1}$ for any large source\footnote{For the Sun, we would find $\Delta d\sim 200$pc whereas $\Lambda^{-1}\sim 10^{-11}$pc.}.
 Therefore, while the two scenarios have many superficial similarities, they are qualitatively different in an important way: galileons and QED generate distance advances which are parametrically larger and smaller than the naive cutoffs of the EFTs in the two cases, respectively.

 For the galileons, and to the extent that they captures infrared modifications of gravity, this would indicate that if it is possible to fix the galileon superluminality, the cure will be of a qualitatively different type than the prescription for QED.  Non-perturbative effects will have to become important at very large distance scales of order the Vainshtein radius, much larger than the naive strong coupling distance scale $\Lambda^{-1}$.  The Vainshtein solutions behave in many ways like black holes in general relativity, with the Vainshtein radius playing the role of the Schwarzschild radius.  From the black hole information paradox, firewall paradox, etc., there are many reasons to believe that quantum effects near the black hole horizon should become important, causing the local effective field theory expectations to fail, despite the fact that curvatures are much smaller than the naive Planck scale cutoff \cite{Mathur:2009hf,Mathur:2005zp,Dvali:2011aa,Almheiri:2012rt}.  The same may be true of the galileons, and there are indications that this is the case \cite{Dvali:2010jz,Keltner:2015xda}, meaning they could potentially serve as a toy model \cite{Klein:2015iud} of the firewall paradox \cite{Almheiri:2012rt}.

Finally, in a theory with gravity, there are strictly speaking no local observables, and it might be objected that local superluminality of the type we have been implicitly discussing is not a sharp observable from which we can draw sharp conclusions.  However, all of the above can be phrased in terms of asymptotic observables, i.e.$\!\,$ cumulative time advances measured by sending a signal in from infinity in an asymptotically flat solution and watching for when it comes out at the other side of infinity.  We will thus consider only scenarios which can in principle be viewed as this kind of asymptotic scattering experiment, and hence represent sharp observables even in the presence of gravity.

\textbf{Conventions}: Our metric and curvature conventions are those of Carroll \cite{Carroll:2004st} (equivalently, Misner, Thorne and Wheeler \cite{Misner:1974qy}): we work in mostly \textit{plus} signature, $\eta_{\mu\nu}=(-,+,+,+)$ and use the curvature conventions
\begin{align}
\Gamma_{\mu\nu}^{\lambda}&= \frac{1}{2}g^{\lambda\sigma}\left [\partial_{\mu }g_{\nu\sigma}+\partial_{\nu}g_{\sigma\mu}-\partial_{\sigma}g_{\mu\nu}\right ]\  \quad
,\nn
R^{\rho}{}_{\sigma\mu\nu}&=\partial_{\mu } \Gamma^{\rho}_{\nu\sigma}-\partial_{\nu} \Gamma^{\rho}_{\mu\sigma}+ \Gamma^{\rho}_{\mu\lambda}\Gamma^{\lambda}_{\nu\sigma}-\Gamma^{\rho}_{\nu\lambda}\Gamma^{\lambda}_{\mu\sigma}\ , \quad
R_{\sigma\nu}=R^{\rho}{}_{\sigma\rho\nu} \, ,
\end{align}
so that $\left [\nabla_{\mu},\nabla_{\nu}\right ]V^{\rho}=R^{\rho}{}_{\sigma\mu\nu}V^{\sigma}\, $.
We symmetrize and anti-symmetrize indices with weight one, i.e.
\begin{align}
T_{(\mu\nu)}=\frac{1}{2}\left [T_{\mu\nu}+T_{\nu\mu}\right ]\ , \quad T_{[\mu\nu]}&=\frac{1}{2}\left [T_{\mu\nu}-T_{\nu\mu}\right ]\ .
\end{align}  Greek indices run over all of spacetime $\mu\in\{0,1,2, 3\}$ and Latin indices run over space $i\in\{1,2,3\}$ (we work in $d=4$ throughout).  The Planck mass conventions are $M_{p}^{2}\equiv 1/l_{p}^{2}\equiv \left (8\pi G_{N}\right )^{-1}$. The Schwarzschild radius for a black hole of mass $M$ is $r_{s}=\frac{M}{4\pi M_{p}^{2}}$.  The distance scale associated to the charge of a charged black hole is defined to be $r_{q}=\frac{Q}{\pi\sqrt{8}M_{p}}$, so that extremal Reissner-Nordstrom black holes satisfy $r_{q}=r_{s}$. The Vainshtein radius for a source of mass $M$ is $r_{V}= \Lambda^{-1}\left (M/M_{p}\right )^{1/3}$. Often we will rewrite the electron mass $m_{e}$ in favor of the length scale $r_{e}\equiv m_{e}^{-1}$, which is (roughly) the cutoff of the EFT.

\section{The QED Effective Theory}

We start with a short review of the QED EFT and discuss its expected regimes of validity.   The EFT is constructed by integrating out the electron from the QED action,
\begin{align}
\exp iS_{\rm eff}\left [g_{\mu\nu}, A_{\mu}\right ]&\equiv \int\mathcal{D}\bar\psi\mathcal{D}\psi\, \exp i S_{\rm QED}\left [g_{\mu\nu}, A_{\mu}, \bar\psi, \psi\right ]\ .\label{IntegratingOut}
\end{align}
Integrating out the electron is a particularly clean procedure in QED since the UV action is strictly quadratic in fermion fields, so the entire contribution of electrons to the low energy effective action can be written as a single one-loop functional determinant,
\begin{align}
S_{\rm eff}&=\int\rd^{4}x\sqrt{-g}\, \left [\frac{M_{p}^{2}}{2}R-\frac{1}{4e^{2}}F_{\mu\nu}^{2}\right ]-i\Tr\ln \left (i\slashed D-m_{e}\right )\label{QEDFunctionalDeterminant}\ .
\end{align}

The effective action is local, expressible as a power series in ${\partial}$.  The precise signs of the various coefficients in the effective action are important for our analysis.  Hence, as a check, we re-derived the effective action using two methods: matching amplitudes and directly expanding the functional determinant (using the technique outlined in Appendix A of \cite{Goon:2016ihr}).  
We find full agreement with the original Drummond-Hathrell result, after accounting for their conventions\footnote{In the literature there appears to be some unstated disagreement about the signs in the effective action.  For instance, the effective action in \cite{Daniels:1993yi} has the same signs as the  Drummond-Hathrell result \cite{Drummond:1979pp} and thus claims to be in agreement with their results. However, \cite{Daniels:1993yi} uses the opposite signature but the same curvature conventions as \cite{Drummond:1979pp} and therefore should have different signs on the $\sim RFF$ terms in $S_{\rm eff}$.  Other references leave these important conventions unstated entirely.  We use the same curvature conventions and opposite metric signature as Drummond-Hathrell.}.

\begin{figure}[h!] 
  \captionsetup{width=0.9\textwidth}
   \centering
     \includegraphics[width=5in]{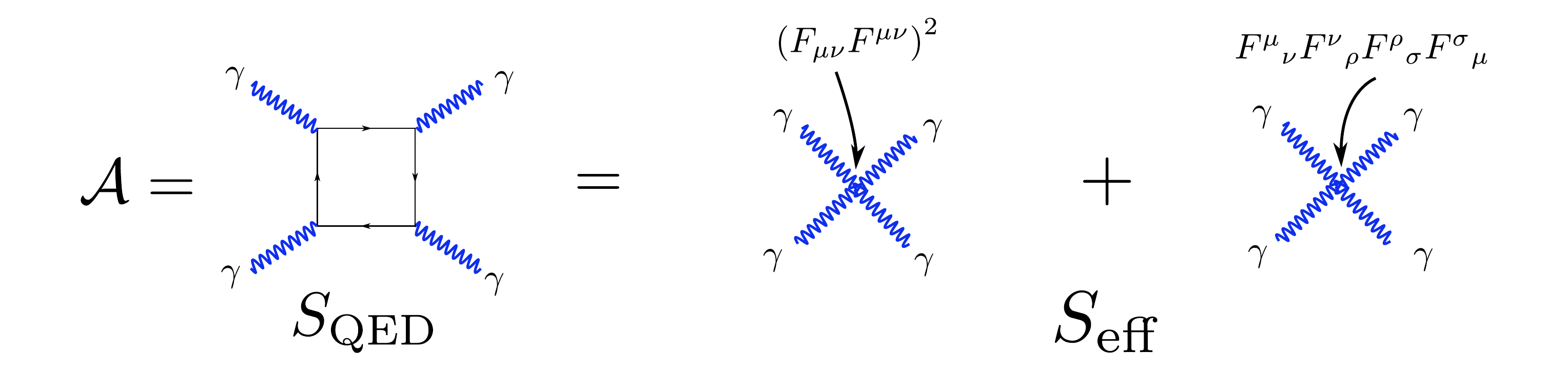}
   \caption{The QED box diagram generates the two $\mathcal{O}(F^{4})$ EFT operators in \eqref{4DEFT}. Throughout the paper, photons are represented by blue, wavy lines.}
   \label{fig:MatchingQEDFFFF}
\end{figure}

\begin{figure}[h!] 
  \captionsetup{width=0.9\textwidth}
   \centering
     \includegraphics[width=5in]{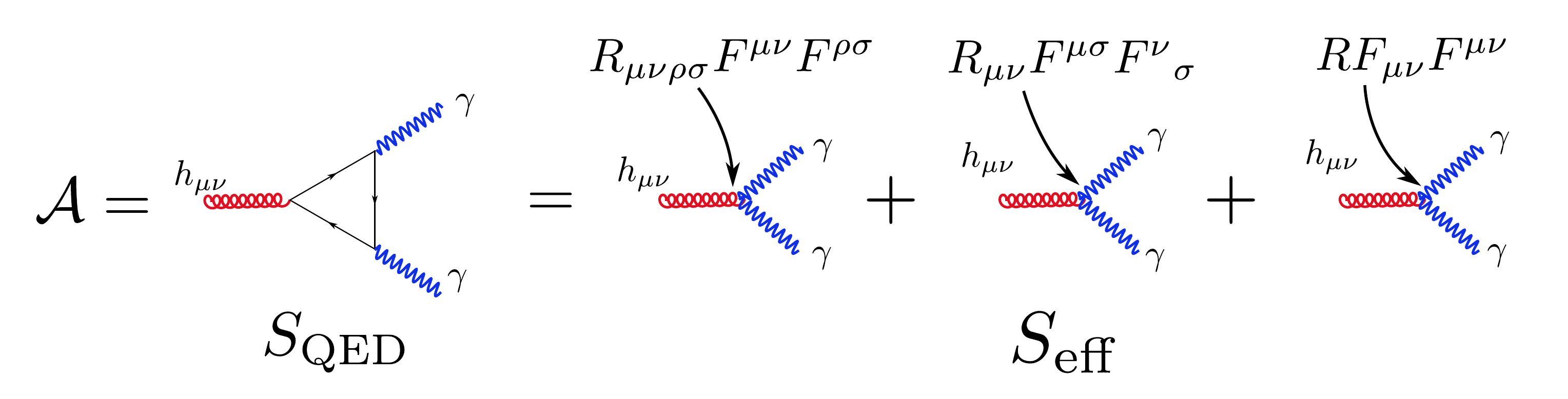}
   \caption{Triangle diagrams generate the three $\sim RFF$ operators in \eqref{4DEFT}. Throughout the paper, gravitons will be represented by red, curly lines.}
   \label{fig:MatchingQEDRFF}
\end{figure}

The effective action contains a finite number of divergent terms, while the remaining terms are finite and unambiguous.
Up to order $\partial^4$, the finite parts of the effective action are
\begin{align}
S_{\rm eff}&=\int\rd^{4}x\sqrt{-g}\,\left (\frac{M_{\rm p}^{2}}{2}R-\frac{1}{4e^{2}}F_{\mu\nu}^{2}\right )\nn
&\quad +\int\rd^{4}x\, \sqrt{-g}\,  \frac{1}{m_{e}^{2}}\left (aRF_{\mu\nu}F^{\mu\nu}+bR_{\mu\nu}F^{\mu\sigma}F^{\nu}{}_{\sigma}+cR_{\mu\nu\sigma\tau	}F^{\mu\nu}F^{\sigma\tau}+d\nabla_{\mu}F^{\mu\nu}\nabla_{\sigma}F^{\sigma}{}_{\nu}\right )
\nn
&\quad +\int\rd^{4}x\, \sqrt{-g}\,\frac{1}{m_{e}^{4}}\left (yF_{\mu\nu}F_{\sigma\tau}F^{\mu\sigma}F^{\nu\tau}+z(F_{\mu\nu}F^{\mu\nu})^{2}\right )\Big] +{\cal O}(\partial^6) ,\label{4DEFT}
\end{align}
where the $\mathcal{O}(1)$ coefficients are\footnote{In an abuse of notation, we will refer to every numerical EFT coefficient in \eqref{EFTCoefficients} as being $\mathcal{O}(1)$ throughout the paper, despite the fact that they're numerically $\mathcal{O}(10^{-4})$ or $\mathcal{O}(10^{-5})$.}
\begin{align}
\begin{pmatrix}
a\\ b\\c\\d\\y\\z
\end{pmatrix}& = \frac{1}{180}\frac{1}{(4\pi)^{2}}\begin{pmatrix}
5\\ -26\\ 2\\ 24 \\14 \\-5
\end{pmatrix}\ .\label{EFTCoefficients}
\end{align}
The two $\sim F^{4}$ operators arise from the matching shown in Fig.$\!$ \ref{fig:MatchingQEDFFFF}.  The $\sim RF^2$ operators arise from the matching in Fig.$\!$ \ref{fig:MatchingQEDRFF}.   

The divergences appear for the operators\footnote{The $\sqrt{-g}R_{\mu\nu\rho\sigma}^{2}$ operator also appears, but we can remove it via the Gauss-Bonnet total derivative.  It's needed, however, as a counterterm in dimensional regularization.}
\be m_e^4 \sqrt{-g},\  m_e^2 \sqrt{-g}R, \ \sqrt{-g}F_{\mu\nu}^{2},\   \sqrt{-g}R^{2},\  \sqrt{-g}R_{\mu\nu}^{2}\, ,\ee
where the coefficients shown reflect the natural scale.
In \eqref{4DEFT} we chose counterterms so as to set to zero the coefficient of $\sqrt{-g}$; this is the usual cosmological constant fine tuning.  The divergences in $\sqrt{-g}R$, $\sqrt{-g}F_{\mu\nu}^{2}$ are absorbed into the definitions of $M_{p}$ and $e$, which are now renormalized quantities.  
 The coefficients of the $R^2$ operators should also be absorbed into renormalized coefficients.  We have not written these operators in the action \eqref{4DEFT} because they play no role in the effects we are interested in as long as its coefficient, $c_{R_{\mu\nu}^{2}}$, obeys $c_{R_{\mu\nu}^{2}}\lesssim \left (\frac{e M_{p}}{m_{e}}\right )^{2}$.  The natural size for $c_{R_{\mu\nu}^{2}}$ is $\mathcal{O}(1)$ and we will assume throughout that $\frac{e M_{p}}{m_{e}}>1$, so no fine tuning is required, given the latter assumption.  The condition $\frac{e M_{p}}{m_{e}}>1$ is (one version of) the Weak Gravity Conjecture (WGC) \cite{ArkaniHamed:2006dz}.  We will come back later to connections between our work and the WGC.

In principle, the full effective action contains all possible information about low energy fields.  For QED, everything we'd ever want to know about processes only involving gravity and light is in $S_{\rm eff}\left [g_{\mu\nu},A_{\mu}\right ]$.  In practice,
we necessarily make an approximation by \textit{truncating} the action: we keep only a few low-dimension operators in $S_{\rm eff}$ and throw everything else away. 
It is therefore clear that the truncated EFT cannot be used to study processes at all possible energies.
Keeping, for instance, $(F_{\mu\nu}F^{\mu\nu})^{2}/m_{e}^{4}$ while neglecting $(F_{\mu\nu}F^{\mu\nu})^{3}/m_{e}^{6}$ is only a good approximation to the extent that $F_{\mu\nu}/m_{e}^{2}\ll 1$, with similar criteria holding for the curvature terms.  Thus, the range of validity of the truncated QED EFT is restricted to regimes in which energies are smaller than $m_{e}$, distances are larger than $m_{e}^{-1}$ and curvatures and field strengths much smaller than $m_{e}^{2}$.  Everything else is below the resolving power of the effective theory.
Therefore, given a superluminal $c_{s}>1$ effect in QED which is unable to generate a distance advance larger than $m_{e}^{-1}$, we cannot exclude the possibility that it is a simple artifact of our approximations.

\section{The Drummond-Hathrell Problem}

The effective theory \eqref{4DEFT}, viewed as a classical theory, admits superluminal propagation around non-trivial backgrounds.   This is known as the Drummond-Hathrell problem.  In this section we review the original Drummond-Hathrell problem and re-derive the appropriate geometric optics equations for describing the propagation of light in the effective theory.

Note that \eqref{4DEFT} is not a classical theory; it incorporates electron loops but graviton and photon loops have not yet been included.  It is not even the one-loop 1PI effective action of the theory \eqref{QEDfullintroa}, because there are one-loop diagrams with internal gravitons and photons which have not been included.  These diagrams become important in some regimes, and we will discuss their effects later on.

In a theory including gravity defined with flat space asymptotics, it is generally asymptotically defined quantities such as the S-matrix which are the cleanest observables to define.  Thus we will ask about superluminality which can in principle be observed asymptotically.  We will stick to backgrounds which are asymtotically flat, and ask about asymptotic observables such as the distance advance by which a superluminal photon overtakes a familiar, \textit{minimally} coupled photon as the two race out to $\infty$ across the asymptotically flat space.

\subsection{Black Hole Setup}

\begin{figure}[h!] 
   \captionsetup{width=0.9\textwidth}
   \centering
     \includegraphics[width=4in]{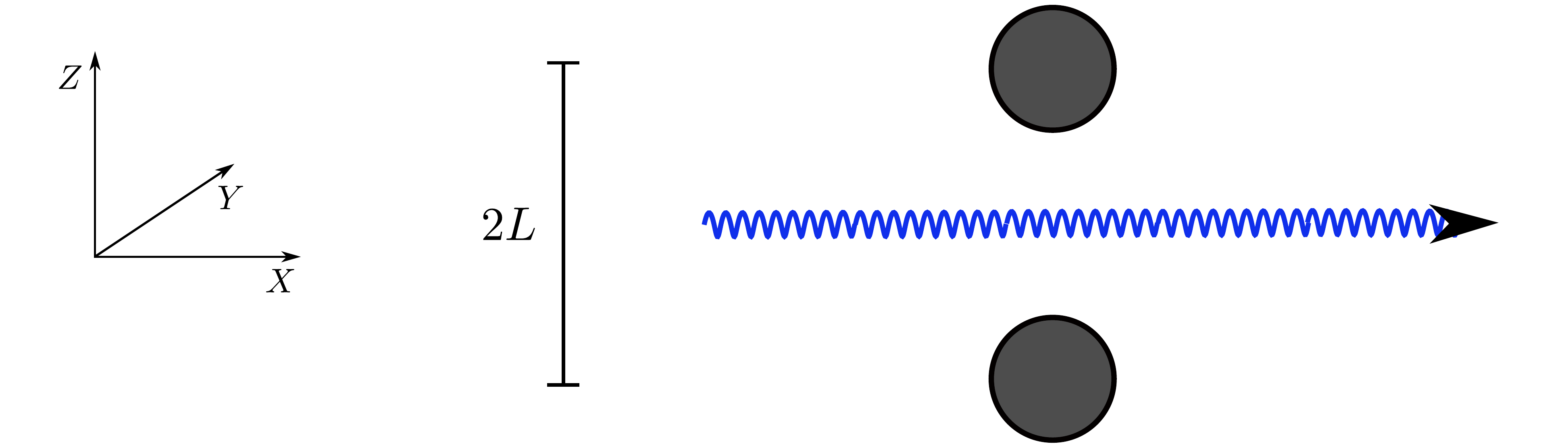}
   \caption{A modified Drummond-Hathrelll setup. The photon passes directly between two black holes a distance $2L$ apart.}
   \label{fig:PhotonBetweenTwoBHs}
\end{figure}

We will start with a slight variation of the Drummond-Hathrell setup: we use two equal sized black holes, instead of one, so that the photon can pass directly between the pair without curving\footnote{This is the scenario used in Appendix A of \cite{Camanho:2014apa} to discuss Shapiro time delay.}.  The black holes are separated by a distance much larger than their Schwarzschild radii so that the spacetime is approximately described by the sum of the metric perturbations from each of the black holes.  We treat the positions of the black holes as constant.  Even though the black holes will attract, the associated time scale is much longer than the time it takes the photon to pass between the pair, so the static approximation is a adequate for our purpose.  See Fig.$\!$ \ref{fig:PhotonBetweenTwoBHs}.

We use isotropic coordinates $x^{\mu}=(t,X,Y,Z)$ and place the two black holes at $\vec{X}_{\pm}=(0,0,\pm L)$, with $L\gg r_{s}$.  Because the black holes are separated by a distance much larger than either of their Schwarzschild radii, the metric in the region between the back holes may be approximated as the sum of the linearized metrics of the two black holes, 
\begin{align}
\rd s^{2}=-\left (1-\frac{r_{s}}{R_{+}}-\frac{r_{s}}{R_-}\right )\rd t^{2}+\left (1+\frac{r_{s}}{R_{+}}+\frac{r_{s}}{R_-}\right )\rd \vec{X}^{2}\ , \label{Isotropic2BHMetric}
\end{align}
where $R_{\pm}^{2}=(X^{i}-X^{i}_{\pm})(X^{j}-X^{j}_{\pm})\delta_{ij}$.   Our photon travels in the $X$ direction along the line $Z=Y=0$, and hence its motion is only sensitive to the following non-trivial Riemann curvature components along this path:
\begin{align}\def\arraystretch{1.3}
\begin{tabular}{|c|c|c|c|c|c|}
\hline $R^{t}{}_{XXt}$ &$R^{t}{}_{YYt}$ & $R^{t}{}_{ZZt}$ & $R^{X}{}_{YYX}$ & $R^{X}{}_{ZZX}$ & $R^{Y}{}_{ZZY}$ \\[.05ex]  \hline
 $\frac{r_{s}(L^{2}-2X^{2})}{\left(L^2+X^2\right)^{5/2}}$ & $\frac{r_{s}}{\left(L^2+X^2\right)^{3/2}}$ & $\frac{r_{s}(X^{2}-2L^{2})}{\left(L^2+X^2\right)^{5/2}}$   &  $\frac{r_{s}(X^{2}-2L^{2})}{\left(L^2+X^2\right)^{5/2}}$ & $\frac{r_{s}}{\left(L^2+X^2\right)^{3/2}}$  &  $\frac{r_{s}(L^{2}-2X^{2})}{\left(L^2+X^2\right)^{5/2}}$  \\[1ex]  \hline
\end{tabular}\label{LinearizedRNCurvatures} \ .
\end{align}

\subsection{Geometric Optics Analysis\label{Sec:DrummondHathrellGeometricOptics}}

Given this background, we may now perform a geometric optics or characteristic analysis to determine the photon trajectories \cite{Misner:1974qy,Preti:2010zz,Benakli:2015qlh}.   Physically, geometric optics is the regime of wave propagation in which the wave's phase varies much more rapidly than the amplitude, and its characteristic wavelength is much smaller than the typical background curvature scale.
 Since we're studying photon propagation in the context of the QED EFT, we have the additional restriction that the characteristic wavelength of the wave be much larger than $m_{e}^{-1}$.  Since the typical length scale associated to the Riemann curvature is $\mathcal{O}(r_{s})$, we are thus working within the wide window between $m_{e}^{-1}$ and $\sim 1/\sqrt{R_{\mu\nu\rho\sigma}}$.  
 
To perform the geometric optics approximation, we take a background solution of \eqref{4DEFT}, $\{g_{\mu\nu},\bar{A}_{\mu}\}$, and introduce a vector potential fluctuation $\delta A_{\mu}$ which is then expanded as a product of a slowly varying amplitude and a rapidly varying phase,
 \begin{align}
 \delta A_{\mu}&=\left (a_{\mu}+\epsilon b_{\mu}+\ldots\right )\exp \left (\frac{i\vartheta(x)}{\epsilon}\right )\label{GeometricOpticsExpansion}\ ,
 \end{align}
where $\epsilon$ is a small, formal constant introduced to keep track of orders in the expansion.  We then derive the equation of motion from the effective action \eqref{4DEFT}, evaluate on $g_{\mu\nu}$ and $A_{\mu}=\bar{A}_{\mu}+\delta A_{\mu}$ and start expanding, keeping only the terms first order in $\delta A$ and lowest non-trivial order in $\epsilon$, all the while working perturbatively in the effective field theory expansion $\frac{\partial}{m_e}$.

The full photon equation of motion is
\begin{align}
\nabla^{\nu}F_{\nu\mu}&=4e^{2}m_{e}^{-2}\left (a\nabla^{\nu}(RF_{\nu\mu})+b\nabla^{\nu}(R_{[\nu|\alpha|}F^{\alpha}{}_{\mu]})+c\nabla^{\nu}(R_{\nu\mu\rho\sigma}F^{\rho\sigma})-d\nabla^{\mu}(\nabla_{[\mu}\nabla_{|\sigma|}F^{\sigma}{}_{\nu]})\right )\nn
&\quad +8e^{2}m_{e}^{-4}\left (y\nabla^{\nu}(F_{\sigma\tau}F_{\nu}{}^{\sigma}F_{\mu}{}^{\tau}+z\nabla^{\nu}(F^{2}F_{\nu\mu}))\right )\ ,\label{PhotonEOM}
\end{align}
and we work in Lorenz gauge 
\be \nabla^{\mu}\delta A_{\mu}=0\, . \ee
Both the gauge condition and equation of motion are expanded in powers of $\epsilon$.  In addition, because we truncated the effective action, we are working perturbatively in $m_e^{-1}$.  Therefore, on the right hand side of \eqref{PhotonEOM} we may use the zero-th order in $m_e^{-1}$ uncorrected black hole solution. 
For Schwarzschild black holes, only the $c$ term in \eqref{PhotonEOM} contributes since all other terms are proportional to $\bar{F}_{\mu\nu}$, $R_{\mu\nu}$ or $R$, all of which are vanishing on the zero-th order solution.  The dispersion relation arises at $\mathcal{O}(\epsilon^{-2})$, stemming from terms in \eqref{PhotonEOM} with two derivatives acting on $\delta A$. 

Defining $k_{\mu}\equiv \nabla_{\mu}\vartheta$, the leading $\mathcal{O}(\epsilon^{-1})$ part of the gauge condition reads
\begin{align}
k^{\mu}a_{\mu}=0\, , \label{Oe1GaugeConditionSchwarzschild} 
\end{align}
which can be used to simplify the $\mathcal{O}(\epsilon^{-2})$ part of the equation of motion \eqref{PhotonEOM} to the form
\begin{align}
k^{\nu}k_{\nu}&=8ce^{2}m_{e}^{-2} R_{\mu\rho\nu\sigma}k^{\mu}k^{\nu}f^{\rho}f^{\sigma} \label{Oe2PhotonEOMSchwarzschild}\ ,
\end{align} after writing $a_{\mu}=a f_{\mu}$ with $f_{\mu}$ a unit vector, $g^{\mu\nu}f_{\mu}f_{\nu}=1$.

The photon propagation is more naturally phrased in terms of an \textit{optical} metric $\tilde{g}_{\mu\nu}$ defined by
\begin{align}
\tilde{g}^{\mu\nu}\equiv g^{\mu\nu}-8ce^{2}m_{e}^{-2} R^{\mu}{}_{\rho}{}^{\nu}{}_{\sigma}f^{\rho}f^{\sigma} \ , \quad \tilde{g}_{\mu\nu}\approx g_{\mu\nu}+8ce^{2}m_{e}^{-2} R_{\mu\rho\nu\sigma}f^{\rho}f^{\sigma} \ . \label{PhotonOpticalMetricSchwarzschild}
\end{align}
Photons are null with respect to this effective metric, $\tilde{g}^{\mu\nu}k_{\mu}k_{\nu}=0$, and follow the geodesics of $\tilde{g}_{\mu\nu}$, not the background metric\footnote{\label{foot:ModifiedGeodesicEquationQED}This is easily proven by defining $\tilde{k}^{\mu}=\tilde{g}^{\mu\nu}k_{\nu}=\tilde{g}^{\mu\nu}\nabla_{\nu}\vartheta$ and taking a covariant derivative (with respect to $\tilde{g}_{\mu\nu}$) of the null condition: $0=\frac{1}{2}\tilde{\nabla}_{\alpha}\left (\tilde{g}^{\mu\nu}k_{\mu}k_{\nu}\right ) =\tilde{k}^{\nu}\tilde{\nabla}_{\nu} k_{\alpha}$ implying $\tilde{k}^{\nu}\tilde{\nabla}_{\nu}\tilde{k}^{\mu}=0$
which is the standard geodesic equation (we used $\tilde{\nabla}_{[\mu}k_{\nu]}=0$ as $k_{\nu}$ is the gradient of a scalar).}.  The tangent vector along the photon worldline, $\frac{\rd x^{\mu }}{\rd\lambda}$, is thus proportional to $\tilde{k}^{\mu}$, defined by
\begin{align}
\tilde{k}^{\mu}\equiv \tilde g^{\mu\nu}k_\nu\label{ktildeDefinition}\ ,
\end{align} \textit{not} $k^{\mu}$ (as was emphasized recently in \cite{Chen:2015bva}).

The interesting question is therefore whether $\tilde{k}^{\mu}$ is spacelike, timelike or null with respect to the background metric $g_{\mu\nu}$, as this is the measure of how different photon propagation in full QED is from naive expectations.  At lowest non-trivial order in $m_e^{-1}$, this test reads
\begin{align}
g_{\mu\nu}\tilde{k}^{\mu}\tilde{k}^{\nu}&\approx -8ce^{2}m_{e}^{-2}R_{\mu\rho\nu\sigma}k^{\mu}f^{\rho}k^{\nu}f^{\sigma}\ .\label{TildekkContraction}
\end{align}

\begin{figure}[h!] 
   \captionsetup{width=0.9\textwidth}
   \centering
     \includegraphics[width=3.3in]{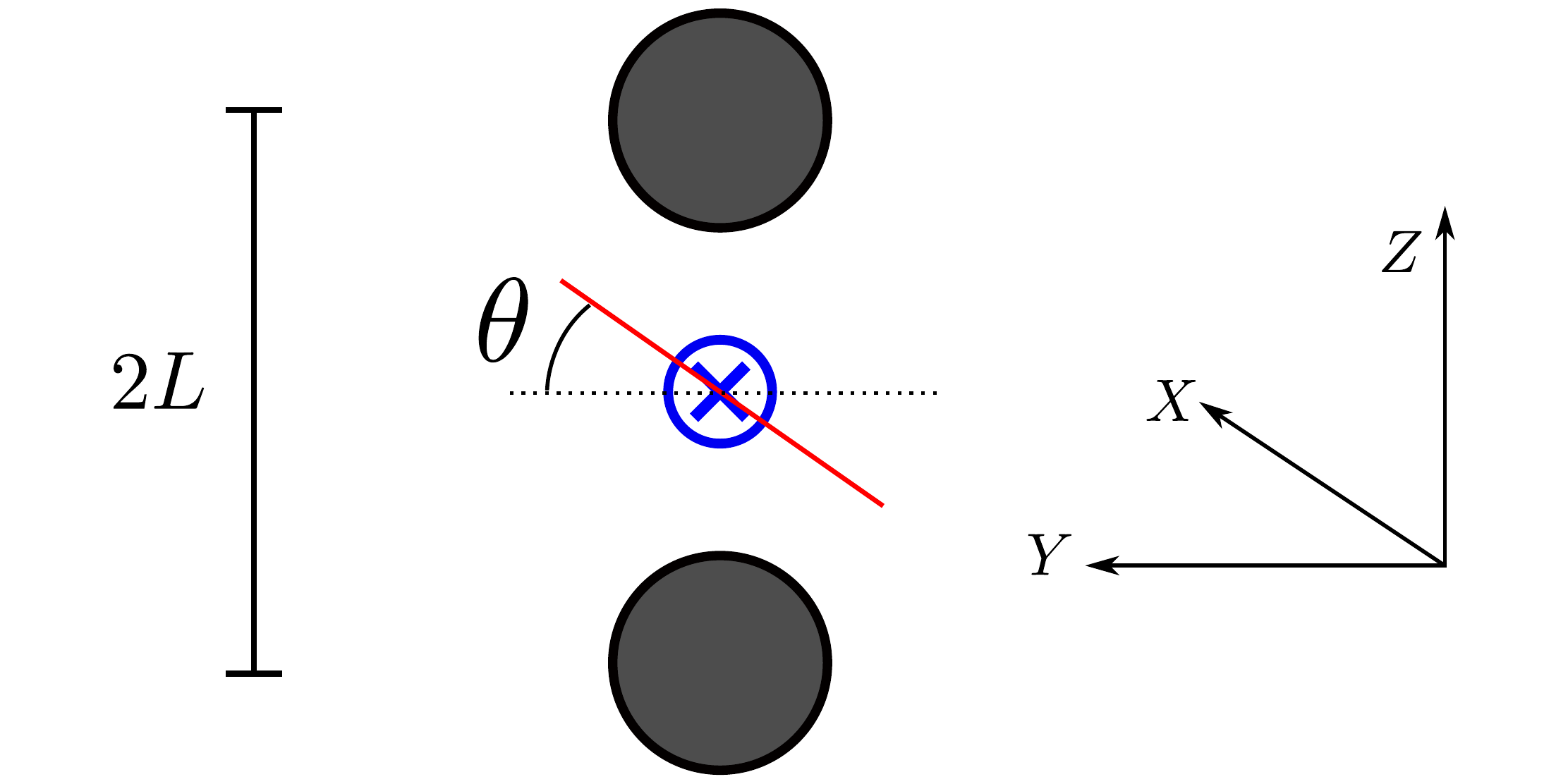}
   \caption{Sketch of the photon traveling into the page. The red line represents the photon polarization vector which makes an angle $\theta$ with the $Z=0$ plane. Photon propagation is maximally timelike if $\theta=0,\pi$ and maximally spacelike if $\theta=\pm \pi/2$ \eqref{PhotonkkContraction}.} 
   \label{fig:PhotonBetweenTwoBHsHeadOnPolarization}
\end{figure}

For our setup in Fig.$\!$ \ref{fig:DrummondHathrell}, we take the photon's polarization vector to make an angle $\theta$ with respect to the positive $Y$ axis, see Fig.$\!$ \ref{fig:PhotonBetweenTwoBHsHeadOnPolarization}, and find:
\begin{align}
g_{\mu\nu}\tilde{k}^{\mu}\tilde{k}^{\nu}&\approx-\frac{24 ce^{2}L^{2}r_{s}\cos 2\theta}{m_{e}^{2}\left (L^{2}+X^{2}\right )^{5/2}}\ ,\label{PhotonkkContraction}
\end{align}
where we used \eqref{LinearizedRNCurvatures} and took $k^{\mu}\approx (1,1,0,0)+\mathcal{O}(r_{s}/\sqrt{L^{2}+X^{2}})$ on the RHS of \eqref{TildekkContraction}.
Since $c>0$ \eqref{EFTCoefficients}, we see that if the polarization vector lies in the plane of the black hole pair, $\theta=\pm \pi/2$, then the propagation is maximally spacelike and superluminal, while if the polarization vector is perpendicular to this plane, $\theta=0, \pi$, then the propagation is maximally timelike and subluminal.

In order to analyze the effect on the photon's path in greater detail, we can perturbatively solve for the altered photon geodesic\footnote{\label{foot:GeodesicWithSourceTerm}The easiest way to do this in practice is to solve for $k^{\mu}$ first, translate the result into $\tilde{k}^{\mu}=\tilde{g}^{\mu\nu}g_{\nu\sigma}k^{\sigma}$ and then integrate to find $x^{\mu}(\lambda)$.  It is straightforward to demonstrate that this is equivalent to solving $\tilde{k}^{\nu}\tilde{\nabla}_{\nu}\tilde{k}_{\mu}=0$ directly.  To solve for $k^{\mu}$, we take a covariant derivative of \eqref{Oe2PhotonEOMSchwarzschild} with respect to $ g_{\mu\nu}$ to derive a modified geodesic equation: $
k^{\nu}\nabla_{\nu}k_{\alpha}=\nabla_{\alpha}\left (4ce^{2}m_{e}^{-2}R_{\mu\rho\nu\sigma}k^{\mu}k^{\nu}f^{\rho}f^{\sigma}\right )$.
Then, $k^{\mu}$ is expanded about a null geodesic of the background metric, $k^{\mu}=\bar{k}^{\mu}+\delta k^{\mu}$ where $\bar{k}^{\mu}$ satisfies $\bar k^{\nu}\nabla_{\nu}\bar{k}_{\mu}=0$, and we solve for $\delta k^{\mu}$ perturbatively.}.
The solution for $x^{\mu}(\lambda)$ is conveniently expressed as an expansion about $\bar{x}^{\mu}(\lambda)$, the geodesic whose tangent vector is $ \bar{k}^{\mu}$.  To lowest non-trivial order, 
\begin{align}
\bar{x}^{\mu}(\lambda)&\approx\begin{pmatrix}
\lambda+2r_{s}\ln\left [\lambda/L+\sqrt{1+\lambda^{2}/L^2}\right ]	\,	 , & \lambda\, , & 0\, , & 0
\end{pmatrix}\ ,\label{BackgroundxbarGeodesic}
\end{align}
where we took $\lambda=0$ to correspond to a photon at the origin.  Writing $x^{\mu}(\lambda)=\bar{x}^{\mu}(\lambda)+\delta x^{\mu}(\lambda)$, we find
\begin{align}
\delta x^{\mu}(\lambda)&\approx\begin{pmatrix}0\,	 , & -\frac{4ce^{2}r_{s}\lambda(3L^{2}+2\lambda^{2})}{m_{e}^{2}L^{2}\left (L^{2}+\lambda^{2}\right )^{3/2}} \cos 2\theta\, , & 0\, , & 0
\end{pmatrix}\ ,\label{QEDGeodesicCorrectionSchwarzschildPair}
\end{align}
to first order in $c$ and lowest order in $r_{s}$.  In \eqref{QEDGeodesicCorrectionSchwarzschildPair}, we've switched to a non-affine parameter in order to simplify the expression and keep $\delta x^{0}(\lambda)=0$ for all $\lambda,\theta$ making the comparison between $x^{\mu}(\lambda)$ and $\bar{x}^{\mu}(\lambda)$ more straightforward.

   By calculating $\delta x^{\mu}(\lambda)$ we are effectively comparing the flight of a \textit{non}-minimally coupled photon to the flight of a minimally coupled ``test" photon on the \textit{same} background in order to understand how much the QED photon's propagation differs from that of a ``normal" photon.  We denote the non-minimally coupled photon by $\gamma_{\rm QED}$ and the minimally coupled photon by $\gamma_{\rm min}$ with the former's motion dictated by \eqref{4DEFT}, while the latter's motion would be described by only a Maxwell term.  We will often refer to $\gamma_{\rm min}$, but if one would rather avoid referring to degrees of freedom not explicitly included in the theory, the entire analysis can be rephrased as a comparison between the strictly luminal QED photon ($\theta=\pi/4$) and the other possible polarizations of $\gamma_{\rm QED}$.

This comparison between photons on the \textit{same} background avoids the complications which would arise if we were to, for example, compare the QED photon's trajectory in a black hole background to the trajectory of a null path in Minkowski space.  Difficulties even arise in attempting to contrast the trajectory of a \textit{minimally} coupled photon, $\gamma_{\rm min}$, in Schwarzschild to a flat space photon as the logarithmic Shapiro time delay term in $\bar{x}^{\mu}$ \eqref{BackgroundxbarGeodesic} causes the Schwarzschild photon to fall behind its flat space counterpart by a diverging amount $\propto r_{s}\ln \lambda$.  See \cite{PenroseCausality,Gao:2000ga} for discussions of related topics.  By comparing trajectories in the same background, we sidestep such issues.

\begin{figure}[h!] 
   \captionsetup{width=0.9\textwidth}
   \centering
     \includegraphics[width=3.3in]{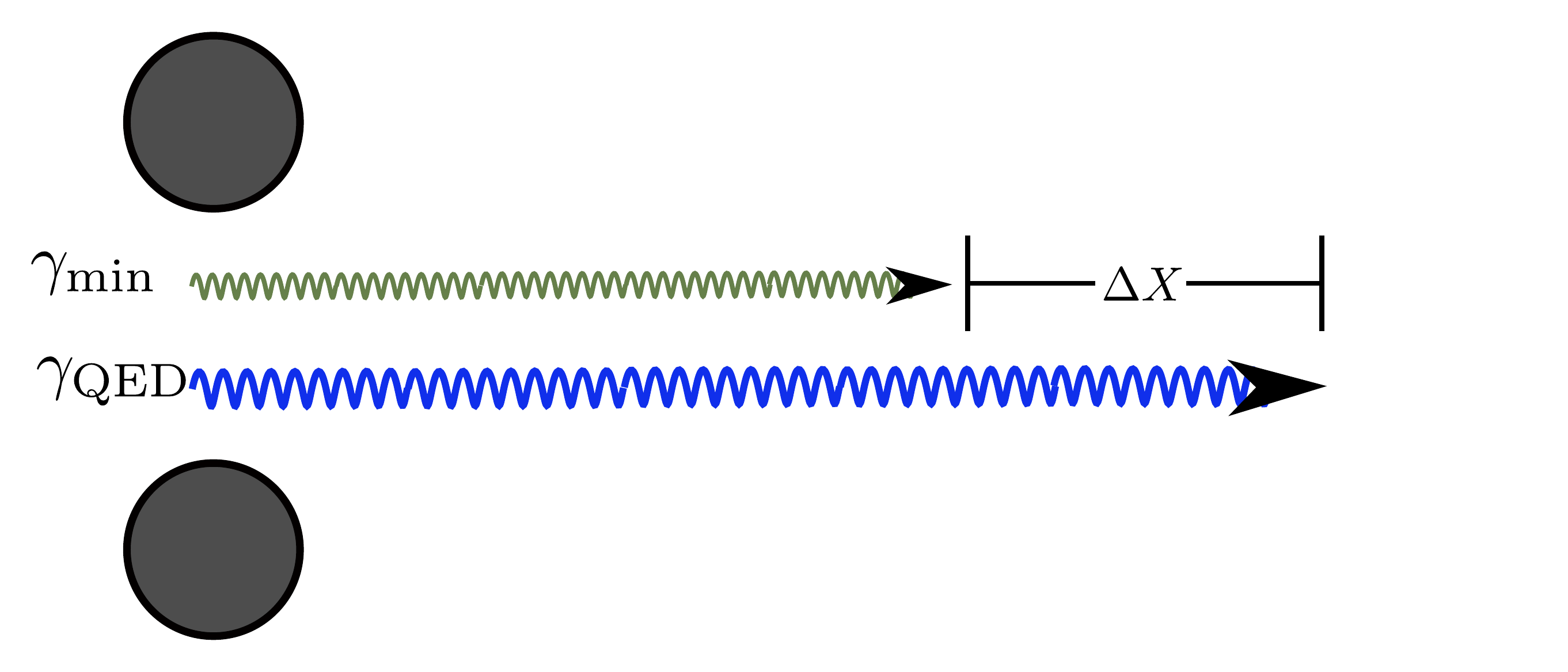}
   \caption{Exaggerated sketch of the race between a maximally superluminal photon, $\gamma_{\rm QED}$ with $\theta=\pm\pi/2$, and a minimally coupled photon, $\gamma_{\rm min}$. The QED photon wins by $\Delta X\approx m_{e}^{-1}\left (\frac{8c e^{2}r_{s}}{m_{e}L^{2}}\right )$.}
   \label{fig:RaceTwoSchwarzschildBHs}
\end{figure}

From \eqref{QEDGeodesicCorrectionSchwarzschildPair}, we can immediately compare the paths of the different photons. If $\gamma_{\rm QED}$ and $\gamma_{\rm min}$ were to race from directly between the black holes out to infinity, the asymptotic difference between the two paths is
\begin{align}
\Delta X\approx\lim _{\lambda\to \infty}\delta x^{1}(\lambda)=-m_{e}^{-1}\left (\frac{8c e^{2}r_{s}}{m_{e}L^{2}}\right )\cos 2\theta\ .\label{QEDDistanceAdvance}
\end{align}
Thus, a maximally subluminal QED photon would lose to $\gamma_{\rm min}$ by a distance $m_{e}^{-1}\left (\frac{8c e^{2}r_{s}}{m_{e}L^{2}}\right )$ which in turn would lose to the maximally superluminal QED photon by the same amount, in agreement with our estimate \eqref{QEDDistanceAdvanceApproximation}.  See Fig.$\!$ \ref{fig:RaceTwoSchwarzschildBHs}.

This distance, for the values $L\gtrsim r_s$, $L\gg m_{e}^{-1}$, $e\sim {\cal O}(1)$, is $\ll { m_e}^{-1}$, well outside the regime of validity of the EFT.  

\section{Building Up QED Superluminality}

In this section we attempt to build up the QED superluminality with the goal of achieving $\Delta X> m_{e}^{-1}$.  We start by discussing two simple attempts which can quickly be shown to fail.  Afterwards, we introduce the main amplifying scenario considered in this paper: a ladder of approximately extremal Reissner-Nordstrom (RN) black holes.

\subsection{Simple Attempts: Large $N_{f}$, Small $r_{s}$ and Photon Orbits}

Examining the expression for the QED distance advance, $\Delta X\approx m^{-1}_{e}\left (\frac{8ce^{2}r_{s}}{m_{e}L^{2}}\right )$, a few methods of amplification immediately come to mind.  Start by noting that the advance is bounded by taking the $L\to r_{s}$ limit of $\Delta X$:
\begin{align}
\Delta X\le m^{-1}_{e}\left (\frac{8ce^{2}}{m_{e}r_{s}}\right ) \ ,\label{QEDAdvanceBound}
\end{align}
corresponding to skipping the photon off of the BH horizon.
We're interested in making $\left (\frac{8ce^{2}}{m_{e}r_{s}}\right )\gg 1$ and the two basic strategies are to either make the numerator large or the denominator small.

The numerator can be made large by considering a new version of the problem where we work with $N_{f}$ flavors of electrons, instead of just one.   In this case, the distance advance formula is changed to
\begin{align}
\Delta X\approx m^{-1}_{e}\left (\frac{8c N_{f}e^{2}r_{s}}{m_{e}L^{2}}\right )
\end{align}
and the prescription is to take $N_{f}e^{2}\gg 1$.  However, this limit cannot be taken while retaining perturbative control of the theory.  The quantity $N_{f}e^{2}$ is the 't Hooft coupling (with $N_{f}$ the number of flavors, rather than the rank of the gauge group) and the one-loop vacuum polarization correction to the photon propagator is $\propto N_{f}e^{2}$.  Large 't Hooft coupling means non-perturbative photon dynamics which implies that we can't trust the approximations we have made in deriving and truncating the effective action. 

The denominator of \eqref{QEDAdvanceBound} can be made small by studying tiny black holes, those for which $r_{s}m_{e}\ll 1$.   However, such miniscule black holes are well outside of the validity of the EFT.  Heuristically, they are objects of size much smaller than the cutoff of the EFT, $r_{s}\ll m_{e}^{-1}$, and hence are not describable.  More quantitatively, the bound \eqref{QEDAdvanceBound} comes from shooting the photons quite close to the horizon of the black hole where the curvature is of order $R_{\mu\nu\rho\sigma}\sim  1/r_{s}^{2}$, meaning that $R_{\mu\nu\rho\sigma}/m_{e}^{2}\sim 1/(m_{e}r_{s})^{2}\gg 1$ and hence our truncation of the EFT \eqref{4DEFT} is invalid for this setup, as we've dropped terms which are higher order in $R_{\mu\nu\rho\sigma}/m_{e}^{2}$ that are in no way suppressed relative to the terms we've kept.

Finally, there is no obvious restriction on building up an integrated macroscopic distance advance by choosing the photon to orbit a large black hole for many cycles.  However, this setup does not permit us to send signals between asymptotic observers any faster than if there were no black hole at all, so it is not the type of sharp asymptotic observable we're interested in.

\subsection{A Ladder of Black Holes: Large $N_{\rm BH}$}

\begin{figure}[h!] 
   \captionsetup{width=0.9\textwidth}
   \centering
     \includegraphics[width=5in]{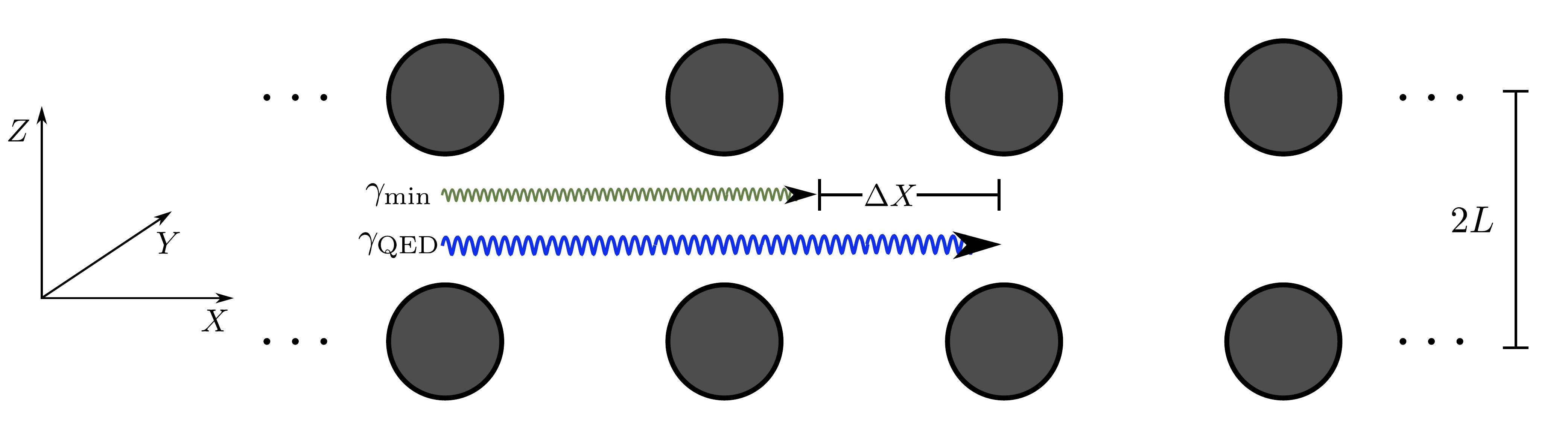}
   \caption{Racing $\gamma_{\rm min}$ and $\gamma_{\rm QED}$ down a ladder of black holes. }
   \label{fig:BlackHoleTunnel}
\end{figure}

A more fruitful direction to push is the limit of \textit{many} black holes.  We consider building a ladder of $N_{\rm BH}$ black holes, arranged in pairs with each pair constituting a rung of the ladder, and racing $\gamma_{\rm QED}$ against $\gamma_{\rm min}$ down the middle of the ladder, see Fig.$\!$ \ref{fig:BlackHoleTunnel}.

If we could construct a ladder of arbitrary length, we could clearly make the distance advance as large as we wish.    We cannot, however, as the multi-black hole solution is not generally static since the black holes mutually attract.   Initially placing the black holes at vertical separation $2L$, the photon race until the separation becomes $\mathcal{O}(r_{s})$, at which point the black holes start to merge and the ladder collapses.

Analytic control of the race is lost when the ladder coalesces, so we should attempt to prolong the lifespan of the setup. One way to accomplish this is to use identical extremal Reissner-Nordstrom black holes.  Famously, the sum of many stationary, extremal RN black holes is also an exact, {stationary} solution to \textit{pure} Einstein-Maxwell theory \cite{Majumdar:1947eu,Papaetrou:1947ib,Hartle:1972ya}, because the electromagnetic repulsion perfectly balances the gravitational attraction.  Our ladder is thus perfectly stable in such a theory. 

 However, since we are working with \textit{full} QED, not just Einstein-Maxwell, these Majumdar-Papapetrou spacetimes are only approximate solutions and the additional operators in the EFT \eqref{4DEFT} introduce new effects.  Further, they are only \textit{classical} solutions of Einstein-Maxwell theory: graviton and photon loops must also be accounted for.  We analyze the new effects in the following sections and determine whether they destabilize the ladder quickly enough to avoid macroscopic superluminality.

\section{Black Hole Ladder Analysis\label{Sec:BlackHoleLadder}}

Here we study the ladder of approximately extremal Reissner-Nordstrom black holes.  First, we recall the exact black hole solutions of pure Einstein-Maxwell theory and their relevant properties.  Next, we discuss how to calculate the perturbative corrections to these solutions, due to the electron-induced operators in the EFT \eqref{4DEFT}.  Effects of photon and graviton loops are then discussed separately, as their treatment is slightly more subtle. 
Finally, we bound the distance advance acquired by $\gamma_{\rm QED}$ in this idealized scenario.

\subsection{Einstein-Maxwell Background}

The pure Einstein-Maxwell action is:
\begin{align}
S_{\rm EM}&=\int\rd^{4}x\, \sqrt{-g}\left [\frac{M_{p}^{2}}{2}R-\frac{1}{4e^{2}}F_{\mu\nu}^{2}\right ]-M\int\rd \tau\, +\frac{Q}{e}\int\rd x^{\mu}A_{\mu}\ ,\label{EinsteinMaxwellAction}
\end{align}
where we've explicitly included the source terms for a single black hole of mass $M$ and charge $Q$.  The background equations of motion from \eqref{EinsteinMaxwellAction} read
\begin{align}
 M_{p}^{2}G_{\mu\nu}&= \frac{1}{e^{2}}\left[ F_{\mu}{}^{\alpha}F_{\nu\alpha}-\frac{1}{4}g_{\mu\nu}F^{2}\right]\ , \quad \nabla^{\nu}F_{\mu\nu}=0\, ,\label{BackgroundRNEOM}
 \end{align} and are satisfied by the Reissner-Nordstrom solution
\begin{align}
\rd s^{2}&= -\Delta(r)\rd t^{2}+\Delta(r)	^{-1}\rd r^{2}+r^{2}\rd\Omega_{2}^{2}\nn
 \Delta(r)&=1-\frac{r_{s}}{r}+\frac{r_{q}^{2}}{4r^{2}} \ ,\quad
F^{tr}=\frac{eM_{p}}{\sqrt{2}}\frac{r_{q}}{r^{2}}\nn  \quad r_{s}&\equiv \frac{M}{4\pi M_{p}^{2} }\ , \quad r_{q}\equiv \frac{Q}{\pi\sqrt{8}M_{p}}\ ,\label{BackgroundExtremal4DRN}
\end{align}
with all other components of $F_{\mu\nu}$ vanishing or related to \eqref{BackgroundExtremal4DRN} by symmetries.  The extremal black hole arises in the limit $r_{q}\to r_{s}$, at which point $\Delta(r)$ factorizes: $\Delta(r)=\left (1-\frac{r_{s}}{2r}\right )^{2}$.

In isotropic coordinates, $x^{\mu}=(t,X,Y,Z)$ with $X^{2}+Y^{2}+Z^{2}=(r-r_{s}/2)^{2}$, \eqref{BackgroundExtremal4DRN} can be generalized to the exact Einstein-Maxwell solution containing $N_{\rm BH}$ equal mass, extremal black holes at locations $\vec{X}_{i}$ \cite{Hartle:1972ya}:
\begin{align}
\rd s^{2}&= -U^{-2}\rd t^{2}+U^{2}\rd\vec{X}^{2}\ , \	\quad  A_{0}=eM_{p}\sqrt{2}U^{-1} \ , \quad  U\equiv 1+\sum_{i=1}^{N_{\rm BH}}\frac{r_{s}}{2|\vec{X}-\vec{X}_{i}|} \ .
\label{BackgroundMultiExtremal4DRNIsotopicCoordinates}
\end{align}

\subsection{Quantum Corrections}

Since we are not working with pure Einstein-Maxwell, but rather the EFT  \eqref{4DEFT}, the configurations of the previous sections are only {approximate} solutions.  There are multiple ways of calculating how the new EFT operators in \eqref{4DEFT} affect the configuration.  We find it most transparent to phrase the calculation in terms of Feynman diagrams for perturbatively solving the equations of motion \cite{Schwartz:2013pla}.  The leading corrections to the single extremal RN solution are calculated and then multiple copies are superimposed to find the approximate multi-black hole metric. Forces between nearby black holes are then calculated using the geodesic equation and the resulting dynamics are calculated in the Newtonian approximation.

Each of these steps involves approximations, but the errors are expected to be small in each case:
\begin{itemize}
\item Black hole solutions cannot generically be added together to form new solutions, due to the non-linearity of GR.  However, as long as the separation between black holes is much larger than their respective horizon sizes, the composite metric should serve as a good approximation.  Our setup falls within this regime.
\item Placing the black holes at some initial separation,  we find the Newtonian potential between a single pair of perturbed black holes and calculate how quickly they come within a distance $\sim r_{s}$ of each other.   This estimate serves as an \textit{upper} bound for the lifetime of the entire ladder, which is all we'll need.   Using a Newtonian description is only valid for weak gravitational forces and velocities much smaller than $c$.  Our setup falls within this regime (as we'll verify).
\item In practice, we've calculated contributions to $\langle h_{\mu\nu}\rangle$ using time-ordered Feynman rules.  Strictly speaking, these ``in-out" matrix elements are only equivalent to expectation values if the system is in equilibrium.  In order to calculate true expectation values and capture non-equilibrium effects such as Hawking radiation (see \cite{Akhmedov:2015xwa} for a recent such study), one would instead need to use the full Schwinger-Keldysh or ``in-in" formalism \cite{KamenevBook,Weinberg:2005vy}. However, because the black holes are nearly extremal, their evaporation rate is miniscule and we expect such effects to be negligible.
\item Photon and graviton loops generate subtle corrections to the metric, which are not immediately interpretable as unambiguous corrections to the Newtonian potential.  They require a more careful treatment, as is discussed in Sec. \ref{Sec:SubtleGaugeLoops}.
\end{itemize}

\subsubsection{Feynman Diagram Estimation: RN Black Holes}

Finding black hole solutions via diagrams has a long history.  Duff pioneered the subject, both reproducing the usual, classical Schwarzschild solution \cite{Duff:1973zz} and finding the leading \textit{quantum} corrections to the metric \cite{Duff:1974ud} from graviton loops (though, this latter subject turns out to be surprisingly subtle, see Sec. \ref{Sec:SubtleGaugeLoops}).   
Though diagrams can be used to find the exact perturbative corrections with all numerical coefficients determined, the primary utility of Feynman diagrams for the present purpose is their efficiency in estimating, comparing and organizing competing corrections to $h_{\mu\nu}$ and $A_{\mu}$.

Consider building up the generic RN solution via Feynman diagrams.  The schematic form of the Einstein-Maxwell action, with source terms included, is
\begin{align}
\mathcal{L}&\sim l_{p}^{-2}h^{n}(\partial h)^{2}-\frac{1}{4e^{2}}h^{n}(\partial A)^{2}-r_{s}l_{p}^{-2} h\delta^{3}(r)+\frac{r_{q}l_{p}^{-1}}{e}A\delta^{3}(r) \ ,
\end{align} 
sufficient for our purposes.  We found it convenient to write all quantities in terms of length scales, $l_{p}=M_{p}^{-1}$, $r_{s}\sim M l_{p}^{2}$ and $r_{q}\sim Ql_{p}$.  The derivation of the Feynman rules is standard and we'll only need their schematic form:
\begin{itemize}
\item All diagrams are drawn as sources feeding into $\langle h_{\mu\nu}\rangle$ or $\langle A_{\mu}\rangle$ from right to left.
\item \gravitonline is a graviton line, appearing with a factor $\sim l_{p}^{2}$.
\item \photonline is a photon line, appearing with a factor $\sim e^{2}$.
\item Any line whose right end is bare has a source attached to that end.
\item If \gravitonline has a source attached, it gets another factor of $\sim r_{s}l_{p}^{-2}$.
\item If \photonline has a source attached, it gets another factor of $\sim r_{q}l_{p}^{-1}e^{-1}$.
\item \classicalEHvertex is an Einstein-Hilbert vertex, appearing with a factor $\sim l_{p}^{-2}$.
\item \classicalFFvertex is a Maxwell vertex, appearing with a factor $\sim e^{-2}$.
 \item Overall dimensions are fixed by inserting factors of $r$.
\end{itemize}
We will refer to any line whose right end is bare as ``external."
These rules allow for fast and easy estimations of the various contributions to the solution.

\begin{figure}[h!] 
   \captionsetup{width=0.9\textwidth}
   \centering
     \includegraphics[width=7in]{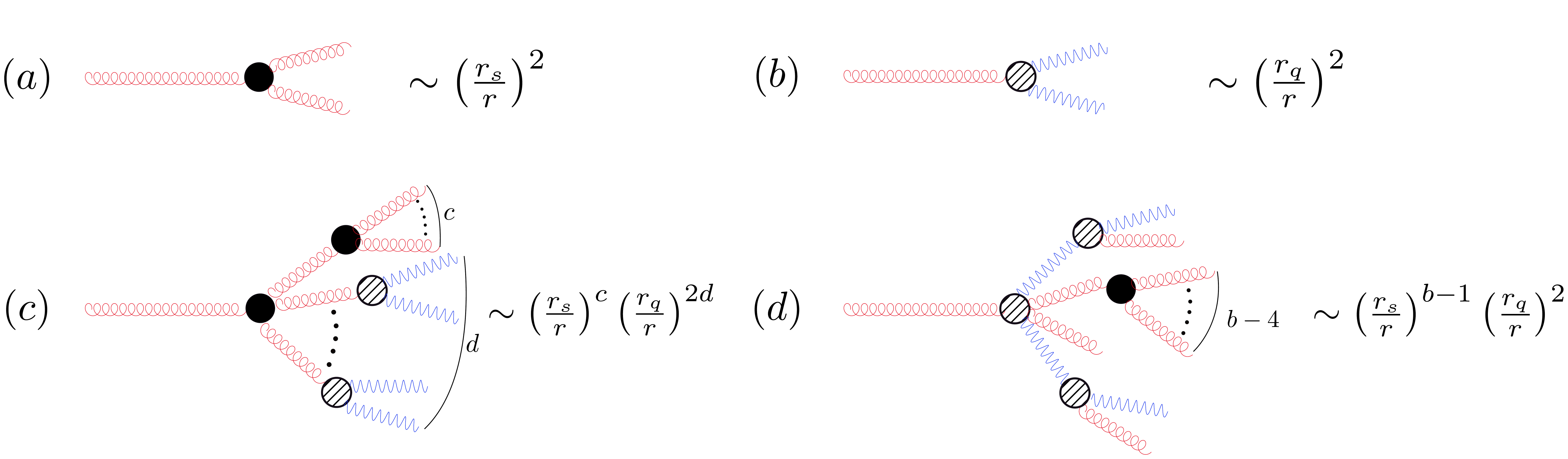}
   \caption{Some examples of Feynman diagrams for the Reissner-Nordstrom metric solution.  Diagrams $(a)$ and $(b)$ are the leading GR correction and charge contribution.  Diagrams $(c)$ and $(d)$ are complicated diagrams whose sizes are easily estimated.}
   \label{fig:ClassicalEinMaxDiagrams}
\end{figure}

For instance, the linear solution for the metric corresponds to a single graviton line: \gravitonline.  The estimation of this diagram is simple. Every graviton line comes with a factor of $\sim l_{p}^{2}$ and since the right end of the single line is bare, there's also a single factor of the source $\sim r_{s}l_{p}^{-2}$.  Feynman rules and dimensional analysis quickly yield the estimate of the standard Newtonian potential,
\begin{align}
\langle h_{\mu\nu}\rangle&= \gravitonline\approx\frac{1}{r}\times l_{p}^{2}\times r_{s}l_{p}^{-2}=\frac{r_{s}}{r}\ .
\end{align}
We similarly estimate that the linearized photon solution is 
\be \langle A_{\mu}\rangle =\photonline\approx \frac{1}{r}\frac{er_{q}}{l_{p}}\,. \ee
The first GR correction and the leading contribution of charge to the metric arise from cubic vertices, as shown in Fig.$\!$ \ref{fig:ClassicalEinMaxDiagrams} $(a)$ and $(b)$. Both are easily estimated and are found to be of the correct form, as can be seen by comparing to the full solution \eqref{BackgroundExtremal4DRN}.  It's also easy to estimate the sizes of more complicated diagrams, see Fig.$\!$ \ref{fig:ClassicalEinMaxDiagrams} $(c)$ and $(d)$.

This diagrammatic language greatly helps to organize the calculation.   In the limiting case of finding the classical, extremal RN solution via Feynman diagrams, the organization is fairly trivial: there's only one scale $r_{s}$ entering the metric solution and a diagram with $n$ external legs and arbitrarily many vertices gives a contribution of size $\sim \left (r_{s}/r\right )^{n}$ to $\langle h_{\mu\nu}\rangle$.   However, when we consider both classical and quantum diagrams built using the full EFT \eqref{4DEFT} several more scales and factors appear ($e$, $m_{e}$ and $l_{p}$) and diagrams become an invaluable organizational tool.

Finally, note that important physics is clearly expressed through these estimates: diagrams tell us the scale at which physics qualitatively changes due to non-linearities and the breakdown of perturbation theory.  For instance, the linear  Schwarzschild BH solution is $\langle h_{\mu\nu}\rangle\sim r_{s}/r$ while the non-linear corrections are all of size $\sim \left (r_{s}/r\right )^{n}$.  These GR corrections are therefore small for $r\gtrsim r_{s}$ and important for $r\lesssim r_{s}$, at which point perturbation theory breaks down and we need to find the full non-linear solution for the metric \eqref{BackgroundExtremal4DRN}. A similar analysis for generic RN black holes demonstrates that non-linearities are important at the whichever scale is largest among $r_{s},r_{q}$.   $r_{s}\gg r_{q}$ corresponds to horizon formation, $r_{s}\ll r_{q}$ corresponds to a naked singularity and $r_{s}\sim r_{q}$ is a special neighborhood containing the extremal black hole.        Generically, physics changes qualitatively when non-linearities become important and Feynman diagrams provide a quick way of determining where these interesting non-linear scales lie.

\subsubsection{Tree Diagrams for  $\langle h_{\mu\nu}\rangle$}

Now we estimate the sizes of the various metric corrections coming from the new EFT operators in \eqref{4DEFT}.  Since the background forces cancel for the approximately extremal RN black holes we're considering, these electron induced corrections provide the leading gravitational forces which destabilize the ladder setup. 
Summing up tree diagrams is equivalent to taking the equations of motion derived from \eqref{4DEFT} and solving perturbatively.  As we'll see, this alone is not sufficient and misses important contributions to the solution.  

The schematic Lagrangian now reads:
\begin{align}
\mathcal{L}&\sim l_{p}^{-2}h^{n}(\partial h)^{2}-\frac{1}{4e^{2}}h^{n}(\partial A)^{2}+r_{e}^{2}h^{n}\partial^{2}h (\partial A)^{2}+r_{e}^{4}h^{n}(\partial A)^{4}-r_{s}l_{p}^{-2} h\delta^{3}(r)+\frac{r_{q}l_{p}^{-1}}{e}A\delta^{3}(r) \ ,\label{SchematicEFTAction}
\end{align}
where we introduced the length scale $r_{e}\equiv m_{e}^{-1}$.  The third term in \eqref{SchematicEFTAction} corresponds to all of the $\sim m_{e}^{-2}RFF$ operators, the fourth corresponds to the $\sim m_{e}^{-4}F^{4}$ operators and we neglected the $d$ operator $\sim m_{e}^{-2}(\partial F)^{2}$ as it's redundant and only ends up providing subleading corrections.  A single new Feynman rule is sufficient for estimating the sizes of the new diagrams:
\begin{itemize}
\item \quantumgravityvertex is an electron induced vertex.
\item If \quantumgravityvertex has two photon lines attached, it corresponds to the third term in \eqref{SchematicEFTAction} and appears with a factor $\sim r_{e}^{2}$.
\item If \quantumgravityvertex has four photon lines attached, it corresponds to the fourth term in \eqref{SchematicEFTAction} and appears with a factor $\sim r_{e}^{4}$.
\end{itemize}

\begin{figure}[h!] 
   \captionsetup{width=0.9\textwidth}
   \centering
     \includegraphics[width=5in]{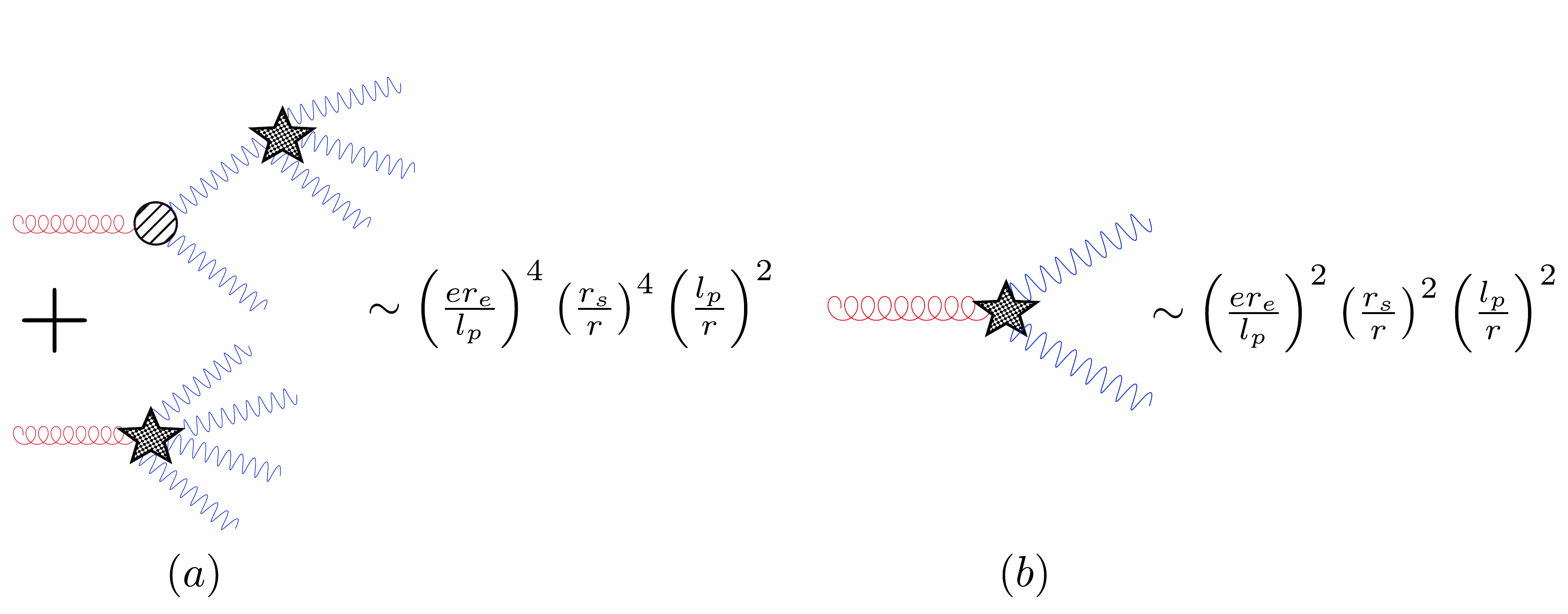}
   \caption{The simplest tree-level EFT corrections to the metric.}
   \label{fig:SimplestEFTMetricDiagrams}
\end{figure}

The simplest EFT corrections to the metric are shown in Fig.$\!$ \ref{fig:SimplestEFTMetricDiagrams}. 
Easy estimates demonstrate that a diagram utilizing both an insertion of the Maxwell term and an $\sim m_{e}^{-4} F^{4}$ operator and a diagram using only a single $\sim m_{e}^{-4} F^{4}$ insertion are of the same order.  The former corresponds to finding the correction to $A_{\mu}$ from the $\sim F^{4}/m_{e}^{4}$ operators and feeding the result into the Einstein-Maxwell $T_{\mu\nu}$ to find how it affects the metric. This correction is just as important as the $\sim F^{4}/m_{e}^{4}$ operators' direct contribution to the metric (the lower diagram in Fig.$\!$ \ref{fig:SimplestEFTMetricDiagrams}).  
 Comparing the two diagrams in Fig.$\!$ \ref{fig:SimplestEFTMetricDiagrams}, one finds that the $(a)$ dominates for $r\lesssim r_{s}\left (\frac{er_{e}}{l_{p}}\right )$ and $(b)$ dominates for $r\gtrsim r_{s}\left (\frac{er_{e}}{l_{p}}\right )$.  The form of the gravitational force law changes depending on how separated the black holes are.

The factor $er_{e}/l_{p}=eM_{p}/m_{e}$ arises often in the calculation and is exactly the quantity that the Weak Gravity Conjecture states should be larger than unity in any theory which can be UV completed into a consistent theory including quantum gravity \cite{ArkaniHamed:2006dz}.  The Standard Model electron satisfies this bound easily, $er_{e}/l_{p}\sim 10^{22}$, and (unless specified otherwise) we proceed assuming that our theory also satisfies this bound, as we wish to stay as close to real world QED as possible.

Finally, we can double check that our calculation is within the validity of the EFT. The existence of electrons shouldn't produce drastic changes to the background solutions, otherwise we're not in the regime where the EFT is valid.  This means that all diagrams using EFT vertices should be small compared to the background solution in the regime of interest, $r\gtrsim r_{s}$.  Equivalently, this implies that any new non-linear scales induced by the new EFT operators should be smaller than $\sim r_{s}$, so that GR's non-linearities always become important first.  

These properties do not hold for all black holes: there is a minimal size black hole below which the EFT description breaks down.
In order to see this, consider evaluating the two diagrams in Fig.$\!$ \ref{fig:SimplestEFTMetricDiagrams} just outside the horizon, $r\sim r_{s}$ where the linear solution and all Einstein-Maxwell corrections are starting to become $\mathcal{O}(1)$.  If electrons aren't very important, then both diagrams should be $\ll \mathcal{O}(1)$ in this regime.  Figs. \ref{fig:SimplestEFTMetricDiagrams} $(a)$ and $(b)$ satisfy this condition only if $r_{s}\gg er_{e}\left (er_{e}/l_{p}\right )$ and $r_{s}\gg l_{p}\left (er_{e}/l_{p}\right )$, respectively, and the first constraint is strongest, due to the WGC assumption.

More stringent bounds come from considering different diagrams.  It turns out that the strongest constraints come from the diagram with a single insertion of an operator $\sim r_{e}^{2n}h(\partial A)^{2(n+1)}$ with $n\to \infty$.  The diagram gives
\begin{align}
\langle h_{\mu\nu}\rangle\sim \left (\frac{er_{e}}{l_{p}}\right )^{n+1}\left (\frac{r_{e}}{r}\right )^{n-1}\left (\frac{r_{s}}{r}\right )^{2(n+1)}\left (\frac{l_{p}}{r}\right )^{2}
\end{align}
and $\langle h_{\mu\nu}\rangle\ll  \mathcal{O}(1)$ at $r\sim r_{s}$ iff $r_{s}\gtrsim r_{e}\left (\frac{er_{e}}{l_{p}}\right )$. If the bound is violated, an infinite tower of operators generate important corrections to the solution.    Again, this analysis is equivalent to identifying the distance scale at which EFT non-linearities become important and then demanding that this scale be smaller than $r_{s}$.

 The bound $r_{s}\gtrsim r_{e}\left (\frac{er_{e}}{l_{p}}\right )$ is nothing but the statement that our EFT description is only valid for field strengths obeying $F^{2}/m_{e}^{2}\ll 1$.  The strongest field strengths we probe are of order
\begin{align}
F_{\mu\nu}\big|_{r=r_{s}}\sim \partial A\big|_{r=r_{s}}\sim \frac{e}{r_{s}l_{p}}\ ,
\end{align} which is smaller than $m_{e}^{2}$ only if\footnote{Funny numerology occurs when this bound is evaluated for the Standard Model.  In terms of the BH mass ($M_{\rm BH}$) and solar masses ($M_{\odot}$), a few fundamental numbers ($e$, $M_{p}$ and $m_{e}$) combine to yield the condition $M_{\rm BH}\gtrsim 10^5M_{\odot}$, as pointed out in \cite{Gibbons:1975kk}, roughly corresponding to the lower mass range of real world supermassive black holes.} $r_{s}\gtrsim r_{e}\left (\frac{er_{e}}{l_{p}}\right )$, which is the condition we found through diagrams.  Physically, we expect rampant $e^{+}$, $e^{-}$ Schwinger pair production when this condition is violated\footnote{For a generic RN BH, a similar analysis gives the condition $r_{s}^{2}\gg r_{e}r_{q}\left (\frac{er_{e}}{l_{p}}\right )$.}, in agreement with the scale found in the detailed pair production analysis of \cite{Gibbons:1975kk}.

The rough estimates given above are fully realized in the precise results of the actual calculation \cite{ToAppear}. 
A similar analysis for the vector potential solution is straightforward and yields the same conclusions.   After finding the leading perturbative corrections, we can simply read off the electron corrections to $h_{00}$ to find the gravitational potential induced by electrons, and similarly for the zero component of the vector potential.

\subsubsection{ $\langle h_{\mu\nu}\rangle$ from Photon and Graviton Loops}

The tree diagrams of the previous section miss an important effect: the contribution of graviton and photon loops.  Not only should these be included for consistency, they also generate the dominant corrections at long distances and qualitatively change the dynamics.  These corrections would be missed entirely if one simply tried to find the metric via perturbatively solving the EFT equations of motion derived from \eqref{4DEFT}.  Instead, they are captured by the 1PI effective action, discussed later.  The use of Feynman diagrams makes it particularly clear that these corrections need to be included and quickly singles out the regime where they dominate.

Physically, it's entirely reasonable that loops of photons and gravitons should compete with the effects of the new EFT operators and that they should dominate at long distances.  Recall that the generated EFT operators all arose via electron loops, as in Fig.$\!$ \ref{fig:MatchingQEDRFF}, and hence the tree diagrams considered in the previous section correspond to loop diagrams in the full theory.    They represent quantum effects.   Since photon and graviton loops represent the quantum corrections from \textit{massless} particles, their effects should be very long ranged, dominating the corrections far from the source, while electron loop effects dominate at shorter distances.

\begin{figure}[h!] 
   \captionsetup{width=0.9\textwidth}
   \centering
     \includegraphics[width=6in]{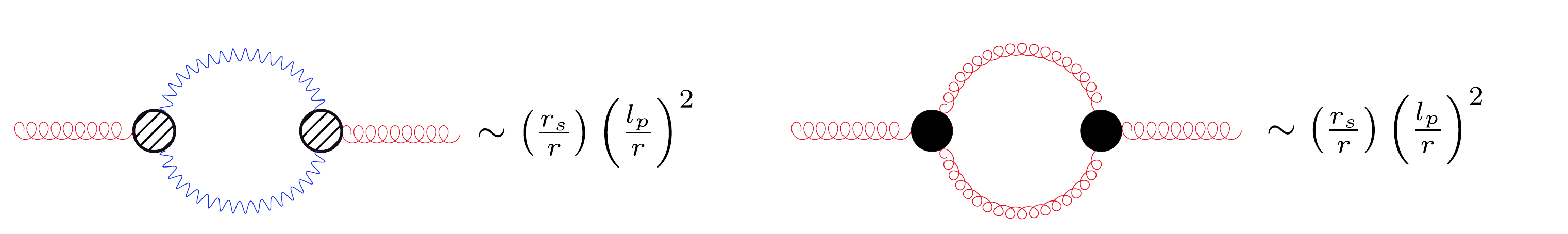}
   \caption{Typical loop corrections to the metric from gravitons and photons.  Here, and in following figures, we leave the necessary ghost diagrams implicit.}
   \label{fig:MetricLoopCorrections}
\end{figure}

Typical loops needed for the calculation are shown in Fig.$\!$  \ref{fig:MetricLoopCorrections}. The full calculation of graviton and photon loops is fairly painful, due to the plethora of indices \cite{'tHooft:1974bx,Goroff:1985th,Capper:1974ed}.  Fortunately, our Feynman rules for approximating diagrams faithfully reproduce the size of these corrections to the metric, first calculated by Duff \cite{Duff:1974ud}.  Very closely related (but not entirely equivalent) ideas were later stated in modern EFT language by Donoghue\footnote{Many, many authors have calculated the correction using a variety of methods.  See, for instance, \cite{Donoghue:2001qc,BjerrumBohr:2002ks,Kirilin:2006en} and \cite{Burgess:2003jk} for a review.  Not all results agree in their precise numerical coefficients, but all find the same order of magnitude as Fig.$\!$ \ref{fig:MetricLoopCorrections}.} \cite{Donoghue:1994dn}.  

The estimate for the correction follows from Fig.$\!$ \ref{fig:MetricLoopCorrections}: $\delta h_{\mu\nu}\sim \frac{r_{s}}{r}\left (\frac{l_{p}}{r}\right )^{2}$.
Comparing these loop corrections to the tree diagrams of the previous section, Fig.$\!$ \ref{fig:SimplestEFTMetricDiagrams} $(a)$ and $(b)$, we find that light loops dominate the metric corrections at distances $r\gtrsim r_{e}\left (\frac{er_{e}}{l_{p}}\right )^{2}$.

\subsubsection{Subtleties of Gauge Loops\label{Sec:SubtleGaugeLoops}}

Unfortunately, turning these \textit{gauge} loop diagrams into a potential is not so straightforward a process.  We can't simply find $\delta h_{00}$ and take this to be the potential because the 1PI action for GR is gauge dependent, which makes the correction $\delta h_{\mu\nu}$ ambiguous.

 Starting from the GR action with a point source,
\begin{align}
 S=\int\rd^{4}x\sqrt{-g}\, \frac{M_{p}^{2}}{2}R-M\int\rd\tau \ ,
 \end{align} we can calculate the 1PI action $\Gamma[\bar{g}_{\mu\nu}]$ via the background field method (BFM) \cite{Abbott:1981ke} by expanding the Einstein-Hilbert action about $g_{\mu\nu}=\bar{g}_{\mu\nu}+\delta g_{\mu\nu}$ and integrating over all 1PI graphs where only $\delta g_{\mu\nu}$ propagates in loops,
 \begin{align}
 \exp	i\Gamma[\bar{g}]&=\int_{\rm 1PI}\mathcal{D}\delta g_{\mu\nu}\exp i S[\bar{g}_{\mu\nu}+\delta g_{\mu\nu}]\label{1PIGRAction}\ .
 \end{align}

Actually performing the calculation \eqref{1PIGRAction} requires gauge fixing for $\delta g_{\mu\nu}$ and we can ensure that $\Gamma[\bar{g}_{\mu\nu}]$ is built from diffeomorphism invariant operators by making a clever choice of gauge fixing functional \cite{'tHooft:1974bx}, for instance $G_{\mu}=\bar{\nabla}^{\nu}\delta g_{\nu\mu}-\frac{1}{2}\bar\nabla_{\mu}\delta g^{\nu}{}_{\nu}$.  A gauge fixing term $\mathcal{L}_{\rm gf}= -\frac{1}{2\xi}G_{\mu}G^{\mu}$ is then added to the action (along with the associated ghost terms) where $\xi$ is an arbitrary parameter.  Performing the necessary integrals, the one-loop 1PI action contains the following non-analytic operators \cite{Dalvit:1997yc}
\begin{align}
\Gamma[\bar{g}_{\mu\nu}]&\supset\int\rd^{4}x\,\sqrt{-\bar g} \left(c_{1}\bar R\log(-\square/\mu ^{2})\bar R+c_{2}\bar R^{\mu\nu}\log(-\square/\mu ^{2})\bar R_{\mu\nu}+c_{3}\bar R^{\mu\nu\rho\sigma}\log(-\square/\mu ^{2})\bar R_{\mu\nu\rho\sigma}\right )\nn
&\quad -c_{4}\frac{M}{M_{p}^{2}}\int\rd\tau\log(-\square/\mu ^{2})\bar  R-c_{5}\frac{M}{M_{p}^{2}}\int\rd\tau\log(-\square/\mu ^{2}) \bar R_{\mu\nu}\frac{\rd x^{\mu }}{\rd\tau}\frac{\rd x^{\nu}}{\rd\tau} \ ,\label{1PIGRActionResult}
\end{align}
with some calculable coefficients $c_{i}$.
The three new operators in the first line correspond to the vacuum polarization diagrams in Fig.$\!$ \ref{fig:MetricLoopCorrections}, along with similar diagrams with more external legs.  The new operators in the final line come from diagrams using insertions of vertices from the point source term.

Using \eqref{1PIGRActionResult}, we can calculate $\langle \delta h_{\mu\nu}\rangle$ in a precise manner: expand $\bar{g}_{\mu\nu}$ about flat space, add a \textit{new} gauge fixing term to make the propagator invertible and compute tree diagrams using the terms in the second line of \eqref{1PIGRActionResult} as the source terms.  This is essentially the method used by Duff \cite{Duff:1974ud}, though the matter corrections $c_{3},c_{4}$ were neglected there.

The problem, then, is that many of the $c_{i}$'s in \eqref{1PIGRActionResult} depend on the choice of the gauge fixing parameter $\xi$ used to fix the background fluctuation in \eqref{1PIGRAction}. The $\xi$ dependence of the $c_{i}$'s then feeds into the metric, which also ends up being $\xi$-dependent.  While the $\xi$-dependence cancels out of the one-loop, BFM result for $\Gamma[\bar{g}_{\mu\nu}]$ in Yang-Mills theories, the analogue statement is not true in GR, a property ascribed to the non-renormalizable nature of GR in \cite{Barvinsky:1985an}.    The background gauge fixing is logically distinct from the gauge fixing required when using $\Gamma[\bar{g}_{\mu\nu}]$ to find $\langle \delta h_{\mu\nu}\rangle$ and represents a true ambiguity.  For instance, the value of the Ricci scalar induced via the one-loop corrections in \eqref{1PIGRActionResult} depend on $\xi$, but not on the parameter used in gauge fixing $\Gamma[\bar{g}_{\mu\nu}]$ to compute the necessary tree diagrams.

This is a general property of the effective action for theories with gauge fields; see, for instance, \cite{Andreassen:2014eha,Nielsen:1975fs,Fukuda:1975di,Aitchison:1983ns}.  The field profiles which extremize the 1PI effective action are generically gauge dependent, since the form of the 1PI action is itself gauge dependent.  Instead of finding the metric, one must use $\Gamma[\bar{g}_{\mu\nu}]$ to calculate physical quantities such as $S$-matrix elements \cite{BjerrumBohr:2002kt,Khriplovich:2002bt} or modified geodesic equations \cite{Dalvit:1997yc} which account for the non-minimal matter coupling in \eqref{1PIGRActionResult}, each of which yield $\xi$-independent predictions.

Despite these subtleties in turning the diagrams of Fig.$\!$ \ref{fig:MetricLoopCorrections} into precise potentials, the figures and power counting rules constitute a good mnemonic for the calculation: the correction of the potential due to massless loops is $\delta V\sim \frac{r_{s}}{r}\left (\frac{l_{p}}{r}\right )^{2}$ \cite{Donoghue:1994dn,Donoghue:1994dn,BjerrumBohr:2002ks,BjerrumBohr:2002kt,Khriplovich:1994qj, Khriplovich:2002bt,Dalvit:1997yc}.  Therefore, we continue to use the diagrams of Fig.$\!$ \ref{fig:MetricLoopCorrections} as a representation of the effect.  The exact one-loop potential is calculated in Appendix \ref{Appendix:OneLoopMasslessPotential} by combining the results of \cite{Holstein:2008sy,BjerrumBohr:2002sx,BjerrumBohr:2002kt}

\subsubsection{Combining All Effects}

\begin{figure}[h!] 
   \captionsetup{width=0.9\textwidth}
   \centering
     \includegraphics[width=6in]{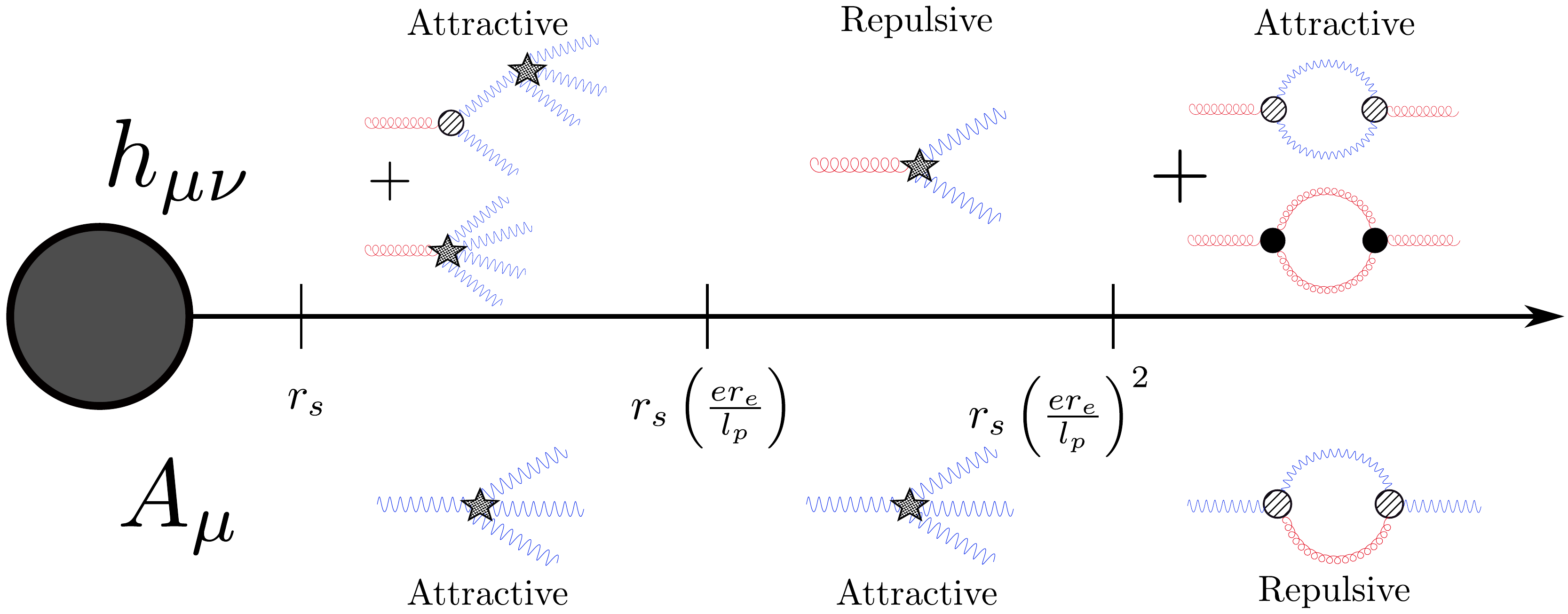}
   \caption{Dominant metric and vector potential corrections at different distances from the black hole.  We indicate whether each diagram corresponds to an attractive or repulsive force between this black hole and a second, identical one, which we imagine is placed in the indicated region. }
   \label{fig:NonSUSYDominantPerturbations}
\end{figure}

Combining the results of the previous two sections, along with the results of the vector potential estimates, we find that the calculation breaks up into three regions where different effects dominate, see Fig.$\!$ \ref{fig:NonSUSYDominantPerturbations}.

If we were to place another, identical black hole in the spacetime, we'd find that the form of the force law depends on the separation: there are three distinct behaviors, depending on which region of Fig.$\!$ \ref{fig:NonSUSYDominantPerturbations} we place the second black hole.  However, no matter where we place the second black hole, it is found that the corrections generate an \textit{attraction} between the black holes.   Not all of the individual diagrams in Fig.$\!$ \ref{fig:NonSUSYDominantPerturbations} generate an attractive perturbative correction, but when all corrections are summed up with their precise signs and coefficients, everything works out such that attraction persists at all scales.

The electron induced effects can be accurately captured as perturbative corrections to the metric and field strength tensor; there are no subtle gauge issues here. Writing the full solution for the metric and vector potential as $g_{\mu\nu}=\bar{g}_{\mu\nu}+\delta g_{\mu\nu}$ and $F_{\mu\nu}=\bar{F}_{\mu\nu}+\delta F_{\mu\nu}$ with $\bar{g}_{\mu\nu},\bar{F}_{\mu\nu}$ the classical, extremal RN solution of \eqref{BackgroundExtremal4DRN}, it is found \cite{ToAppear} that electrons induce the corrections:
\begin{align}
\delta g_{tt}&=\left (c-2a\right )\left (\frac{er_{e}}{l_{p}}\right )^{2}\left (\frac{r_{s}}{r}\right )^{2}\left( \frac{l_{p}}{r}\right )^{2}+\frac{(y+2z)}{10}\left (\frac{er_{e}}{l_{p}}\right )^{4}\left (\frac{r_{s}}{r}\right )^{4}\left( \frac{l_{p}}{r}\right )^{2}\nn
\delta g_{rr}&=\left (8a+3b+4c\right )\left (\frac{er_{e}}{l_{p}}\right )^{2}\left (\frac{r_{s}}{r}\right )^{2}\left( \frac{l_{p}}{r}\right )^{2}+\frac{(y+2z)}{10}\left (\frac{er_{e}}{l_{p}}\right )^{4}\left (\frac{r_{s}}{r}\right )^{4}\left( \frac{l_{p}}{r}\right )^{2}\nn
\delta F^{tr}&=\frac{e}{rl_{p}}\left [4\sqrt{2}c\left (\frac{er_{e}}{l_{p}}\right )^{2}\left (\frac{r_{s}}{r}\right )^{2}\left (\frac{l_{p}}{r}\right )^{2}-2\sqrt{2}\left (y+2z\right )\left (\frac{er_{e}}{l_{p}}\right )^{4}\left (\frac{r_{s}}{r}\right )^{3}\left (\frac{l_{p}}{r}\right )^{2}\right ] \label{FullEFTMetricAndVectorCorrections}
\end{align}
and all other perturbations are vanishing or trivially related to the above.

 Again, the result \eqref{FullEFTMetricAndVectorCorrections} only represents the \textit{dominant} long distance corrections to the metric due to electrons; many subleading terms are neglected.  For instance, for every diagram used in building the above, we could attach $n$ more external graviton lines to create a related diagram which is down by a factor of $\sim (r_{s}/r)^{n}$, relative to the original.  These are all negligible for the interests of this paper, but are necessary for understanding the near horizon region, calculating how the fermion field affects the Hawking temperature, etc.  Re-summing these subleading terms requires solving the fully non-linear EOM, while still working only to leading order in EFT coefficients \eqref{EFTCoefficients}.  This is done in \cite{ToAppear}.

 Massless loops dominate for $r\gtrsim r_{s}\left (\frac{er_{e}}{l_{p}}\right )^{2}$ and writing their representation as a contribution to $\delta g_{\mu\nu}$ and $\delta F^{\mu\nu}$ is misleading due to the gauge loop subtleties covered in Sec. \ref{Sec:SubtleGaugeLoops}.  The precise one-loop potential generated by massless loops is calculated in Appendix \ref{Appendix:OneLoopMasslessPotential}, using the work of \cite{Holstein:2008sy,BjerrumBohr:2002sx,BjerrumBohr:2002kt}, and is found to be of the expected, attractive $\delta V\sim \frac{r_{s}}{r}\left (\frac{l_{p}}{r}\right )^{2}$ form.

\begin{figure}[h!] 
   \captionsetup{width=0.9\textwidth}
   \centering
     \includegraphics[width=6in]{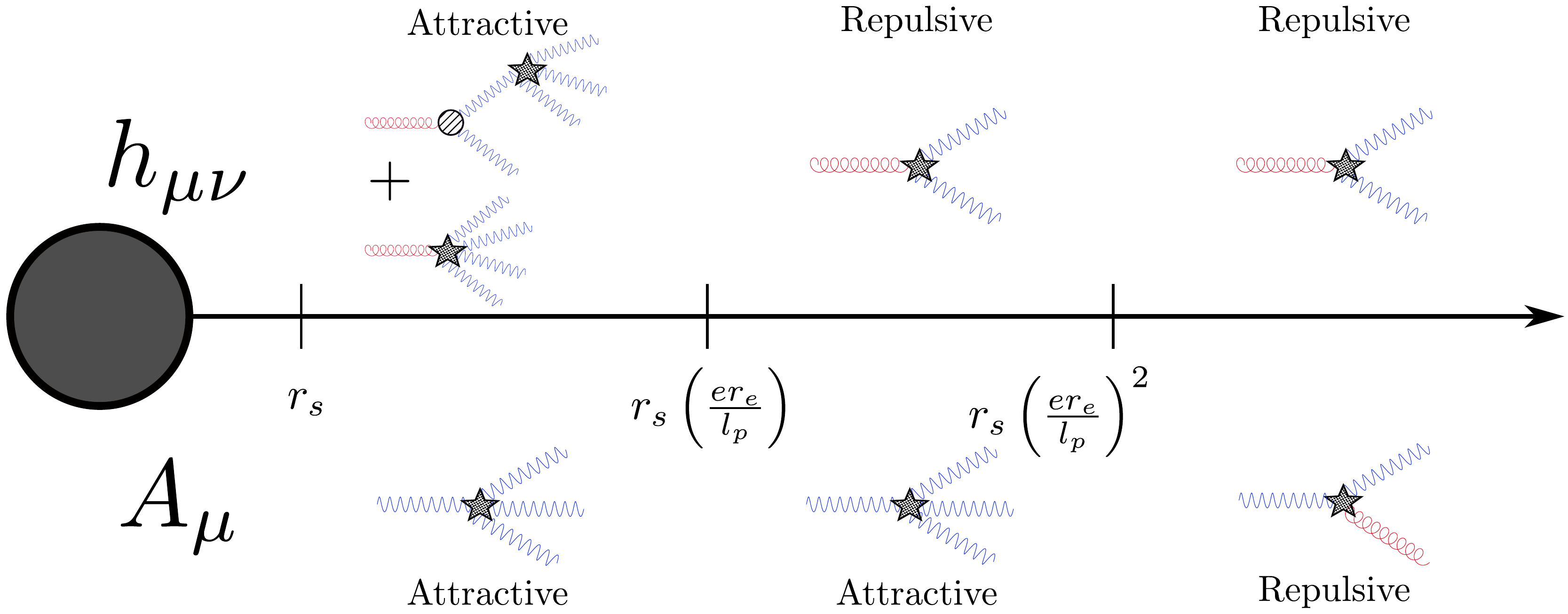}
   \caption{Dominant metric and vector potential corrections at different distances from the black hole \textit{when massless loops are neglected}.  By ignoring massless loops, qualitatively wrong dynamics are found.}
   \label{fig:SUSYDominantPerturbations}
\end{figure}

Before we analyze the dynamics of the black hole ladder, we wish to quickly emphasize the importance of including gauge loops. Had they been neglected, we'd find qualitatively wrong physics.  Including only the effects of electrons in the analysis, the sketch of the system would be changed from Fig.$\!$ \ref{fig:NonSUSYDominantPerturbations} to Fig.$\!$ \ref{fig:SUSYDominantPerturbations}.  The result is a hilltop potential which generates an attraction between extremal RN black holes separated by distances $r\lesssim r_{e}\left (\frac{er_{e}}{l_{p}}\right )^{2}$ and a repulsion for those separated by $r\gtrsim r_{e}\left (\frac{er_{e}}{l_{p}}\right )^{2}$.    This is the behavior one would find by only perturbatively solving the equations of motion arising from \eqref{4DEFT}.

\subsection{Tunnel Dynamics and Distance Advance}

From the perturbative corrections \eqref{FullEFTMetricAndVectorCorrections}, we can calculate the forces which act on the tunnel and ask whether is collapses before we are able to up a distance advance which \textit{parametrically} violates the macroscopic superluminality bound, $\Delta X\gtrsim m_{e}^{-1}$.  We find that no such violation is possible: our setup only approaches this bound from below and always remains a parametric distance away from saturation. Precisely, we find $\Delta X\lesssim e\times m_{e}^{-1}$.  We ignore Hawking radiation and assume that the black holes retain a fixed charge-to-mass ratio throughout the process.

\subsubsection{Tunnel Dynamics}

Consider the dynamics of a single pair of black holes.  The entire tunnel would coalesce at least as quickly as this pair would, hence as a conservative estimate we need only look at the dynamics of this single pair.  
 A particle of charge $q$ and mass $m$ traveling in some charged spacetime obeys the geodesic equation sourced by the Lorentz force law,
 \begin{align}
 \frac{\rd x^{\nu}}{\rd\tau}\nabla_\nu\frac{\rd x^{\mu }}{\rd \tau}&=\frac{q}{em}F^{\mu}{}_{\nu}\frac{\rd x^{\nu}}{\rd\tau}\ .\label{ModifiedGeodesicEquationCharged}
  \end{align}
  We model the motion of the far separated black holes via the above relation.
  In the Newtonian limit, the spatial components of the above reduce to their familiar form,
  \begin{align}
  \frac{\rd^{2}x^{i}}{\rd t^{2}}&=\frac{1}{2}\partial_{i}h_{00}+\frac{q}{em}F^{i}{}_{0}\, ,\label{NewtonianLimitModifiedGeodesicEqn}
  \end{align}
where $g_{\mu\nu}=\eta_{\mu\nu}+h_{\mu\nu}$. 

We model the two black holes by two coupled Newtonian equations, each of the form \eqref{NewtonianLimitModifiedGeodesicEqn}.    The two body problem can be reduced to an effective one-body problem for the separation between the black holes, in the usual way.  Letting $r=|\vec{x}_{1}-\vec{x}_{2}|$ be the separation between the pair, the relation for the BH's becomes,
\begin{align}
\frac{\rd^{2} r}{\rd t^{2}}&=\partial_{i}h_{00}+\frac{\sqrt{2}}{eM_{p}}F^{i}{}_{0}\ .\label{NewtonianLimitModifiedGeodesicEqn2body}
\end{align}
Since all background forces cancel (see Appendix \ref{Appendix:ExtremalRN}), the leading force terms in \eqref{NewtonianLimitModifiedGeodesicEqn2body} arise from the perturbative corrections found in the previous section \eqref{FullEFTMetricAndVectorCorrections} and those due to massless loops, as calculated in Appendix \ref{Appendix:OneLoopMasslessPotential}.  The problem is therefore efficiently recast in terms of a conserved energy and effective (dimensionless) potential,
\begin{align}
E=\frac{1}{2}\dot{r}^{2}+V_{\rm eff}(r)\ ,\label{NonRelativisticEnergy}
\end{align}
where $V_{\rm eff}(r)$ descends from \eqref{FullEFTMetricAndVectorCorrections}.  We might also worry about subleading velocity dependent forces, but these can be  neglected, as is justified in Appendix \ref{Appendix:ExtremalRN}.

From the explicit form of the corrections discussed in the previous section, it can be determined that the effective potential $V_{\rm eff}(r)$ has two distinct types of behavior, depending on the value of $r$:
\begin{align}
V_{\rm eff}(r)&=\begin{dcases} -C_{1}\left (\frac{r_{s}}{r}\right )\left (\frac{l_{p}}{r}\right )^{2}\ , &  r_{s}\left (\frac{er_{e}}{l_{p}}\right )^{2} \lesssim r\\
-C_{2}\left (\frac{er_{e}}{l_{p}}\right )^{4}\left (\frac{r_{s}}{r}\right )^{3}\left (\frac{l_{p}}{r}\right )^{2}\ ,& r_{s}\lesssim r \lesssim r_{s}\left (\frac{er_{e}}{l_{p}}\right )^{2} \label{EffectivePotentialPiecewiseDescription}
\end{dcases}\ .
\end{align}
At long distances, $r\gtrsim r_{s}\left (\frac{er_{e}}{l_{p}}\right )^{2} $, massless loops generate the leading forces, while the $\sim F^{4}/m^{4}$ EFT operators generate the dominate forces at shorter scales.
The $C_{i}$'s are linear combinations of the $\mathcal{O}(1)$ coefficients which describe the effects of either electron  \eqref{EFTCoefficients} or gauge loops.  Their precise values are not needed.  Instead, it's only important we know they're \textit{positive} so that the black hole pair is attracted at all distances.

It's easy to check that the dynamics stay non-relativistic throughout the collapse, justifying the use of \eqref{NonRelativisticEnergy}. From \eqref{EffectivePotentialPiecewiseDescription},  magnitude of potential is bounded by
\begin{align}
|V_{\rm max}|\lesssim \left (\frac{er_{e}}{l_{p}}\right )^{4}\left (\frac{l_{p}}{r_{s}}\right )^{2}\ \label{MaxVMagnitude}
\end{align} for $r\gtrsim r_{s}$.  We previously found that the radius of the extremal black holes must satisfy $r_{s}\gg r_{e}\left (\frac{er_{e}}{l_{p}}\right )$ to fall within the validity of the EFT description.  Plugging this fact into \eqref{MaxVMagnitude} we find that the potential is bound by $|V_{\rm max}| \ll e^{2}\ll 1$, and hence the velocities which obey $v^{2}\sim V$ are also much smaller than unity, as we wanted to show. 
The pair's dynamics can then be tracked using Newtonian dynamics until the separation becomes $r\sim \mathcal{O}(r_{s})$, at which point the perturbative treatment breaks down.  

\subsubsection{Distance Advance}

We now estimate the total distance advance acquired by $\gamma_{\rm QED}$ as it passes through each region.  Start by placing the pairs at rest with $r\to \infty$ and track the net distance advance gained by the photon.

 If the black holes were all Schwarzschild, the velocity of the maximally superluminal photon would be similar to what we found in the Drummond-Hathrell section, schematically: 
\begin{align}
c_{s}\approx 1+ C_{4}\frac{e^{2}R_{\mu\nu\rho\sigma}}{m_{e}^{2}}\ .\label{Schematicdeltav}\ 
\end{align}
In \eqref{Schematicdeltav}, $C_{4}$ is a positive, $\mathcal{O}(1)$ number directly proportional to the EFT coefficient $c$ which also takes into account the geometry of the tunnel and $R_{\mu\nu\rho\sigma}$ represents the typical curvature felt by the photon when placed between a black hole pair.

  The expression \eqref{Schematicdeltav} only comes from considering the $\sim RFF/m_{e}^{2}$ terms in the EFT.  When there are non-trivial electromagnetic sources, as in the present case, the $\sim F^{4}/m_{e}^{4}$ terms can also affect propagation, generically \cite{Daniels:1993yi,Adler:1971wn}. These operators \textit{decrease} $c_{s}$  and generate physically relevant effects for pulsar physics\footnote{We thank Sam Gralla for bringing this fact to our attention.} of $\mathcal{O}\left (10\%\right )$, see Sec. 4 of the review \cite{Harding:2006qn} and references therein.  However, in our current, highly symmetric scenario where the photon is sent directly between the black hole pair, the effects from each $\sim F^{2n}/m_{e}^{4n-4}$ operator vanishes  due to symmetry, as shown in Appendix \ref{Appendix:FFEffectsOnCs}.
  In many ways, the scenario we're considering is ideal for enhancing the superluminality, since these operators serve only to decrease $c_{s}$ in more generic setups.

In \eqref{SchematicEFTAction}, the leading contribution to $R_{\mu\nu\rho\sigma}$ is of the form $R_{\mu\nu\rho\sigma}\sim \frac{r_{s}}{D^{3}}$ where $D$ is the distance between the photon and the \textit{nearest} black hole pair.  Therefore, the $\delta v$ is well-approximated by
\begin{align}
\delta v&\approx C_{4}e^{2}\frac{r_{e}^{2}r_{s}}{r^{3}}\, ,\label{LessSchematicdeltav}
\end{align}
with $r$ the black hole separation appearing in \eqref{EffectivePotentialPiecewiseDescription}.  Replacing $D$ by $r$ is a {conservative} estimate which approximates the setup by assuming that there's always a black hole directly on either side of the photon.  For an appropriate choice of $C_{4}\sim \mathcal{O}(1)$, which accounts for both the EFT coefficient $c$ and the geometry of the tunnel, \eqref{LessSchematicdeltav} serves as an {upper} bound on the velocity boost gained by the photon.

We now calculate the distance advanced gained by the maximally superluminal QED photon, relative to a minimally coupled photon, as it passes through the two regions described by \eqref{EffectivePotentialPiecewiseDescription}:
\begin{itemize}
\item In the outer region, $r\gtrsim r_{s}\left (\frac{er_{e}}{l_{p}}\right )^{2}$, the distance advance gained is:
\begin{align}
\Delta X&\approx \int_{t_{i}}^{t_{f}}\rd t\, \delta v\approx\int_{r_{s}\left (\frac{er_{e}}{l_{p}}\right )^{2}}^{\infty}\rd r\, \frac{\rd t	}{\rd r}\delta v\approx\int_{r_{s}\left (\frac{er_{e}}{l_{p}}\right )^{2}}^{\infty}\rd r\, \frac{\delta v}{\sqrt{-V_{\rm eff}}}\approx\int_{r_{s}\left (\frac{er_{e}}{l_{p}}\right )^{2}}^{\infty}\rd r\, \frac{C_{4}e^{2}\frac{r_{e}^{2}r_{s}}{r^{3}}}{\sqrt{C_{1}\left (\frac{r_{s}}{r}\right )\left (\frac{l_{p}}{r}\right )^{2} }}\nn
&\approx \frac{C_4}{\sqrt{C_1}}e m_{e}^{-1}\ .\label{DistanceAdvanceOuterRegion}
\end{align}
Here, and below, we drop $\mathcal{O}(1)$ numerical prefactors.
The distance advance acquired is parametrically smaller than the cutoff of the EFT by a factor of the gauge coupling, $\Delta X\sim e\times m_{e}^{-1}$.
\item The calculation is similar for the inner region, $ r_{s}\lesssim r\lesssim r_{s}\left (\frac{er_{e}}{l_{p}}\right )^{2}$:
\begin{align}
\Delta X\approx \int_{r_{s}}^{r_{s}\left (\frac{er_{e}}{l_{p}}\right )^{2}}\rd r\,  \frac{C_{4}e^{2}\frac{r_{e}^{2}r_{s}}{r^{3}}}{\sqrt{C_{2}\left (\frac{er_{e}}{l_{p}}\right )^{4}\left (\frac{r_{s}}{r}\right )^{3}\left (\frac{l_{p}}{r}\right )^{2}}}\approx \frac{C_4}{\sqrt{C_2}}e m_{e}^{-1}\ .\label{DistanceAdvanceInnerRegion}
\end{align}
Again, the distance advance is again parametrically smaller than the cutoff, $\Delta X\sim e\times m_{e}^{-1}$. 
\end{itemize}
Therefore, the total distance advance is parametrically smaller than the cutoff of the EFT.

The QED EFT appears to conspire in such a way that macroscopic is superluminality is avoided.  For instance, the transition between the two force laws behaviors occurs at \textit{exactly} the scale it must to keep the distance advance parametrically suppressed.  Had massless loops dominated down to, say, a distance scale $\sim r_{s}\left (\frac{er_{e}}{l_{p}}\right )$ instead of $\sim r_{s}\left (\frac{er_{e}}{l_{p}}\right )^{2}$, the distance advance gained by the photon in the outer region would have been $\Delta X\sim  e\left ( \frac{er_{e}}{l_{p}}\right )^{1/2} m_{e}^{-1} $ which can be consistently taken much larger than $m_{e}^{-1}$.  For example, $e\left ( \frac{er_{e}}{l_{p}}\right )^{1/2}\sim 10^{12}$ in the Standard Model.

\subsubsection{Variations}

The above analysis can be refined and extended to variations of this scenario, but the conclusion remains the same: at worst, $\Delta X\approx e\times m_{e}^{-1}$.  

 For instance, one could give the black holes an initial outward velocity so that the tunnel expands out to infinity and then collapses again, but this only leads to a factor of two improvement.  Alternatively, since $\delta v$ grows as the black holes get closer together, we could initially place the black holes at a distance $r\sim r_{e}\left (\frac{er_{e}}{l_{p}}\right )^{2}$, for instance.  This way, the BHs pass less quickly through regions where $\delta v$ is relatively large.  However the improvement is again only characterized by factors of two. 

A more interesting possibility comes from \textit{overcharging} the black hole.  That is, in pure Einstein-Maxwell the black hole charge is bounded so that the inequality $r_{q}\le r_{s}$ is satisfied.  Otherwise, there's a naked singularity.  However, in full QED where there are also fermionic fields, the bound is altered so that the black hole can carry slightly \textit{more} charge\footnote{This expression assumes the WGC, $er_{e}/l_{p}\gg 1$.  If the WGC is violated, then the $\sim R^{2}$ terms we've neglected in the action instead provide the leading corrections to this bound \cite{ToAppear}.}\cite{ToAppear,Ruffini:2013hia} (see \cite{Yajima:2000kw}, also):
\begin{align}
 r_{q}\lesssim r_{s}+\left (\frac{er_{e}}{l_{p}}\right )^{4}\frac{l_{p}^{2}}{225\pi^{2}r_{s}}-\left (\frac{er_{e}}{l_{p}}\right )^{2}\frac{3l_{p}^{2}}{225\pi^{2}r_{s}}\label{NewBHExtremalityBound}\ .
 \end{align} 
 This is expected to be a generic property of theories which obey the WGC \cite{ArkaniHamed:2006dz,Kats:2006xp}: black holes should allow for a maximum charge to mass ratio, $\max(r_{q}/r_{s})$, which is slightly larger than unity and, further, smaller black holes should be able to carry proportionally more charge, $\frac{\rd}{\rd r_{s}}\max(r_{q}/r_{s})<0$.  Such properties prevent the existence of unnatural, exactly stable remnants whose stability doesn't follow from any symmetry principle \cite{Kats:2006xp}.
 
By overcharging the black holes we can set up a hilltop type potential for the black holes which is attractive at short distances and then repulsive at large separation where the small $\sim 1/r^{2}$ force due to overcharging begins to dominate. However, an analysis entirely analogous to that of the previous section demonstrates that we cannot use this effect to our advantage.  The ladder either collapses too quickly, as before, or gets blown apart too fast, depending on the initial setup. 

 For instance, assuming $er_{e}/l_{p}\gg 1$, extremal black holes in QED obey
\begin{align}
r_{q}\approx r_{s}\left (1+\frac{1}{225\pi^{2}}\left (\frac{er_{e}}{l_{p}}\right )^{4}\left (\frac{l_{p}}{r_{s}}\right )^{2}\right )\label{ExtremalQEDBHCondition}\ .
\end{align}
The repulsive potential generated from overcharging dominates at all distances and is given by
\begin{align}
V&\sim \frac{r_{s}}{r}\left (\frac{er_{e}}{l_{p}}\right )^{4}\left (\frac{l_{p}}{r_{s}}\right )^{2}\ . \label{ExtremalQEDBHPotential} 
\end{align}
Releasing the ladder from an initial separation of $\mathcal{O}(r_{s})$, it's found that we generate an asymptotic distance advance $\Delta X\sim l_{p}\ll m_{e}^{-1}$.  
There is no parametric win in any of these scenarios.

  \subsubsection{Weak Gravity Conjecture}

  One might wonder whether WGC-violating theories can achieve $\Delta d>m_{e}^{-1}$.  It appears not to be the case.  Assuming that $er_{e}/l_{p}\ll 1$ and that the extremality bound for black holes still allows for $r_{q}=r_{s}$ (only true for certain coefficients on the $\sim R^{2}$ terms in the action), then graviton/photon loops generate the dominant large-distance corrections to the force law, $V\sim \frac{r_{s}}{r}\frac{l_{p}^{2}}{r^{2}}$.  Releasing the black holes from infinity, the distance advance is
  \begin{align}
  \Delta d\sim \int^{\infty}_{r_{s}}\rd r\, \frac{\delta v}{-\sqrt{V}}\sim \int^{\infty}_{r_{s}}\rd r\, \frac{ \frac{e^{2}}{m_{e}^{2}}\frac{r_{s}}{r^{3}}}{\sqrt{\frac{r_{s}}{r}\frac{l_{p}^{2}}{r^{2}}}}\sim \left (\frac{er_{e}}{l_{p}}\right )^{2}l_{p}\ll l_{p}\ll r_{e}\ .
  \end{align}
  While the force law is depressed in certain regions relative to that in WGC-obeying theories, the superluminal boost $\delta c_{s} \sim \left (\frac{er_{e}}{l_{p}}\right )^{2}\left (\frac{l_{p}^{2}r_{s}}{r^{3}}\right )$ is also diminished.  The WGC violating scenario is actually better behaved.

\subsubsection{Polarization Rotation\label{Sec:ProtectionMechanisms}}

A final effect which fights against the generation of macroscopic distance advances in generic setup is the fact that in full QED photon polarizations rotate due to the different velocities for different polarization states in anisotropic backgrounds.    While there may be some discrete polarization eigen-directions which travel with fixed polarization, a photon initially placed in a generic state will rotate into other ones as it propagates.  This has been known for the case of electromagnetic backgrounds for some time \cite{Adler:1971wn}, but is also true in gravitational backgrounds.  The rotation tends to wash out any superluminal effects.  

We now sketch how the effect arises, exploring it in more detail in \cite{ToAppear}.  It is found by pushing the geometric optics analysis to the next order, $\mathcal{O}(\epsilon^{-1})$.  The $\mathcal{O}(\epsilon^{-2})$ geometric optics relation determined the dynamics of the wave vector through modifications of the geodesic equation.  In pure Einstein-Maxwell, we'd have found $k^{\nu}\nabla_{\nu}k^{\mu}=0$, but when electrons are included the relation is changed to $k^{\nu}\nabla_{\nu}k^{\mu}=\mathcal{F}^{\mu}$ for some source term (see Footnote \ref{foot:GeodesicWithSourceTerm}).  This was rephrased as a true geodesic equation along an \textit{effective} metric in Footnote \ref{foot:ModifiedGeodesicEquationQED}. Similarly, in the absence of electrons, the $\mathcal{O}(\epsilon^{-1})$ relation would read $k^{\nu}\nabla_{\nu}f_{\mu}=0$, where $f_{\mu}$ is the unit vector proportional to the polarization $a_{\mu}$, meaning that polarization is covariantly constant along the photon's trajectory.  When electrons are included in the theory, we instead find
\begin{align}
k^{\nu}\nabla_{\nu}f_{\mu}&=\Pi_{\mu}{}^{\nu}\mathcal{S}_{\nu}\ .\label{PolarizationDynamics}
\end{align}
In \eqref{PolarizationDynamics} $\Pi_{\mu}{}^{\nu}=\delta^{\nu}_{\mu}-f_{\nu}f^{\mu}$ is the projection tensor constructed from $f_{\mu}$ and $\mathcal{S}_{\nu}$ is a source term depending on background curvatures, field strengths and properties of the wave whose form is given in \cite{ToAppear}.

Outside of a single black hole, only radially polarized photons travel superluminally.  We find that the effects represented in \eqref{PolarizationDynamics} make this polarization unstable, while the azimuthal, subluminal polarization state is stable.  That is, a photon which is initially polarized in a nearly (but not exactly) radial direction far from the black hole will have its polarization vector rotated further and further into the azimuthal direction as it nears the black hole.  In contrast, a nearly azimuthally polarized photon becomes even further azimuthally polarized as it approaches the black hole.   The gravitational field breaks the symmetry between the two polarization states and induces a preferred direction for the vector.   To our knowledge, the effect of the Dirac field on the polarization of a propagating photon due to \textit{gravitational} fields has not been studied before.

\begin{figure}[h!] 
   \captionsetup{width=0.9\textwidth}
   \centering
     \includegraphics[width=5in]{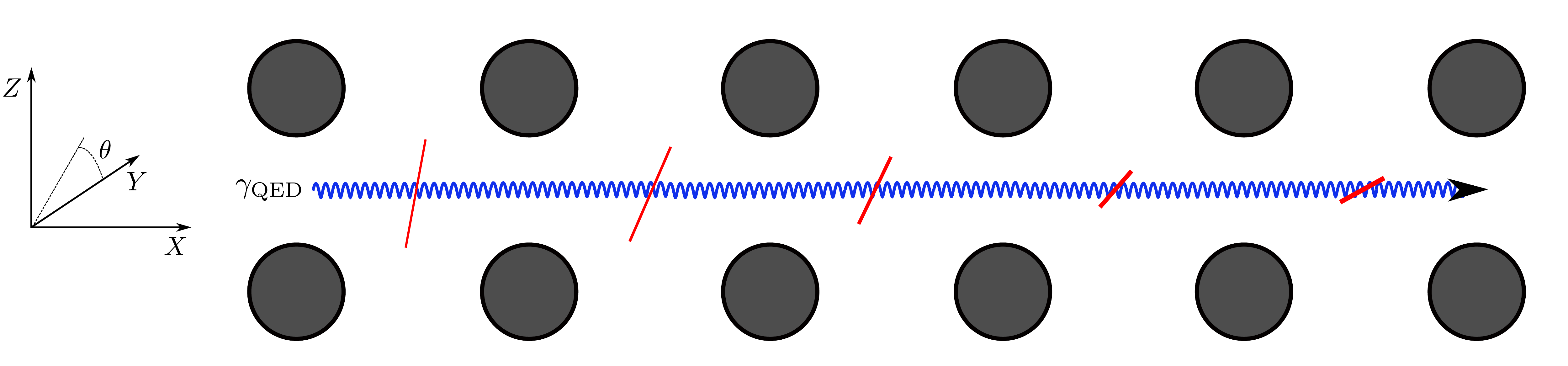}
   \caption{An exaggerated cartoon of polarization rotation for a photon traveling between a black hole ladder.  The polarization vector is indicated by the red lines with the segments drawn shorter, thicker and more highly angled as the vectors point further and further into the page.  The polarization angle starts near $\theta=\pi/2$, corresponding  a superluminal photon.  This polarization is unstable and slowly rotates back down towards $\theta=0$.}
   \label{fig:BlackHoleTunnelPolarizationRotation}
\end{figure}

The rotation is miniscule, but it could certainly become relevant in thought experiments like the black hole ladder of Fig.$\!$ \ref{fig:BlackHoleTunnel}.  Here, if the QED photon started with a nearly maximally superluminal polarization vector, $\theta_{0}=\pi/2-\delta$ with $\delta>0$, then as it passes through the first black hole pair, the angle would be slightly rotated down to some $\theta_{1}<\theta_{0}$.  The difference between the two angles would be tiny, but it sets the initial condition for $\theta$ as $\gamma_{\rm QED}$ passes through the next pair, after which the polarization angle will be again rotated down to some $\theta_{2}<\theta_{1}$.  This process continues and $\gamma_{\rm QED}$ gets smaller and smaller superluminal kicks as the travel continues, with the velocity turning subluminal at some point.  This is sketched in Fig.$\!$ \ref{fig:BlackHoleTunnelPolarizationRotation}.

For emphasis, this rotation  is \textit{not} being generated by the existence of free charged particles.  It is an effect which persists in vacuum and is the result of having ``virtual" electrons and positrons which exist because of the Dirac field.

\section{Galileon Superluminality}

We now turn to the superluminality which arises in the simplest galileon model \eqref{CubicGalileon}.  First the background is derived, then the geometric optics analysis is carried out and, finally, we race a galileon perturbation against a photon, showing that the superluminality is of a qualitatively different magnitude.  Gravity is ignored in this section.

\subsection{Background Solution}

Consider the cubic galileon coupled to a point mass \eqref{CubicGalileon},
\begin{align}
 \mathcal{L}&=-\frac{1}{2}(\partial\pi)^{2}-\frac{1}{\Lambda^{3}}(\square\pi)(\partial\pi)^{2}+\frac{1}{M_{p}}\pi T^{\mu}{}_{\mu} \ , \quad T_{\mu}{}^{\mu}=-M\delta^{3}(r)\ .
 \end{align}   The galileon equations of motion are particularly simple as they admit a first integral \cite{Nicolis:2008in}:
 \begin{align}
0= \frac{\delta\mathcal{L}}{\delta\pi}&=\frac{1}{r^{2}}\frac{\rd }{\rd r}\left (r^{2}\pi'\right )+\frac{4}{\Lambda^{3}r^{2}}\frac{\rd}{\rd r}\left (r\pi'\right )+T^{\mu}{}_{\mu}/M_{p}\ , 
 \end{align}
 where $\pi'=\frac{\rd\pi}{\rd r}$. Solving, one finds two distinct behaviors for $\pi(r)$, depending on whether $r$ is much larger or smaller than the Vainshtein radius of the source $r_{V}=\Lambda^{-1}\left (M/M_{p}\right )^{1/3}$:
 \begin{align}
 \pi(r)&\approx\begin{dcases}
 -\frac{1}{l_{p}}\frac{r_{s}}{r} & r\gg r_{V}\\
\frac{1}{\sqrt{4\pi}}\frac{1}{l_{p}}\frac{r_{s}}{r}\left (\frac{r}{r_{V}}\right )^{3/2} & r\ll r_{V}
 \end{dcases}\label{BackgroundGalileonSolution}\ .
 \end{align}

\subsection{Geometric Optics Analysis\label{Sec:GeomOpticsAnalysis}}

Now we apply geometric optics to the propagation of perturbations about the background solution. 
 We let $\pi=\bar{\pi}+\delta\pi$ with $\delta\pi(x)=\left (a+\epsilon b+\ldots\right )\exp i\vartheta(x)/\epsilon$ and expand the equation of motion to first order in $\delta \pi$.  The background equation of motion is
\begin{align}
\square\pi-\frac{2}{\Lambda^{3}}\left (\partial^{\mu}\partial^{\nu}\pi\partial_{\mu}\partial_{\nu}\pi-(\square	\pi)^{2}\right )&=-\frac{T^{\mu}{}_{\mu }}{M_{p}}\, ,\ 
\end{align}
and the $\mathcal{O}(\delta\pi)$ piece is
\begin{align}
\square\delta\pi=\frac{4}{\Lambda^{3}}\left (\partial^{\mu}\partial^{\nu}\bar{\pi}\partial_{\mu}\partial_{\nu}\delta\pi-\square\bar{\pi}\square\delta\pi\right )\ .
\end{align}
Therefore, the leading term in the geometric optics EOM is
\begin{align}
\tilde{g}^{\mu\nu}k_{\mu}k_{\nu}&=0\ ,\quad \tilde{ g}^{\mu\nu}=\eta^{\mu\nu}+\frac{4}{\Lambda^{3}}\square\bar{\pi}\eta^{\mu\nu}-\frac{4}{\Lambda^{3}}\partial^{\mu}\partial^{\nu}\bar{\pi}\ .
\end{align}

Following the same steps as the previous section, we can study the galileon trajectory by finding the geodesics of the optical metric $\tilde{g}_{\mu\nu}$.  Working at distances $r\gtrsim r_{V}$ and parameterizing the geodesic such that $X=X_0\gtrsim r_{V}$ at $\lambda	=0$, we find
\begin{align}
x^{\mu}(\lambda)&\approx\begin{pmatrix}
\lambda	\,	 , & X_{0}+\lambda+\frac{r_{V}^{3}\lambda(\lambda+2X_{0})}{2\pi X_{0}^{2}\left (\lambda+X_{0}\right )^{2}}\, , & 0\, , & 0
\end{pmatrix}\ .\label{GalileonGeodesic}
\end{align}

\subsection{Racing $\delta\pi$ Against a Photon}

We can now compare the galileon geodesic to that of a test photon which also travels from $X_{0}$ to infinity.  The photon's geodesics are manifestly unaffected by the galileon field, since the $\sim\pi T^{\mu}{}_{\mu}$ coupling vanishes for the Maxwell term\footnote{Other massless species also end up traveling along their Minkowski geodesics because the $\sim\pi T^{\mu}{}_{\mu}$ coupling results in an effective optical metric $\tilde{g}_{\mu\nu}$ which is conformally flat, but the argument is especially simple for photons.}.
Racing from $X_{0}\sim \mathcal{O}(r_{V})$ out to infinity, it's  found from \eqref{GalileonGeodesic} that the galileon perturbation beats the photon by a distance
\begin{align}
\Delta X&=\lim _{\lambda\to \infty	}x^{1}(\lambda)=\frac{r_{V}}{2\pi}\left (\frac{r_{V}}{X_{0}}\right )^{2}\ ,\label{GalileonDistanceAdvance}
\end{align}
which is $\mathcal{O}(r_{V})\gg \Lambda^{-1}$, as previously estimated \eqref{GalileonDistanceAdvanceApproximation} (see \cite{Creminelli:2014zxa}, also).

\section{Conclusions}

We have studied the Drummond-Hathrell superluminality present in the low energy effective field theory obtained by integrating the electron out of QED coupled to gravity.  This effective field theory has a cutoff at a distance scale corresponding to the Compton wavelength of the electron, $m_{e}^{-1}$.  If the full QED theory does not allow for superluminality, and no strong coupling effects, new particles or other non-perturbative effects come in before $m_{e}^{-1}$, then any distance advance $\Delta X$ generated by a superluminal photon along any trajectory in any background of the effective theory must not be resolvable in the EFT, so we must have $\Delta X \lesssim m_{e}^{-1}$.  We have tested this assertion by attempting to contrive various backgrounds to amplify the superluminal effects, and indeed in all the cases we try the distance advance is smaller than $m_{e}^{-1}$.  

The main scenario we consider is building distance advances via a ladder of approximately extremal, Reissner-Nordstrom black holes.  In order to account for all relevant effects, we not only needed to find the perturbative corrections to the RN solution to the higher-derivative, electron-induced operators in the QED EFT, but also needed to consider the subtle effects of graviton and photon loops. In the end, the distance advance we were capable of generating was parametrically bounded by $\Delta X\lesssim e\times m_{e}^{-1}$ with $e$ the QED gauge coupling. 

We then compared this to the analogous story for the galileons.  Unlike the QED case, we do not know of a local weakly coupled UV completion for the galileon (and there exist argument against \textit{any} such completion \cite{Adams:2006sv}).  All we have is the low energy effective theory, which comes with a strong coupling distance scale $\Lambda^{-1}$.  The superluminal distance advances in the galileon case can easily be made much larger than $\Lambda^{-1}$, and are typically as large as the Vainshtein radius, $r_V$, associated with the background.  It should, however, be noted that it proves very difficult to generate distance advances parametrically larger than $r_{V}$.

If the underlying UV theory for the galileons is indeed subluminal, then the UV completion must proceed in a qualitatively different way than it does in the QED case.  It cannot be simply be a new weakly coupled particle coming in at the scale $\Lambda$.  In order to cure the superluminality, there must instead be strong coupling effects or strong quantum effects coming in at the background-dependent scale $r_V$.

This kind of situation is also thought to occur in GR.  In GR, the Schwarzschild radius is the scale at which non-linearities become important, and plays the role of the Vainshtein radius of the galileon theories.  The black hole information paradox, along with the assumptions of unitarity and the equivalence principle, tell us that strict locality must break down at the scale of the horizon, so that the information may escape from the black hole.  Quantum gravity effects, which are completely invisible from the point of view of the low energy local effective field theory, must come in at the scale of the horizon and mediate these non-localities \cite{Marolf:2013dba}.  

A similar picture could hold true for the galileons, consistent with the findings of \cite{Keltner:2015xda}, and with the classicalization ideas of \cite{Dvali:2010jz}.  If so, then the true physics of galileon-like theories is highly non-perturbative in the quantum sense (not just classically non-linear), at all scales within the Vainshtein regime, which includes essentially all scales of phenomenological interest.  This of course does not mean it's ruled out, only that it is difficult to calculate anything with it.

 Alternatively, one can impose boundary conditions on the theory so that only backgrounds which do not possess large scale superluminality are available, e.g. \cite{Gabadadze:2014gba}.  In this way, the above conclusions can be avoided, without sacrificing UV subluminality.  However, we should also keep in mind that it is also logically possible to simply withdraw the demand that the UV theory be subluminal, in which case the above does not have to apply, and the superluminality of the low energy galileons is physical.

\noindent
{\bf Acknowledgments}:
It is our pleasure to thank Daniel Baumann, David Chernoff, Claudia de Rham, Ian Drummond, Siavash Golkar, Austin Joyce, Justin Khoury, Guilherme Pimentel, Rachel Rosen and Mark Trodden for discussions and comments on the draft. GG gratefully acknowledges support from a Starting Grant of the European Research Council (ERC StG grant 279617).  We thank the participants and organizers of the Superluminality in Effective Field Theories for Cosmology workshop at Perimeter Institute where this work originated. This work was also performed in part at the Aspen Center for Physics, which is supported by National Science Foundation grant PHY-1066293 (GG). 

\appendix

 \section{Velocity Dependent Forces\label{Appendix:ExtremalRN}}

 In this appendix, we estimate the sizes of various velocity dependent effects and verify that they're negligible.
 
 First, we explore the stability of the multi-extremal RN solution \eqref{BackgroundMultiExtremal4DRNIsotopicCoordinates} by adding a new extremal black hole to the spacetime and calculating the forces acting upon it.  If the new black hole is placed at rest relative to the other black holes, then the system is perfectly stable, but if it's in motion, forces are generated.

Placing the new black hole far from the others, we can analyze its motion with the modified geodesic equation appropriate for a particle with charge $Q$ and mass $M$:
\begin{align}
\frac{\rd x^{\nu }}{\rd \tau}\nabla_{\nu}\frac{\rd x^{\mu }}{\rd\tau}&=\frac{Q}{eM}F^{\mu}{}_{\nu}\frac{\rd x^{\nu}}{\rd \tau}\label{ModifiedGeodesicEq}\ .
\end{align}
The spacetime has a timelike Killing vector $K=\partial_{t}$, implying that the following is conserved:
\begin{align}
C\equiv -K_{\mu}\left (\frac{\rd x^{\mu }}{\rd\tau}+\frac{Q}{eM}A^{\mu}\right )\label{ConservedEMultiRN}\ .
\end{align}  Evaluating the spatial components of the modified geodesic equation \eqref{ModifiedGeodesicEq} at $\frac{\rd X^{i}}{\rd\tau}=0$ and using \eqref{ConservedEMultiRN} and $-1=g_{\mu\nu}\frac{\rd x^{\mu }}{\rd\tau}\frac{\rd x^{\nu }}{\rd\tau}$ yields the acceleration for the initially stationary probe particle:
\begin{align}
\frac{\rd^{2}X^{i}}{\rd \tau^{2}}&=-\frac{1}{2}\left (1-\frac{\sqrt{2}M_{p}Q}{M}\right )\partial_{i}\left (U^{-2}\right )\ .
\end{align}
This is vanishing \textit{only} if the new BH is extremal and carries the same sign charge as the original BHs: $Q=\frac{M}{\sqrt{2}M_{p}}$.

Next, we can calculate the forces which act on the new, extremal BH if it were moving with some velocity $\frac{\rd X^{i}}{\rd \tau}\neq 0$.  If the instantaneous velocity is $v^{2}=\delta_{ij}\frac{\rd X^{i}}{\rd \tau}\frac{\rd X^{j}}{\rd \tau}$, the spatial geodesic equations read (to first order in $v^{2}$):
\begin{align}
\frac{\rd^{2} X^{i}}{\rd \tau^{2}}&=\frac{v^{2}}{2}U^{-2} \partial_{i}U\ .\label{VelocityDependentRNForces}
\end{align}
Recalling that $U=1+\sum_{i}\frac{r_{s}}{2|\vec{X}-\vec{X}_{i}|}$ \eqref{BackgroundMultiExtremal4DRNIsotopicCoordinates}, it's found that \eqref{VelocityDependentRNForces} correspond to a small \textit{attraction} between the probe particle and the original BHs.  The origin of this attraction is clear: when the new BH is stationary, the gravitational attraction generated by its energy is perfectly tuned to cancel off the electromagnetic repulsion due to the BH's charge.  Therefore, when in motion, the BH carries some additional kinetic energy, leading to a slightly increased gravitational (and therefore overall) attraction.

For our interests, \eqref{VelocityDependentRNForces} is important because it justifies the neglect of such velocity-dependent forces in our analysis of the black hole ladder in Sec. \ref{Sec:BlackHoleLadder}.  At any given moment, the velocity of a black hole in the ladder is of order the potential generated from electron-induced EFT corrections to the gravitational and electromagnetic background, $v^{2}\sim\ V_{\rm EFT}$, schematically. The velocity-dependent force is thus of size $F_{\rm v.d.}\sim v^{2}\frac{r_{s}}{r^{2}} \sim \frac{r_{s}}{r^{2}}V_{\rm EFT}$, while the EFT forces are of size $F_{\rm EFT}\sim \partial_{r}V_{\rm EFT}\sim \frac{1}{r}V_{\rm EFT}$.  The velocity dependent force is therefore suppressed relative to the EFT forces by a factor of $r_{s}/r$ and are thus dominated in our regime of interest. 

Next, we can also consider radiation reaction forces.  We will show they they are also negligible, meaning that the black holes don't radiate significantly as they accelerate towards one another.  The Abraham-Lorentz law corresponds to a force $F_{\rm AL}\sim Q^{2}\dot{a}$ where $a$ is the acceleration of the charged object.  As $ M a\sim \frac{1}{ r}\sqrt{V_{\rm EFT}}$, we have
\begin{align}
\dot{a}\sim\frac{1}{M}\frac{\rd}{\rd t}\left (\frac{1}{r}\sqrt{V_{\rm EFT}}\right )\sim  \frac{v}{Mr^{2}}\sqrt{V_{\rm EFT}}\sim \frac{1}{Mr^{2}}V_{\rm EFT}\ ,
\end{align}
and hence the size of this effect is
\begin{align}
F_{\rm AL}\sim \frac{Q^{2}}{Mr} \frac{1}{r}V_{\rm EFT}\sim \frac{Q^{2}}{Mr}  F_{\rm EFT}\ .
\end{align}  For extremal objects, $\frac{Q^{2}}{Mr}\sim \frac{r_{s}}{r}$ and hence the radiation reaction force is, again, smaller than the leading forces by a factor of $r_{s}/r\ll 1$ and is negligible.

\section{Effects of $F^{2n}/m_{e}^{4n-4}$ Operators on $c_{s}$\label{Appendix:FFEffectsOnCs}}  

In this appendix we demonstrate that none of the $F^{2n}/m_{e}^{4n-4}$ operators in the EFT affect $c_{s}$ in our very specific setup.

    The EOM for photon fluctuations is of the form $\nabla_{\nu}\delta F^{\nu}{}_{\mu}=\nabla_{\nu}\frac{\delta\mathcal{L}_{\rm eff}}{\delta F_{\nu}{}^{\mu}}$ and if we work in Lorenz gauge for the fluctuation (implying $a_{\mu}k^{\mu}=0$ \eqref{Oe1GaugeConditionSchwarzschild}), then the geometric optics dispersion relation follows from $k^{2}\propto a^{\mu}\nabla_{\nu}\delta F^{\nu}{}_{\mu}=a^{\mu}\nabla_{\nu}\frac{\delta\mathcal{L}_{\rm eff}}{\delta F_{\nu}{}^{\mu}}$, where only the $\mathcal{O}(\epsilon^{-2})$ parts\footnote{See Sec. \ref{Sec:GeomOpticsAnalysis} for the review of geometric optics and the definition of $\epsilon$.} of $a^{\mu}\nabla_{\nu}\delta F^{\nu}$ and $a^{\mu}\nabla_{\nu}\frac{\delta\mathcal{L}_{\rm eff}}{\delta F_{\nu}{}^{\mu}}$ are kept.

    Consider the terms in $\mathcal{L}_{\rm eff}$ of the form $\sim F^{2n}/m_{e}^{4n-4}$.  From the explicit expressions for the Euler-Heisenberg action \cite{Heisenberg:1935qt}, each term can be put in to the form $\sim  (F_{\mu\nu}F^{\mu\nu})^{i}(F_{\mu\nu}\tilde{F}^{\mu\nu})^{j}$ where, by parity conservation, $j$ is even.  Because $F\tilde{F}\sim E\cdot B$ and $B=0$ along the photon's path (by the symmetry of the problem), terms with $j\ge 4$ have no effect on the dispersion relation, as their contribution is proportional to a power of $ F\tilde{F}$.  The $j=2$ case needs to be treated separately since it can yield a nontrivial term:
\begin{align}
  a_{\mu}\nabla_{\nu}\frac{\delta\mathcal{L}_{\rm eff}}{\delta F_{\nu}{}^{\mu}}\sim (FF)^{i}  \left (a_{\mu}k_{\nu}\tilde{F}^{\mu\nu}\right )^{2} \ ,
  \end{align} where we've dropped other vanishing contributions $\propto F\tilde{F}$.  On the line $Z=Y=0$, the only non-vanishing components of $F^{\mu\nu}$ is $F^{tX}=-F^{Xt}$, and hence the only non-vanishing component of $\tilde{F}^{\mu\nu}$ is $\tilde{F}^{YZ}=-\tilde{F}^{ZY}$.  Therefore, $a_{\mu}k_{\nu}\tilde{F}^{\mu\nu}$ vanishes for the trajectory we're interested in, since $k^{\mu}\sim(1,1,0,0)$, to the order needed for this calculation.

  The remaining $\sim (FF)^{i}$ terms yield contributions of the form $a^{\mu}\nabla_{\nu}\frac{\delta\mathcal{L}_{\rm eff}}{\delta F_{\nu}{}^{\mu}}\sim a_{\mu}\nabla_{\nu}\left [(FF)^{i-1}F^{\mu\nu}\right ]$.  These generate two different types of expressions which are either proportional to $a_{\mu}k_{\nu}a^{[\mu}k^{\nu]}$ or $a_{\mu}k_{\nu}F^{\mu\nu}$.  The first expression either vanishes by the gauge condition or is $\propto k^{2}$ which represents a higher order effect (quadratic in EFT coefficients).  The second combination, $a_{\mu}k_{\nu}F^{\mu\nu}$, vanishes along the photon's path, due to the fact that $a_{\mu}$ points in the $Y-Z$ plane\footnote{More precisely, to zeroth order in EFT coefficients, which is all we need for this calculation, the gauge condition only determines $a_{\mu}$ up to an equivalence class: $a_{\mu}\sim a_{\mu}+k_{\mu}$.  All elements in the class lead to the same result for $a_{\mu}k_{\nu}F^{\mu\nu}$, due to the antisymmetry of $F^{\mu\nu}$, and it's possible to work with a representative element, $a_{\mu}$, which lies in the $Y$-$Z$ plane.}, while only $F^{tX}=-F^{Xt}$ is non-zero along the trajectory.
  
  Therefore, none of the $\sim F^{2n}/m_{e}^{4n-4}$ operators affect $c_{s}$ for this very tuned scenario, as claimed.

  \section{One-loop Potential from Massless Fields\label{Appendix:OneLoopMasslessPotential}}
  
  In this Appendix, we use the results of \cite{Holstein:2008sy,BjerrumBohr:2002sx,BjerrumBohr:2002kt} to calculate the one-loop correction to the potential due to massless graviton and photon loops.
  
  First, \cite{BjerrumBohr:2002sx,Holstein:2008sy} calculated the one-loop, $\mathcal{O}(Q^{2}/M_{p}^{2})$ contribution to the non-relativistic potential in scalar QED (where $Q$ is the charge of $\phi$) due to mixed photon-graviton scattering diagrams.  For equal charges and masses, the result is a repulsive potential:
  \begin{align}
  \mathcal{V}_{\mathcal{O}(Q^{2}/M_{p}^{2})}&= \frac{5e^{2}Q^{2}}{48M_{p}^{2}\pi^{3}r^{3}} \ ,\label{OrderQ2Mp2OneLoopPotentialAppendix}
  \end{align}
  where we took the equal mass, equal charge limit of the $\mathcal{O}(\hbar)$ part of eq.$\!$ (41) in \cite{BjerrumBohr:2002sx}, or eq.$\!$ (58) in \cite{Holstein:2008sy}, and translated conventions.  We use the symbol $\mathcal{V}$ for the dimensionful potential to differentiate it from the dimensionless potential $V$ used in the body of the paper, as in \eqref{NonRelativisticEnergy} and following expressions.
  
  Next, \cite{BjerrumBohr:2002kt} found the one-loop correction to the non-relativistic potential between two masses.  This was calculated in the context of pure GR where the following $\mathcal{O}(M^{2}/M_{p}^{4})$ attractive correction between two equal-mass particles was found
  \begin{align}
  \mathcal{V}^{\rm GR}_{\mathcal{O}(m^{2}/M_{p}^{4})}&=-\frac{41M^{2}}{640 M_{p}^{4}\pi^{3}r^{3}} \label{Orderm2Mp4OneLoopPotentialPureGRAppendix}\ ,
  \end{align}
  in our conventions, from their eq.$\!$ (44).
    
  However, because \eqref{Orderm2Mp4OneLoopPotentialPureGRAppendix} was obtained in GR, it's not immediately applicable to the scenario considered in this paper.  We also need to include the $\mathcal{O}(M^{2}/M_{p}^{4})$ contribution from photon loops.  Fortunately, this is a simple fix: one only needs to change the vacuum polarization diagram so that both gravitons \textit{and} photons (along with their associated ghosts) run the loop, Figs.$\!$ $6(a)$ and $6(b)$ in \cite{BjerrumBohr:2002kt}.
  
  Vacuum polarization diagrams involving massless loops generate non-analytic terms in the 1PI effective action of the form\footnote{Strictly speaking, we'd need to also have a $\sim \bar{R}^{\mu\nu\rho\sigma}\log(-\square/\mu ^{2})\bar R_{\mu\nu\rho\sigma}$, but at $\mathcal{O}(h_{\mu\nu}^{2})$, sufficient for the present calculation, the three $\mathcal{O}(R^{2})$ operators are degenerate and the form \eqref{1PIActionR2TermsAppendix} is adequate.} \eqref{1PIGRActionResult}  
\begin{align}\Gamma[\bar{g}_{\mu\nu}]&\supset\int\rd^{4}x\,\sqrt{-\bar g} \left(c_{1}\bar R\log(-\square/\mu ^{2})\bar R+c_{2}\bar R^{\mu\nu}\log(-\square/\mu ^{2})\bar R_{\mu\nu}\right )\label{1PIActionR2TermsAppendix}\ .
\end{align}
The contribution of these operators to the potential \eqref{Orderm2Mp4OneLoopPotentialPureGRAppendix} is:
\begin{align}
\mathcal{V}_{R\log \square R}&=\left (c_{1}+c_{2}\right )\frac{M^{2}}{M_{p}^{4}\pi r^{3}} \label{PotentialFromRLogBoxRTermsAppendix}\ ,
\end{align}
and \cite{BjerrumBohr:2002kt} used the 't Hooft-Veltman \cite{'tHooft:1974bx}, pure GR result 
\begin{align}
\begin{pmatrix}
c_{1}^{\rm GR}\\
c_{2}^{\rm GR}
\end{pmatrix}&=\frac{1}{(4\pi)^{2}}\begin{pmatrix}
 -\frac{1}{120} \\ -\frac{7}{20} 
\end{pmatrix}\ ,\label{tHooftVeltmansGRCsAppendox}
\end{align}
as calculated in Feynman gauge, which provided the following contribution to \eqref{Orderm2Mp4OneLoopPotentialPureGRAppendix}
\begin{align}
\mathcal{V}^{\rm GR}_{R\log \square R}&=\frac{-43}{1920}\frac{M^{2}}{M_{p}^{4}\pi^{3} r^{3}}\ , 
\end{align}
see eq.$\!$ (43) of \cite{BjerrumBohr:2002kt}.

When photons are included, they also contribute an amount $c_{i}^{\gamma}$ to the $c_{i}$'s, with the total result $c_{i}=c_{i}^{\rm GR}+c_{i}^{\gamma}$.  Calculating the necessary vacuum polarization diagrams, we find
\begin{align}
\begin{pmatrix}
c_{1}^{\gamma}\\
c_{2}^{\gamma}
\end{pmatrix}&=\frac{1}{(4\pi)^{2}}\begin{pmatrix}
\frac{1}{30}\\ -\frac{1}{10}
\end{pmatrix}\ ,\label{CapperDuffPhotonCsAppendix}
\end{align}
(in agreement with \cite{Capper:1974ed})
which generates the extra, attractive potential
\begin{align}
\mathcal{V}^{\gamma}_{R\log \square R}&=\frac{-131}{1920}\frac{M^{2}}{M_{p}^{4}\pi^{3} r^{3}}\ .\label{ExtraPotentialFromPhotonsAppendix}
\end{align}

Adding \eqref{OrderQ2Mp2OneLoopPotentialAppendix}, \eqref{Orderm2Mp4OneLoopPotentialPureGRAppendix} and \eqref{ExtraPotentialFromPhotonsAppendix} together and taking the extremal limit $eQ=\frac{M}{M_{p}\sqrt{2}}$, the total one-loop contribution to the potential due to graviton and photon loops is found to be
\begin{align}
\mathcal{V}^{{\rm GR}+\gamma}_{\rm total}&=\frac{-23}{1920}\frac{M^{2}}{M_{p}^{4}\pi^{3} r^{3}}\ ,
\end{align}
which is attractive and of the claimed form \eqref{EffectivePotentialPiecewiseDescription} (when turned into a dimensionless potential via $ V\sim \mathcal{V}/M$).

\bibliographystyle{utphys}
\bibliography{SuperluminalityDraftarXiv2}

\providecommand{\href}[2]{#2}\begingroup\raggedright\begin{thebibliography}{100}

\bibitem{Babichev:2007dw}
E.~Babichev, V.~Mukhanov, and A.~Vikman, ``{k-Essence, superluminal
  propagation, causality and emergent geometry},''
  \href{http://dx.doi.org/10.1088/1126-6708/2008/02/101}{{\em JHEP} {\bf 02}
  (2008)  101},
\href{http://arxiv.org/abs/0708.0561}{{\tt arXiv:0708.0561 [hep-th]}}.

\bibitem{Geroch:2010da}
R.~Geroch, ``{Faster Than Light?},''
\href{http://arxiv.org/abs/1005.1614}{{\tt arXiv:1005.1614 [gr-qc]}}.

\bibitem{Burrage:2011cr}
C.~Burrage, C.~de~Rham, L.~Heisenberg, and A.~J. Tolley, ``{Chronology
  Protection in Galileon Models and Massive Gravity},''
  \href{http://dx.doi.org/10.1088/1475-7516/2012/07/004}{{\em JCAP} {\bf 1207}
  (2012)  004},
\href{http://arxiv.org/abs/1111.5549}{{\tt arXiv:1111.5549 [hep-th]}}.

\bibitem{Papallo:2015rna}
G.~Papallo and H.~S. Reall, ``{Graviton time delay and a speed limit for small
  black holes in Einstein-Gauss-Bonnet theory},''
  \href{http://dx.doi.org/10.1007/JHEP11(2015)109}{{\em JHEP} {\bf 11} (2015)
  109},
\href{http://arxiv.org/abs/1508.05303}{{\tt arXiv:1508.05303 [gr-qc]}}.

\bibitem{PhysRevD.42.1915}
J.~Friedman, M.~S. Morris, I.~D. Novikov, F.~Echeverria, G.~Klinkhammer, K.~S.
  Thorne, and U.~Yurtsever,
  \href{http://dx.doi.org/10.1103/PhysRevD.42.1915}{``Cauchy problem in
  spacetimes with closed timelike curves,''{\em Phys. Rev. D} {\bf 42} (Sep,
  1990)  1915--1930}. \url{http://link.aps.org/doi/10.1103/PhysRevD.42.1915}.

\bibitem{Drummond:1979pp}
I.~Drummond and S.~Hathrell, ``{QED Vacuum Polarization in a Background
  Gravitational Field and Its Effect on the Velocity of Photons},''
\href{http://dx.doi.org/10.1103/PhysRevD.22.343}{{\em Phys.Rev.} {\bf D22}
  (1980)  343}.

\bibitem{Dubovsky:2007ac}
S.~Dubovsky, A.~Nicolis, E.~Trincherini, and G.~Villadoro, ``{Microcausality in
  curved space-time},''
  \href{http://dx.doi.org/10.1103/PhysRevD.77.084016}{{\em Phys.Rev.} {\bf D77}
  (2008)  084016},
\href{http://arxiv.org/abs/0709.1483}{{\tt arXiv:0709.1483 [hep-th]}}.

\bibitem{Shore:1995fz}
G.~Shore, ``{'Faster than light' photons in gravitational fields: Causality,
  anomalies and horizons},''
  \href{http://dx.doi.org/10.1016/0550-3213(95)00646-X}{{\em Nucl.Phys.} {\bf
  B460} (1996)  379--396},
\href{http://arxiv.org/abs/gr-qc/9504041}{{\tt arXiv:gr-qc/9504041 [gr-qc]}}.

\bibitem{Shore:2003zc}
G.~M. Shore, ``{Quantum gravitational optics},''
  \href{http://dx.doi.org/10.1080/00107510310001617106}{{\em Contemp.Phys.}
  {\bf 44} (2003)  503--521},
\href{http://arxiv.org/abs/gr-qc/0304059}{{\tt arXiv:gr-qc/0304059 [gr-qc]}}.

\bibitem{Shore:2007um}
G.~Shore, ``{Superluminality and UV completion},''
  \href{http://dx.doi.org/10.1016/j.nuclphysb.2007.03.034}{{\em Nucl.Phys.}
  {\bf B778} (2007)  219--258},
\href{http://arxiv.org/abs/hep-th/0701185}{{\tt arXiv:hep-th/0701185
  [hep-th]}}.

\bibitem{Khriplovich:1994qj}
I.~B. Khriplovich, ``{Superluminal velocity of photons in a gravitational
  background},'' \href{http://dx.doi.org/10.1016/0370-2693(94)01679-7}{{\em
  Phys. Lett.} {\bf B346} (1995)  251--254},
\href{http://arxiv.org/abs/gr-qc/9411052}{{\tt arXiv:gr-qc/9411052 [gr-qc]}}.

\bibitem{Mohanty:1998qq}
S.~Mohanty and A.~R. Prasanna, ``{Photon propagation in Einstein and higher
  derivative gravity},''
  \href{http://dx.doi.org/10.1016/S0550-3213(98)00275-2}{{\em Nucl. Phys.} {\bf
  B526} (1998)  501--508},
\href{http://arxiv.org/abs/gr-qc/9804017}{{\tt arXiv:gr-qc/9804017 [gr-qc]}}.

\bibitem{Preti:2010zz}
G.~Preti, ``{A note on the geometrical-optics solution to the Maxwell tensor
  wave equation in curved spacetime},''
\href{http://dx.doi.org/10.1016/j.nuclphysb.2010.03.028}{{\em Nucl. Phys.} {\bf
  B834} (2010)  390--394}.

\bibitem{Akhoury:2010hi}
R.~Akhoury and A.~D. Dolgov, ``{On the Possibility of Super-luminal Propagation
  in a Gravitational Background},''
\href{http://arxiv.org/abs/1003.6110}{{\tt arXiv:1003.6110 [hep-th]}}.

\bibitem{Daniels:1993yi}
R.~Daniels and G.~Shore, ``{'Faster than light' photons and charged black
  holes},'' \href{http://dx.doi.org/10.1016/0550-3213(94)90291-7}{{\em
  Nucl.Phys.} {\bf B425} (1994)  634--650},
\href{http://arxiv.org/abs/hep-th/9310114}{{\tt arXiv:hep-th/9310114
  [hep-th]}}.

\bibitem{Daniels:1995yw}
R.~D. Daniels and G.~M. Shore, ``{'Faster than light' photons and rotating
  black holes},'' \href{http://dx.doi.org/10.1016/0370-2693(95)01468-3}{{\em
  Phys. Lett.} {\bf B367} (1996)  75--83},
\href{http://arxiv.org/abs/gr-qc/9508048}{{\tt arXiv:gr-qc/9508048 [gr-qc]}}.

\bibitem{Hollowood:2007kt}
T.~J. Hollowood and G.~M. Shore, ``{Causality and Micro-Causality in Curved
  Spacetime},'' \href{http://dx.doi.org/10.1016/j.physletb.2007.08.073}{{\em
  Phys.Lett.} {\bf B655} (2007)  67--74},
\href{http://arxiv.org/abs/0707.2302}{{\tt arXiv:0707.2302 [hep-th]}}.

\bibitem{Hollowood:2007ku}
T.~J. Hollowood and G.~M. Shore, ``{The Refractive index of curved spacetime:
  The Fate of causality in QED},''
  \href{http://dx.doi.org/10.1016/j.nuclphysb.2007.11.034}{{\em Nucl.Phys.}
  {\bf B795} (2008)  138--171},
\href{http://arxiv.org/abs/0707.2303}{{\tt arXiv:0707.2303 [hep-th]}}.

\bibitem{Hollowood:2008kq}
T.~J. Hollowood and G.~M. Shore, ``{The Causal Structure of QED in Curved
  Spacetime: Analyticity and the Refractive Index},''
  \href{http://dx.doi.org/10.1088/1126-6708/2008/12/091}{{\em JHEP} {\bf 12}
  (2008)  091},
\href{http://arxiv.org/abs/0806.1019}{{\tt arXiv:0806.1019 [hep-th]}}.

\bibitem{Hollowood:2009qz}
T.~J. Hollowood, G.~M. Shore, and R.~J. Stanley, ``{The Refractive Index of
  Curved Spacetime II: QED, Penrose Limits and Black Holes},''
  \href{http://dx.doi.org/10.1088/1126-6708/2009/08/089}{{\em JHEP} {\bf 08}
  (2009)  089},
\href{http://arxiv.org/abs/0905.0771}{{\tt arXiv:0905.0771 [hep-th]}}.

\bibitem{Hollowood:2010bd}
T.~J. Hollowood and G.~M. Shore, ``{The Effect of Gravitational Tidal Forces on
  Vacuum Polarization: How to Undress a Photon},''
  \href{http://dx.doi.org/10.1016/j.physletb.2010.07.006}{{\em Phys. Lett.}
  {\bf B691} (2010)  279--284},
\href{http://arxiv.org/abs/1006.0145}{{\tt arXiv:1006.0145 [hep-th]}}.

\bibitem{Hollowood:2010xh}
T.~J. Hollowood and G.~M. Shore, ``{`Superluminal' Photon Propagation in QED in
  Curved Spacetime is Dispersive and Causal},''
\href{http://arxiv.org/abs/1006.1238}{{\tt arXiv:1006.1238 [hep-th]}}.

\bibitem{Hollowood:2011yh}
T.~J. Hollowood and G.~M. Shore, ``{The Effect of Gravitational Tidal Forces on
  Renormalized Quantum Fields},''
  \href{http://dx.doi.org/10.1007/JHEP02(2012)120}{{\em JHEP} {\bf 02} (2012)
  120},
\href{http://arxiv.org/abs/1111.3174}{{\tt arXiv:1111.3174 [hep-th]}}.

\bibitem{Hollowood:2012as}
T.~J. Hollowood and G.~M. Shore, ``{The Unbearable Beingness of Light, Dressing
  and Undressing Photons in Black Hole Spacetimes},''
  \href{http://dx.doi.org/10.1142/S0218271812410039}{{\em Int.J.Mod.Phys.} {\bf
  D21} (2012)  1241003},
\href{http://arxiv.org/abs/1205.3291}{{\tt arXiv:1205.3291 [hep-th]}}.

\bibitem{Hollowood:2015elj}
T.~J. Hollowood and G.~M. Shore, ``{Causality Violation, Gravitational
  Shockwaves and UV Completion},''
\href{http://arxiv.org/abs/1512.04952}{{\tt arXiv:1512.04952 [hep-th]}}.

\bibitem{Hollowood:2016ryc}
T.~J. Hollowood and G.~M. Shore, ``{Causality, Renormalizability and Ultra-High
  Energy Gravitational Scattering},''
\href{http://arxiv.org/abs/1601.06989}{{\tt arXiv:1601.06989 [hep-th]}}.

\bibitem{Simon:1990ic}
J.~Z. Simon, ``{Higher Derivative Lagrangians, Nonlocality, Problems and
  Solutions},''
\href{http://dx.doi.org/10.1103/PhysRevD.41.3720}{{\em Phys. Rev.} {\bf D41}
  (1990)  3720}.

\bibitem{Jaen:1986iz}
X.~Jaen, J.~Llosa, and A.~Molina, ``{A Reduction of order two for infinite
  order lagrangians},''
\href{http://dx.doi.org/10.1103/PhysRevD.34.2302}{{\em Phys. Rev.} {\bf D34}
  (1986)  2302}.

\bibitem{Burgess:2014lwa}
C.~P. Burgess and M.~Williams, ``{Who You Gonna Call? Runaway Ghosts, Higher
  Derivatives and Time-Dependence in EFTs},''
  \href{http://dx.doi.org/10.1007/JHEP08(2014)074}{{\em JHEP} {\bf 08} (2014)
  074},
\href{http://arxiv.org/abs/1404.2236}{{\tt arXiv:1404.2236 [gr-qc]}}.

\bibitem{BrillouinBook}
L.~Brillouin, {\em {Wave Propagation and Group Velocity (Series in Pure \&
  Applied Physics)}}.
\newblock 1960.

\bibitem{MilonniBook}
P.~Milonni, {\em {Fast Light, Slow Light and Left-Handed Light, (Series in
  Optics and Optoelectronics)}}.
\newblock 2004.

\bibitem{deRham:2014zqa}
C.~de~Rham, ``{Massive Gravity},''
  \href{http://dx.doi.org/10.12942/lrr-2014-7}{{\em Living Rev. Rel.} {\bf 17}
  (2014)  7},
\href{http://arxiv.org/abs/1401.4173}{{\tt arXiv:1401.4173 [hep-th]}}.

\bibitem{Nicolis:2008in}
A.~Nicolis, R.~Rattazzi, and E.~Trincherini, ``{The Galileon as a local
  modification of gravity},''
  \href{http://dx.doi.org/10.1103/PhysRevD.79.064036}{{\em Phys. Rev.} {\bf
  D79} (2009)  064036},
\href{http://arxiv.org/abs/0811.2197}{{\tt arXiv:0811.2197 [hep-th]}}.

\bibitem{Dvali:2000hr}
G.~R. Dvali, G.~Gabadadze, and M.~Porrati, ``{4-D gravity on a brane in 5-D
  Minkowski space},''
  \href{http://dx.doi.org/10.1016/S0370-2693(00)00669-9}{{\em Phys. Lett.} {\bf
  B485} (2000)  208--214},
\href{http://arxiv.org/abs/hep-th/0005016}{{\tt arXiv:hep-th/0005016
  [hep-th]}}.

\bibitem{Luty:2003vm}
M.~A. Luty, M.~Porrati, and R.~Rattazzi, ``{Strong interactions and stability
  in the DGP model},''
  \href{http://dx.doi.org/10.1088/1126-6708/2003/09/029}{{\em JHEP} {\bf 09}
  (2003)  029},
\href{http://arxiv.org/abs/hep-th/0303116}{{\tt arXiv:hep-th/0303116
  [hep-th]}}.

\bibitem{deRham:2010kj}
C.~de~Rham, G.~Gabadadze, and A.~J. Tolley, ``{Resummation of Massive
  Gravity},'' \href{http://dx.doi.org/10.1103/PhysRevLett.106.231101}{{\em
  Phys. Rev. Lett.} {\bf 106} (2011)  231101},
\href{http://arxiv.org/abs/1011.1232}{{\tt arXiv:1011.1232 [hep-th]}}.

\bibitem{Hinterbichler:2011tt}
K.~Hinterbichler, ``{Theoretical Aspects of Massive Gravity},''
  \href{http://dx.doi.org/10.1103/RevModPhys.84.671}{{\em Rev.Mod.Phys.} {\bf
  84} (2012)  671--710},
\href{http://arxiv.org/abs/1105.3735}{{\tt arXiv:1105.3735 [hep-th]}}.

\bibitem{deRham:2012az}
C.~de~Rham, ``{Galileons in the Sky},''
  \href{http://dx.doi.org/10.1016/j.crhy.2012.04.006}{{\em Comptes Rendus
  Physique} {\bf 13} (2012)  666--681},
\href{http://arxiv.org/abs/1204.5492}{{\tt arXiv:1204.5492 [astro-ph.CO]}}.

\bibitem{deRham:2010eu}
C.~de~Rham and A.~J. Tolley, ``{DBI and the Galileon reunited},''
  \href{http://dx.doi.org/10.1088/1475-7516/2010/05/015}{{\em JCAP} {\bf 1005}
  (2010)  015},
\href{http://arxiv.org/abs/1003.5917}{{\tt arXiv:1003.5917 [hep-th]}}.

\bibitem{Hinterbichler:2010xn}
K.~Hinterbichler, M.~Trodden, and D.~Wesley, ``{Multi-field galileons and
  higher co-dimension branes},''
  \href{http://dx.doi.org/10.1103/PhysRevD.82.124018}{{\em Phys. Rev.} {\bf
  D82} (2010)  124018},
\href{http://arxiv.org/abs/1008.1305}{{\tt arXiv:1008.1305 [hep-th]}}.

\bibitem{Goon:2011qf}
G.~Goon, K.~Hinterbichler, and M.~Trodden, ``{Symmetries for Galileons and DBI
  scalars on curved space},''
  \href{http://dx.doi.org/10.1088/1475-7516/2011/07/017}{{\em JCAP} {\bf 1107}
  (2011)  017},
\href{http://arxiv.org/abs/1103.5745}{{\tt arXiv:1103.5745 [hep-th]}}.

\bibitem{Goon:2011uw}
G.~Goon, K.~Hinterbichler, and M.~Trodden, ``{A New Class of Effective Field
  Theories from Embedded Branes},''
  \href{http://dx.doi.org/10.1103/PhysRevLett.106.231102}{{\em Phys. Rev.
  Lett.} {\bf 106} (2011)  231102},
\href{http://arxiv.org/abs/1103.6029}{{\tt arXiv:1103.6029 [hep-th]}}.

\bibitem{Vainshtein:1972sx}
A.~I. Vainshtein, ``{To the problem of nonvanishing gravitation mass},''
\href{http://dx.doi.org/10.1016/0370-2693(72)90147-5}{{\em Phys. Lett.} {\bf
  B39} (1972)  393--394}.

\bibitem{Babichev:2013usa}
E.~Babichev and C.~Deffayet, ``{An introduction to the Vainshtein mechanism},''
  \href{http://dx.doi.org/10.1088/0264-9381/30/18/184001}{{\em Class. Quant.
  Grav.} {\bf 30} (2013)  184001},
\href{http://arxiv.org/abs/1304.7240}{{\tt arXiv:1304.7240 [gr-qc]}}.

\bibitem{Goon:2016ihr}
G.~Goon, K.~Hinterbichler, A.~Joyce, and M.~Trodden, ``{Aspects of Galileon
  Non-Renormalization},''
\href{http://arxiv.org/abs/1606.02295}{{\tt arXiv:1606.02295 [hep-th]}}.

\bibitem{Goon:2011xf}
G.~Goon, K.~Hinterbichler, and M.~Trodden, ``{Galileons on Cosmological
  Backgrounds},'' \href{http://dx.doi.org/10.1088/1475-7516/2011/12/004}{{\em
  JCAP} {\bf 1112} (2011)  004},
\href{http://arxiv.org/abs/1109.3450}{{\tt arXiv:1109.3450 [hep-th]}}.

\bibitem{Trodden:2011xh}
M.~Trodden and K.~Hinterbichler, ``{Generalizing Galileons},''
  \href{http://dx.doi.org/10.1088/0264-9381/28/20/204003}{{\em Class. Quant.
  Grav.} {\bf 28} (2011)  204003},
\href{http://arxiv.org/abs/1104.2088}{{\tt arXiv:1104.2088 [hep-th]}}.

\bibitem{Goon:2012dy}
G.~Goon, K.~Hinterbichler, A.~Joyce, and M.~Trodden, ``{Galileons as
  Wess-Zumino Terms},'' \href{http://dx.doi.org/10.1007/JHEP06(2012)004}{{\em
  JHEP} {\bf 1206} (2012)  004},
\href{http://arxiv.org/abs/1203.3191}{{\tt arXiv:1203.3191 [hep-th]}}.

\bibitem{Dvali:2002vf}
G.~Dvali, A.~Gruzinov, and M.~Zaldarriaga, ``{The Accelerated universe and the
  moon},'' \href{http://dx.doi.org/10.1103/PhysRevD.68.024012}{{\em Phys. Rev.}
  {\bf D68} (2003)  024012},
\href{http://arxiv.org/abs/hep-ph/0212069}{{\tt arXiv:hep-ph/0212069
  [hep-ph]}}.

\bibitem{Adams:2006sv}
A.~Adams, N.~Arkani-Hamed, S.~Dubovsky, A.~Nicolis, and R.~Rattazzi,
  ``{Causality, analyticity and an IR obstruction to UV completion},''
  \href{http://dx.doi.org/10.1088/1126-6708/2006/10/014}{{\em JHEP} {\bf 10}
  (2006)  014},
\href{http://arxiv.org/abs/hep-th/0602178}{{\tt arXiv:hep-th/0602178
  [hep-th]}}.

\bibitem{Goon:2010xh}
G.~L. Goon, K.~Hinterbichler, and M.~Trodden, ``{Stability and superluminality
  of spherical DBI galileon solutions},''
  \href{http://dx.doi.org/10.1103/PhysRevD.83.085015}{{\em Phys. Rev.} {\bf
  D83} (2011)  085015},
\href{http://arxiv.org/abs/1008.4580}{{\tt arXiv:1008.4580 [hep-th]}}.

\bibitem{Andrews:2010km}
M.~Andrews, K.~Hinterbichler, J.~Khoury, and M.~Trodden, ``{Instabilities of
  Spherical Solutions with Multiple Galileons and SO(N) Symmetry},''
  \href{http://dx.doi.org/10.1103/PhysRevD.83.044042}{{\em Phys. Rev.} {\bf
  D83} (2011)  044042},
\href{http://arxiv.org/abs/1008.4128}{{\tt arXiv:1008.4128 [hep-th]}}.

\bibitem{Evslin:2011vh}
J.~Evslin and T.~Qiu, ``{Closed Timelike Curves in the Galileon Model},''
  \href{http://dx.doi.org/10.1007/JHEP11(2011)032}{{\em JHEP} {\bf 11} (2011)
  032},
\href{http://arxiv.org/abs/1106.0570}{{\tt arXiv:1106.0570 [hep-th]}}.

\bibitem{Curtright:2012gx}
T.~L. Curtright and D.~B. Fairlie, ``{A Galileon Primer},''
\href{http://arxiv.org/abs/1212.6972}{{\tt arXiv:1212.6972 [hep-th]}}.

\bibitem{deFromont:2013iwa}
P.~de~Fromont, C.~de~Rham, L.~Heisenberg, and A.~Matas, ``{Superluminality in
  the Bi- and Multi- Galileon},''
  \href{http://dx.doi.org/10.1007/JHEP07(2013)067}{{\em JHEP} {\bf 07} (2013)
  067},
\href{http://arxiv.org/abs/1303.0274}{{\tt arXiv:1303.0274 [hep-th]}}.

\bibitem{Garcia-Saenz:2013gya}
S.~Garcia-Saenz, ``{Behavior of perturbations on spherically symmetric
  backgrounds in multi-Galileon theory},''
  \href{http://dx.doi.org/10.1103/PhysRevD.87.104012}{{\em Phys. Rev.} {\bf
  D87} (2013) no.~10, 104012},
\href{http://arxiv.org/abs/1303.2905}{{\tt arXiv:1303.2905 [hep-th]}}.

\bibitem{Berezhiani:2013dw}
L.~Berezhiani, G.~Chkareuli, and G.~Gabadadze, ``{Restricted Galileons},''
  \href{http://dx.doi.org/10.1103/PhysRevD.88.124020}{{\em Phys. Rev.} {\bf
  D88} (2013)  124020},
\href{http://arxiv.org/abs/1302.0549}{{\tt arXiv:1302.0549 [hep-th]}}.

\bibitem{Gabadadze:2014gba}
G.~Gabadadze, R.~Kimura, and D.~Pirtskhalava, ``{Vainshtein Solutions Without
  Superluminal Modes},''
  \href{http://dx.doi.org/10.1103/PhysRevD.91.124024}{{\em Phys. Rev.} {\bf
  D91} (2015) no.~12, 124024},
\href{http://arxiv.org/abs/1412.8751}{{\tt arXiv:1412.8751 [hep-th]}}.

\bibitem{Hinterbichler:2009kq}
K.~Hinterbichler, A.~Nicolis, and M.~Porrati, ``{Superluminality in DGP},''
  \href{http://dx.doi.org/10.1088/1126-6708/2009/09/089}{{\em JHEP} {\bf 09}
  (2009)  089},
\href{http://arxiv.org/abs/0905.2359}{{\tt arXiv:0905.2359 [hep-th]}}.

\bibitem{Deser:2015wta}
S.~Deser, A.~Waldron, and G.~Zahariade, ``{Propagation peculiarities of mean
  field massive gravity},''
  \href{http://dx.doi.org/10.1016/j.physletb.2015.07.055}{{\em Phys. Lett.}
  {\bf B749} (2015)  144--148},
\href{http://arxiv.org/abs/1504.02919}{{\tt arXiv:1504.02919 [hep-th]}}.

\bibitem{Creminelli:2014zxa}
P.~Creminelli, M.~Serone, G.~Trevisan, and E.~Trincherini, ``{Inequivalence of
  Coset Constructions for Spacetime Symmetries},''
  \href{http://dx.doi.org/10.1007/JHEP02(2015)037}{{\em JHEP} {\bf 02} (2015)
  037},
\href{http://arxiv.org/abs/1403.3095}{{\tt arXiv:1403.3095 [hep-th]}}.

\bibitem{Mathur:2009hf}
S.~D. Mathur, ``{The Information paradox: A Pedagogical introduction},''
  \href{http://dx.doi.org/10.1088/0264-9381/26/22/224001}{{\em Class. Quant.
  Grav.} {\bf 26} (2009)  224001},
\href{http://arxiv.org/abs/0909.1038}{{\tt arXiv:0909.1038 [hep-th]}}.

\bibitem{Mathur:2005zp}
S.~D. Mathur, ``{The Fuzzball proposal for black holes: An Elementary
  review},'' \href{http://dx.doi.org/10.1002/prop.200410203}{{\em Fortsch.
  Phys.} {\bf 53} (2005)  793--827},
\href{http://arxiv.org/abs/hep-th/0502050}{{\tt arXiv:hep-th/0502050
  [hep-th]}}.

\bibitem{Dvali:2011aa}
G.~Dvali and C.~Gomez, ``{Black Hole's Quantum N-Portrait},''
  \href{http://dx.doi.org/10.1002/prop.201300001}{{\em Fortsch. Phys.} {\bf 61}
  (2013)  742--767},
\href{http://arxiv.org/abs/1112.3359}{{\tt arXiv:1112.3359 [hep-th]}}.

\bibitem{Almheiri:2012rt}
A.~Almheiri, D.~Marolf, J.~Polchinski, and J.~Sully, ``{Black Holes:
  Complementarity or Firewalls?},''
  \href{http://dx.doi.org/10.1007/JHEP02(2013)062}{{\em JHEP} {\bf 02} (2013)
  062},
\href{http://arxiv.org/abs/1207.3123}{{\tt arXiv:1207.3123 [hep-th]}}.

\bibitem{Dvali:2010jz}
G.~Dvali, G.~F. Giudice, C.~Gomez, and A.~Kehagias, ``{UV-Completion by
  Classicalization},'' \href{http://dx.doi.org/10.1007/JHEP08(2011)108}{{\em
  JHEP} {\bf 08} (2011)  108},
\href{http://arxiv.org/abs/1010.1415}{{\tt arXiv:1010.1415 [hep-ph]}}.

\bibitem{Keltner:2015xda}
L.~Keltner and A.~J. Tolley, ``{UV properties of Galileons: Spectral
  Densities},''
\href{http://arxiv.org/abs/1502.05706}{{\tt arXiv:1502.05706 [hep-th]}}.

\bibitem{Klein:2015iud}
R.~Klein, M.~Ozkan, and D.~Roest, ``{Galileons as the Scalar Analogue of
  General Relativity},''
  \href{http://dx.doi.org/10.1103/PhysRevD.93.044053}{{\em Phys. Rev.} {\bf
  D93} (2016) no.~4, 044053},
\href{http://arxiv.org/abs/1510.08864}{{\tt arXiv:1510.08864 [hep-th]}}.

\bibitem{Carroll:2004st}
S.~M. Carroll, {\em {Spacetime and geometry: An introduction to general
  relativity}}.
\newblock 2004.
\newblock
\url{http://www.slac.stanford.edu/spires/find/books/www?cl=QC6:C37:2004}.
\newblock

\bibitem{Misner:1974qy}
C.~W. Misner, K.~S. Thorne, and J.~A. Wheeler, {\em {Gravitation}}.
\newblock W. H. Freeman, San Francisco,
1973.
\newblock

\bibitem{ArkaniHamed:2006dz}
N.~Arkani-Hamed, L.~Motl, A.~Nicolis, and C.~Vafa, ``{The String landscape,
  black holes and gravity as the weakest force},''
  \href{http://dx.doi.org/10.1088/1126-6708/2007/06/060}{{\em JHEP} {\bf 0706}
  (2007)  060},
\href{http://arxiv.org/abs/hep-th/0601001}{{\tt arXiv:hep-th/0601001
  [hep-th]}}.

\bibitem{Camanho:2014apa}
X.~O. Camanho, J.~D. Edelstein, J.~Maldacena, and A.~Zhiboedov, ``{Causality
  Constraints on Corrections to the Graviton Three-Point Coupling},''
\href{http://arxiv.org/abs/1407.5597}{{\tt arXiv:1407.5597 [hep-th]}}.

\bibitem{Benakli:2015qlh}
K.~Benakli, S.~Chapman, L.~DarmŽ, and Y.~Oz, ``{On Swift Gravitons},''
\href{http://arxiv.org/abs/1512.07245}{{\tt arXiv:1512.07245 [hep-th]}}.

\bibitem{Chen:2015bva}
J.~J.~S. Chen and Q.~Pan, ``{Geometric optics for a coupling model of the
  electromagnetic and gravitational fields},''
\href{http://arxiv.org/abs/1510.03316}{{\tt arXiv:1510.03316 [gr-qc]}}.

\bibitem{PenroseCausality}
R.~Penrose, ``{On Schwarzschild Causality-- A Problem for ``Lorentz Covariant"
  General Relativity},'' {\em Essays in General Relativity: A Festschrift for
  Abraham Taub} (1980)  .

\bibitem{Gao:2000ga}
S.~Gao and R.~M. Wald, ``{Theorems on gravitational time delay and related
  issues},'' \href{http://dx.doi.org/10.1088/0264-9381/17/24/305}{{\em
  Class.Quant.Grav.} {\bf 17} (2000)  4999--5008},
\href{http://arxiv.org/abs/gr-qc/0007021}{{\tt arXiv:gr-qc/0007021 [gr-qc]}}.

\bibitem{Majumdar:1947eu}
S.~D. Majumdar, ``{A class of exact solutions of Einstein's field equations},''
\href{http://dx.doi.org/10.1103/PhysRev.72.390}{{\em Phys. Rev.} {\bf 72}
  (1947)  390--398}.

\bibitem{Papaetrou:1947ib}
A.~Papaetrou, ``{A Static solution of the equations of the gravitational field
  for an arbitrary charge distribution},''
{\em Proc. Roy. Irish Acad.(Sect. A)} {\bf A51} (1947)  191--204.

\bibitem{Hartle:1972ya}
J.~B. Hartle and S.~W. Hawking, ``{Solutions of the Einstein-Maxwell equations
  with many black holes},''
\href{http://dx.doi.org/10.1007/BF01645696}{{\em Commun. Math. Phys.} {\bf 26}
  (1972)  87--101}.

\bibitem{Schwartz:2013pla}
M.~D. Schwartz, {\em {Quantum Field Theory and the Standard Model}}.
\newblock Cambridge University Press, 2014.
\newblock
\url{http://www.cambridge.org/us/academic/subjects/physics/theoretical-physics-and-mathematical-physics/quantum-field-theory-and-standard-model}.
\newblock

\bibitem{Akhmedov:2015xwa}
E.~T. Akhmedov, H.~Godazgar, and F.~K. Popov, ``{Hawking radiation and
  secularly growing loop corrections},''
\href{http://arxiv.org/abs/1508.07500}{{\tt arXiv:1508.07500 [hep-th]}}.

\bibitem{KamenevBook}
A.~Kamenev, {\em {Field Theory of Non-Equilibrium Systems}}.
\newblock Cambridge University Press, 2011.

\bibitem{Weinberg:2005vy}
S.~Weinberg, ``{Quantum contributions to cosmological correlations},''
  \href{http://dx.doi.org/10.1103/PhysRevD.72.043514}{{\em Phys. Rev.} {\bf
  D72} (2005)  043514},
\href{http://arxiv.org/abs/hep-th/0506236}{{\tt arXiv:hep-th/0506236
  [hep-th]}}.

\bibitem{Duff:1973zz}
M.~J. Duff, ``{Quantum Tree Graphs and the Schwarzschild Solution},''
\href{http://dx.doi.org/10.1103/PhysRevD.7.2317}{{\em Phys. Rev.} {\bf D7}
  (1973)  2317--2326}.

\bibitem{Duff:1974ud}
M.~J. Duff, ``{Quantum corrections to the schwarzschild solution},''
\href{http://dx.doi.org/10.1103/PhysRevD.9.1837}{{\em Phys. Rev.} {\bf D9}
  (1974)  1837--1839}.

\bibitem{Gibbons:1975kk}
G.~W. Gibbons, ``{Vacuum Polarization and the Spontaneous Loss of Charge by
  Black Holes},''
\href{http://dx.doi.org/10.1007/BF01609829}{{\em Commun. Math. Phys.} {\bf 44}
  (1975)  245--264}.

\bibitem{ToAppear}
G.~Goon and K.~Hinterbichler, ``{In Preparation},''.

\bibitem{'tHooft:1974bx}
G.~'t~Hooft and M.~J.~G. Veltman, ``{One loop divergencies in the theory of
  gravitation},''
{\em Annales Poincare Phys. Theor.} {\bf A20} (1974)  69--94.

\bibitem{Goroff:1985th}
M.~H. Goroff and A.~Sagnotti, ``{The Ultraviolet Behavior of Einstein
  Gravity},''
\href{http://dx.doi.org/10.1016/0550-3213(86)90193-8}{{\em Nucl. Phys.} {\bf
  B266} (1986)  709}.

\bibitem{Capper:1974ed}
D.~M. Capper, M.~J. Duff, and L.~Halpern, ``{Photon corrections to the graviton
  propagator},''
\href{http://dx.doi.org/10.1103/PhysRevD.10.461}{{\em Phys. Rev.} {\bf D10}
  (1974)  461--467}.

\bibitem{Donoghue:2001qc}
J.~F. Donoghue, B.~R. Holstein, B.~Garbrecht, and T.~Konstandin, ``{Quantum
  corrections to the Reissner-Nordstrom and Kerr-Newman metrics},''
  \href{http://dx.doi.org/10.1016/S0370-2693(02)01246-7}{{\em Phys. Lett.} {\bf
  B529} (2002)  132--142},
\href{http://arxiv.org/abs/hep-th/0112237}{{\tt arXiv:hep-th/0112237
  [hep-th]}}.

\bibitem{BjerrumBohr:2002ks}
N.~E.~J. Bjerrum-Bohr, J.~F. Donoghue, and B.~R. Holstein, ``{Quantum
  corrections to the Schwarzschild and Kerr metrics},''
  \href{http://dx.doi.org/10.1103/PhysRevD.68.084005,
  10.1103/PhysRevD.71.069904}{{\em Phys. Rev.} {\bf D68} (2003)  084005},
  \href{http://arxiv.org/abs/hep-th/0211071}{{\tt arXiv:hep-th/0211071
  [hep-th]}}.
[Erratum: Phys. Rev.D71,069904(2005)].

\bibitem{Kirilin:2006en}
G.~G. Kirilin, ``{Quantum corrections to the Schwarzschild metric and
  reparametrization transformations},''
  \href{http://dx.doi.org/10.1103/PhysRevD.75.108501}{{\em Phys. Rev.} {\bf
  D75} (2007)  108501},
\href{http://arxiv.org/abs/gr-qc/0601020}{{\tt arXiv:gr-qc/0601020 [gr-qc]}}.

\bibitem{Burgess:2003jk}
C.~P. Burgess, ``{Quantum gravity in everyday life: General relativity as an
  effective field theory},'' \href{http://dx.doi.org/10.12942/lrr-2004-5}{{\em
  Living Rev. Rel.} {\bf 7} (2004)  5--56},
\href{http://arxiv.org/abs/gr-qc/0311082}{{\tt arXiv:gr-qc/0311082 [gr-qc]}}.

\bibitem{Donoghue:1994dn}
J.~F. Donoghue, ``{General relativity as an effective field theory: The leading
  quantum corrections},''
  \href{http://dx.doi.org/10.1103/PhysRevD.50.3874}{{\em Phys. Rev.} {\bf D50}
  (1994)  3874--3888},
\href{http://arxiv.org/abs/gr-qc/9405057}{{\tt arXiv:gr-qc/9405057 [gr-qc]}}.

\bibitem{Abbott:1981ke}
L.~F. Abbott, ``{Introduction to the Background Field Method},''
{\em Acta Phys. Polon.} {\bf B13} (1982)  33.

\bibitem{Dalvit:1997yc}
D.~A.~R. Dalvit and F.~D. Mazzitelli, ``{Geodesics, gravitons and the gauge
  fixing problem},'' \href{http://dx.doi.org/10.1103/PhysRevD.56.7779}{{\em
  Phys. Rev.} {\bf D56} (1997)  7779--7787},
\href{http://arxiv.org/abs/hep-th/9708102}{{\tt arXiv:hep-th/9708102
  [hep-th]}}.

\bibitem{Barvinsky:1985an}
A.~O. Barvinsky and G.~A. Vilkovisky, ``{The Generalized Schwinger-Dewitt
  Technique in Gauge Theories and Quantum Gravity},''
\href{http://dx.doi.org/10.1016/0370-1573(85)90148-6}{{\em Phys. Rept.} {\bf
  119} (1985)  1--74}.

\bibitem{Andreassen:2014eha}
A.~Andreassen, W.~Frost, and M.~D. Schwartz, ``{Consistent Use of Effective
  Potentials},'' \href{http://dx.doi.org/10.1103/PhysRevD.91.016009}{{\em Phys.
  Rev.} {\bf D91} (2015) no.~1, 016009},
\href{http://arxiv.org/abs/1408.0287}{{\tt arXiv:1408.0287 [hep-ph]}}.

\bibitem{Nielsen:1975fs}
N.~K. Nielsen, ``{On the Gauge Dependence of Spontaneous Symmetry Breaking in
  Gauge Theories},''
\href{http://dx.doi.org/10.1016/0550-3213(75)90301-6}{{\em Nucl. Phys.} {\bf
  B101} (1975)  173--188}.

\bibitem{Fukuda:1975di}
R.~Fukuda and T.~Kugo, ``{Gauge Invariance in the Effective Action and
  Potential},''
\href{http://dx.doi.org/10.1103/PhysRevD.13.3469}{{\em Phys. Rev.} {\bf D13}
  (1976)  3469}.

\bibitem{Aitchison:1983ns}
I.~J.~R. Aitchison and C.~M. Fraser, ``{Gauge Invariance and the Effective
  Potential},''
\href{http://dx.doi.org/10.1016/0003-4916(84)90209-4}{{\em Annals Phys.} {\bf
  156} (1984)  1}.

\bibitem{BjerrumBohr:2002kt}
N.~E.~J. Bjerrum-Bohr, J.~F. Donoghue, and B.~R. Holstein, ``{Quantum
  gravitational corrections to the nonrelativistic scattering potential of two
  masses},'' \href{http://dx.doi.org/10.1103/PhysRevD.71.069903,
  10.1103/PhysRevD.67.084033}{{\em Phys. Rev.} {\bf D67} (2003)  084033},
  \href{http://arxiv.org/abs/hep-th/0211072}{{\tt arXiv:hep-th/0211072
  [hep-th]}}.
[Erratum: Phys. Rev.D71,069903(2005)].

\bibitem{Khriplovich:2002bt}
I.~B. Khriplovich and G.~G. Kirilin, ``{Quantum power correction to the Newton
  law},'' \href{http://dx.doi.org/10.1134/1.1537290}{{\em J. Exp. Theor. Phys.}
  {\bf 95} (2002)  981--986}, \href{http://arxiv.org/abs/gr-qc/0207118}{{\tt
  arXiv:gr-qc/0207118 [gr-qc]}}.
[Zh. Eksp. Teor. Fiz.95,1139(2002)].

\bibitem{Holstein:2008sy}
B.~R. Holstein and A.~Ross, ``{Long Distance Effects in Mixed
  Electromagnetic-Gravitational Scattering},''
\href{http://arxiv.org/abs/0802.0717}{{\tt arXiv:0802.0717 [hep-ph]}}.

\bibitem{BjerrumBohr:2002sx}
N.~E.~J. Bjerrum-Bohr, ``{Leading quantum gravitational corrections to scalar
  QED},'' \href{http://dx.doi.org/10.1103/PhysRevD.66.084023}{{\em Phys. Rev.}
  {\bf D66} (2002)  084023},
\href{http://arxiv.org/abs/hep-th/0206236}{{\tt arXiv:hep-th/0206236
  [hep-th]}}.

\bibitem{Adler:1971wn}
S.~L. Adler, ``{Photon splitting and photon dispersion in a strong magnetic
  field},''
\href{http://dx.doi.org/10.1016/0003-4916(71)90154-0}{{\em Annals Phys.} {\bf
  67} (1971)  599--647}.

\bibitem{Harding:2006qn}
A.~K. Harding and D.~Lai, ``{Physics of Strongly Magnetized Neutron Stars},''
  \href{http://dx.doi.org/10.1088/0034-4885/69/9/R03}{{\em Rept. Prog. Phys.}
  {\bf 69} (2006)  2631},
\href{http://arxiv.org/abs/astro-ph/0606674}{{\tt arXiv:astro-ph/0606674
  [astro-ph]}}.

\bibitem{Ruffini:2013hia}
R.~Ruffini, Y.-B. Wu, and S.-S. Xue, ``{Einstein-Euler-Heisenberg Theory and
  charged black holes},''
  \href{http://dx.doi.org/10.1103/PhysRevD.88.085004}{{\em Phys. Rev.} {\bf
  D88} (2013)  085004},
\href{http://arxiv.org/abs/1307.4951}{{\tt arXiv:1307.4951 [hep-th]}}.

\bibitem{Yajima:2000kw}
H.~Yajima and T.~Tamaki, ``{Black hole solutions in Euler-Heisenberg theory},''
  \href{http://dx.doi.org/10.1103/PhysRevD.63.064007}{{\em Phys. Rev.} {\bf
  D63} (2001)  064007},
\href{http://arxiv.org/abs/gr-qc/0005016}{{\tt arXiv:gr-qc/0005016 [gr-qc]}}.

\bibitem{Kats:2006xp}
Y.~Kats, L.~Motl, and M.~Padi, ``{Higher-order corrections to mass-charge
  relation of extremal black holes},''
  \href{http://dx.doi.org/10.1088/1126-6708/2007/12/068}{{\em JHEP} {\bf 0712}
  (2007)  068},
\href{http://arxiv.org/abs/hep-th/0606100}{{\tt arXiv:hep-th/0606100
  [hep-th]}}.

\bibitem{Marolf:2013dba}
D.~Marolf and J.~Polchinski, ``{Gauge/Gravity Duality and the Black Hole
  Interior},'' \href{http://dx.doi.org/10.1103/PhysRevLett.111.171301}{{\em
  Phys. Rev. Lett.} {\bf 111} (2013)  171301},
\href{http://arxiv.org/abs/1307.4706}{{\tt arXiv:1307.4706 [hep-th]}}.

\bibitem{Heisenberg:1935qt}
W.~Heisenberg and H.~Euler, ``{Consequences of Dirac's theory of positrons},''
  \href{http://dx.doi.org/10.1007/BF01343663}{{\em Z. Phys.} {\bf 98} (1936)
  714--732},
\href{http://arxiv.org/abs/physics/0605038}{{\tt arXiv:physics/0605038
  [physics]}}.

\end{thebibliography}\endgroup

\end{document}